\pdfoutput=1

\documentclass[%
 aip,
 amsmath,amssymb,
 reprint,%
]{revtex4-1}
\usepackage[normalem]{ulem}
\usepackage{filecontents}
\usepackage[version=3]{mhchem}
\usepackage{color} 
\usepackage{siunitx}

\newcommand{\mr}[1]{\mathrm{#1}}
\newcommand{\bv}[1]{\mathbf{#1}}
\usepackage{multirow}
\usepackage{makecell}
\usepackage{subcaption}
\usepackage{graphicx}
\usepackage{dcolumn}
\usepackage{bm}

\usepackage[utf8]{inputenc}
\usepackage[T1]{fontenc}
\usepackage{mathptmx}
\usepackage{xr}
\makeatletter
\newcommand*{\addFileDependency}[1]{
  \typeout{(#1)}
  \@addtofilelist{#1}
  \IfFileExists{#1}{}{\typeout{No file #1.}}
}
\makeatother

\newcommand*{\myexternaldocument}[1]{%
    \externaldocument{#1}%
    \addFileDependency{#1.tex}%
    \addFileDependency{#1.aux}%
}
\myexternaldocument{./xsupp} 
\usepackage{prettyref}
\newrefformat{supp-fig}{S\ref{#1}}
\newrefformat{supp-tbl}{S\ref{#1}}
\DeclareUnicodeCharacter{2212}{-}

\begin{document}

\preprint{AIP/123-QED}

\title[]{Role of Water Model on Ion Dissociation at Ambient Conditions}

\author{Alec Wills}
\email{alec.wills@stonybrook.edu}

\author{Marivi Fern\'{a}ndez-Serra}%
 \email{maria.fernandez-serra@stonybrook.edu}
\affiliation{ 
Physics and Astronomy Department, Stony Brook University. Stony Brook, New York 11794-3800, United States
}%
\affiliation{Institute for Advanced Computational Science, Stony Brook, New York 11794-3800, United States}

\date{29 January 2021}

\begin{abstract}

We study ion pair dissociation in water at ambient conditions
using a combination of classical and ab initio approaches.
The goal of this study is to disentangle the sources of discrepancy observed in computed potentials of mean force.
In particular we aim to understand why some models favor
the stability of solvent-separated ion pairs versus
contact ion pairs.
We found that some observed differences can be explained
by non-converged simulation parameters.
However, we also unveil that for some models, small changes in the solution density can have  significant effects on modifying the equilibrium balance between the two configurations.
We conclude that the thermodynamic stability of contact and solvent-separated
ion pairs is very sensitive to the dielectric properties of the underlying simulation model.
In general, classical models are very robust on providing a
similar estimation of the contact ion pair stability, while
this is much more variable in density functional theory-based models.
The barrier to transition from solvent-separated to contact ion pair is fundamentally dependent on the balance between electrostatic potential energy and entropy. 
This reflects the importance of water intra- and inter-molecular polarizability in obtaining an accurate description of the screened ion-ion interactions.

\end{abstract}

\maketitle

\section{\label{sec:level1} Introduction}

Electrolyte solutions are ubiquitous in nature, and are of great importance in many biological and industrial processes.
As such, it is important to understand the nature of ion solvation and the effects that the ions have on the solvent.
In so doing, more effective and efficient applications can make use of accurate solvation knowledge to improve a variety of already-existing technologies, including drug delivery mechanisms and crystallization methods.

Despite their prevalence, it is not completely understood the role that ions have in regards to structural changes to solvent particles, in particular water.
X-ray spectroscopic results show that the alkali cations are generally chaotropic (structure-breaking), having effects on water similar to that of increasing the solution temperature.\cite{xraychaoWaluyo2014}
Neutron diffraction studies similarly show chaotropic effects, where the ions act by collapsing the second hydration shell inwards as pressurizing the solution would do.\cite{khalideSoper2006,na_k_cl_Mancinelli2007}
However, vibrational relaxation studies show increased mode lifetimes for Li$^+$ and Na$^+$ cations, indicating some level of cosmotropic (structure-making) effects for smaller alkali cations.\cite{nacosmoKropman2004}
A better understanding of the explicit solvation process will be beneficial in explaining the apparent discrepancies between experimental interpretations, and yield insight into what exactly structure ``making" and ``breaking" entails.

As the solvent is a critical component in any electrolyte solution, the choice of model for water alone is also a factor that will influence any simulated dynamics.
Classical, empirically derived potentials for water molecules have been shown to be inadequate in reproducing experimental results for nonpolar, charged solute molecules, and the choice of charge localization in the model results in affects orientational dynamics in the solvation process.\cite{lonepairRemsing2018}
X-ray and neutron diffraction studies disagree on the \ce{Na-O} distance of hydrated sodium ions by 0.1 \AA, and \textit{ab initio} simulations have a wide range of predictions for this value that typically predict values an additional 0.1 \AA\ longer than experimental results.\cite{naradFifen2019}
Coordination number has an effect on these predicted distances, from which ionic radii are estimated.
Charge localization in classical water models will affect stable configuration geometry, thereby affecting coordination number.
The inter-connectivity of all these model dependent processes lead to an ever more complex web of model choices and parameters that prevents reproducibility in simulations and can obscure physical results.

\begin{figure}
    \centering
    \begin{subfigure}[b]{0.3\textwidth}
    \includegraphics[scale=0.25]{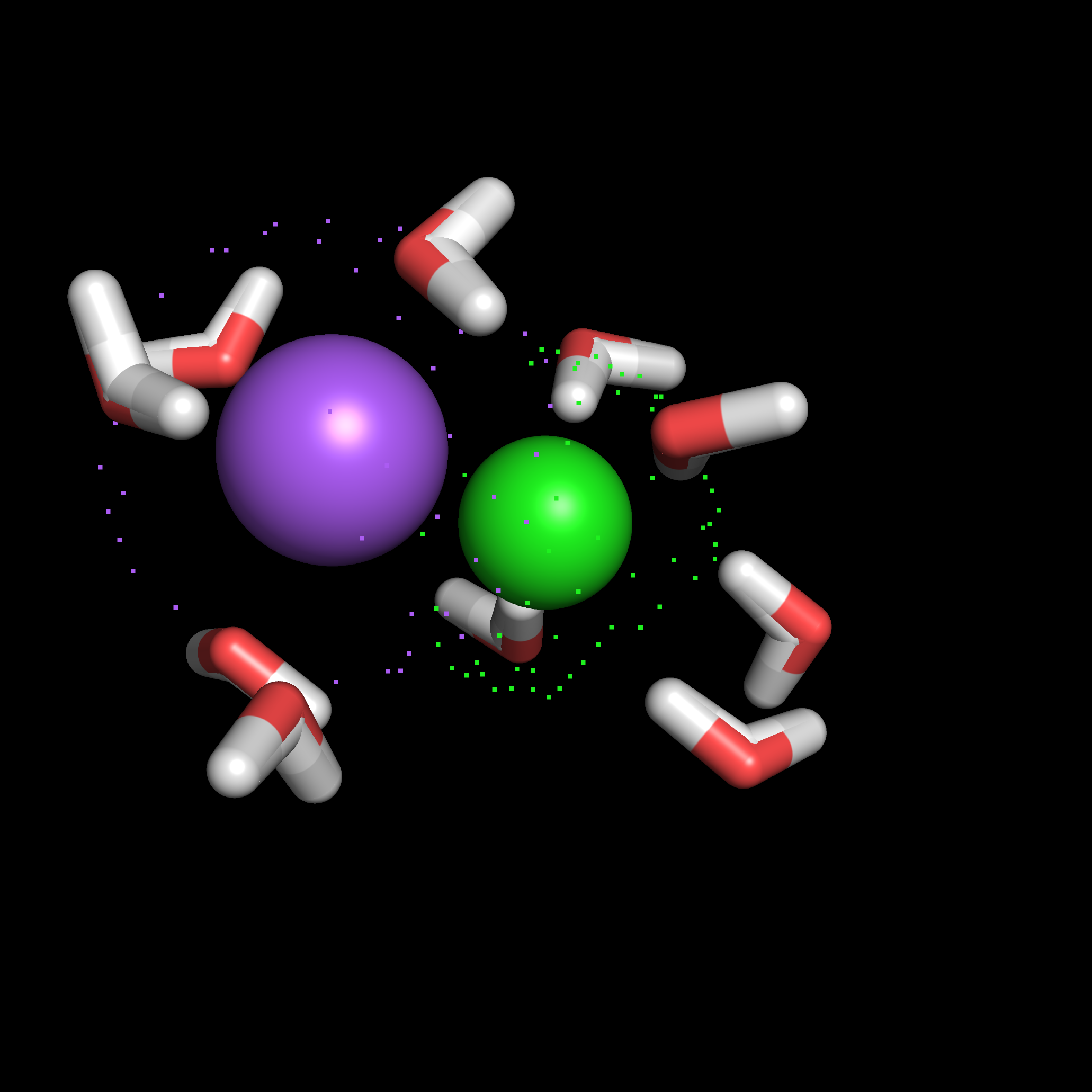}
        \caption{}
        \label{fig:solvstatescip}
    \end{subfigure}
    ~ 
    \begin{subfigure}[b]{0.3\textwidth}
    \includegraphics[scale=0.25]{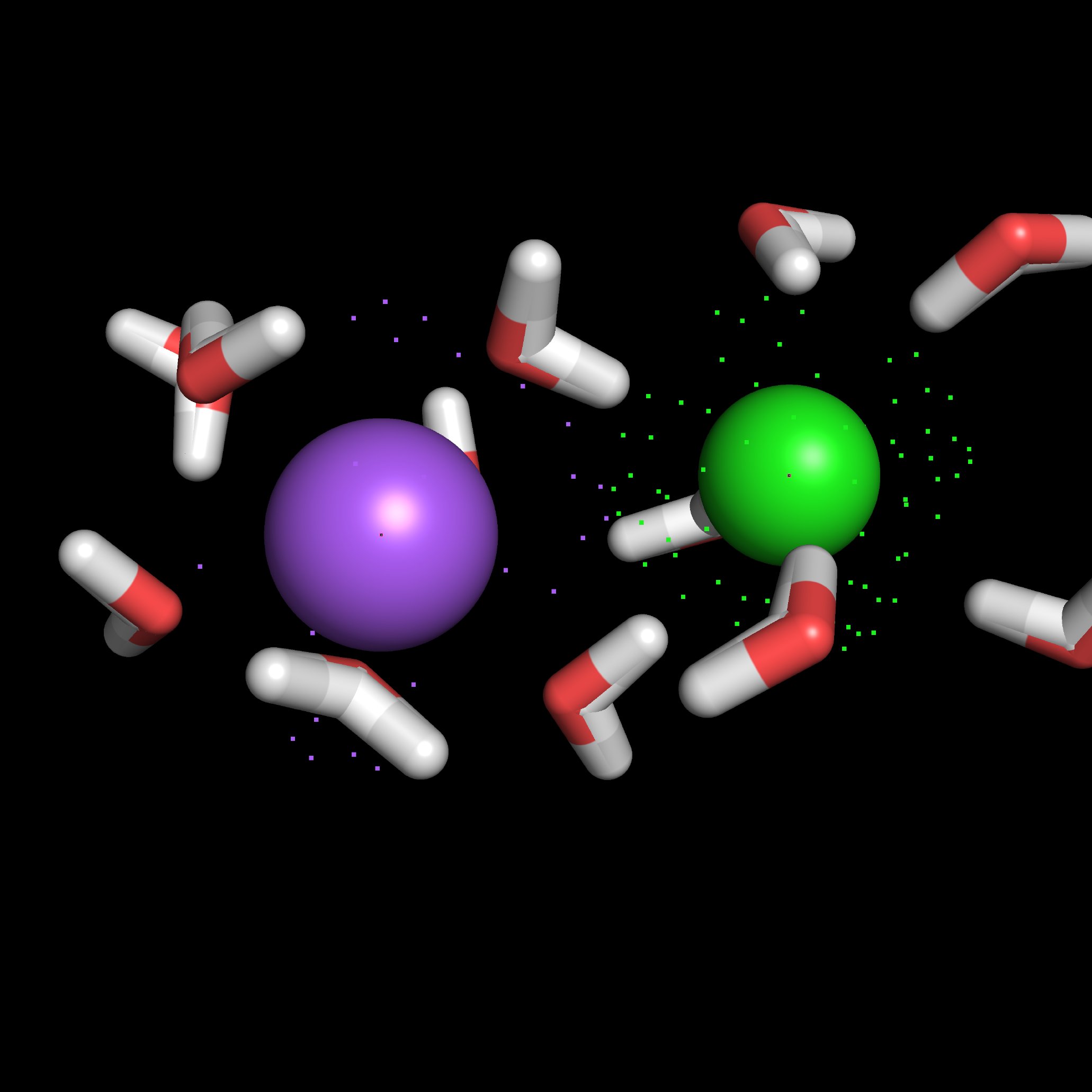}
        \caption{}
        \label{fig:solvstatesssip}
    \end{subfigure}
    ~
    \begin{subfigure}[b]{0.3\textwidth}
    \includegraphics[scale=0.25]{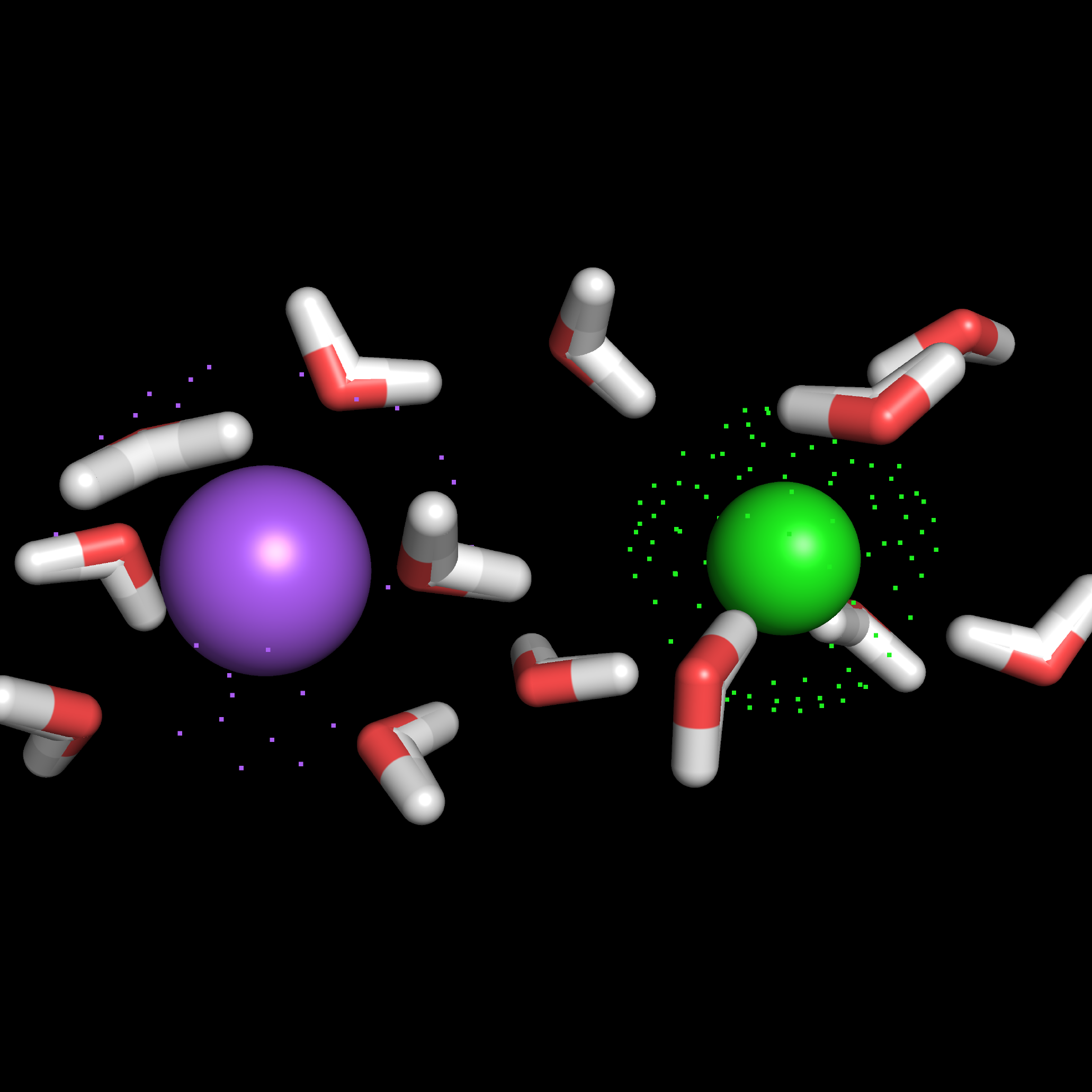}
        \caption{}
        \label{fig:solvstatesdiss}
    \end{subfigure}
    \caption{\textbf{(a)} The contact-ion pair (CIP) solvation state. Here, the ions have no intermediate buffer molecules between them. \textbf{(b)} The solvent-separated ion pair (SSIP) state, where the ions share a solvent hydration shell. \textbf{(c)} The dissociated ion state, where the ions each have their own hydration shells.}\label{fig:solvstates}
\end{figure}

As it is the \textit{de facto} method for examining system characteristics along a variable reaction coordinate, potentials of mean force (PMFs) are widely used.
In particular, \ce{NaCl} is frequently used as the model system due to the simplicity (monovalent constituent ions) and prevalence of the salt and its components in biological, industrial, and civilian usage.
For NaCl PMFs in particular, there are three main features that characterize the solvation process, shown in Fig.~\ref{fig:solvstates}.
The contact ion pair (CIP, shown in Fig.~\ref{fig:solvstatescip}), is where the two ions are Coulombically bound and are surrounded by solvent molecules.
The solvent-separated ion pair (SSIP, shown in Fig.~\ref{fig:solvstatesssip}), is a stable state of solvation wherein the ions share a coordination shell of solvent between them.
A transition barrier exists between these two states, typically attributed to the work required to separate (bring together) the ions and insert (remove) the intermediate solvent molecules.
The height of this barrier allows us to characterize the relative stability of the separate solvation states, as we do in Tab.~\ref{tbl:hdldaimd}.

There are a variety of studies that have been done to reconstruct the \ce{NaCl} PMF with the reaction coordinate as the inter-ionic distance in different classical force fields.\cite{pmf1charmmggahfbKhavrutskii2008,pmf2_classical_cpmd_Timko2010,pmf3_classical_ihs_qmefp_Ghosh2013,pmf4_classical_scancpmd_Yao2018,pmf5_classical_marcusthry_Roy2017}
\textit{Ab initio} treatments of systems vary, ranging from Car-Parrinello molecular dynamics (CPMD)\cite{cpmdCar1985} treatments of the nuclear motions,\cite{pmf2_classical_cpmd_Timko2010} to QM/EFP dynamics where fixed geometry fragments are assigned electrostatic parameters based on \textit{ab initio} simulations,\cite{pmf3_classical_ihs_qmefp_Ghosh2013} to a mix of both BOMD and CPMD for different XC functionals in the same study.\cite{pmf4_classical_scancpmd_Yao2018}
Two-dimensional PMFs have also been examined, using coordination number as the second reaction coordinate.\cite{pmf5_classical_marcusthry_Roy2017, galli_Zhang2020}
Classical force-field generated PMFs typically predict more stable CIP states than the SSIP state that precedes dissociation (shown in Fig.~\ref{fig:solvstatesdiss}), although some conditions show the SSIP to be more favorable.\textcolor{black}{\cite{galli_Zhang2020,response1_Guardia1991}}
The results are not as clear for \textit{ab initio} simulations, with results depending on the choice of XC functional, amongst other parameters.\cite{galli_Zhang2020, pmf4_classical_scancpmd_Yao2018}

Beyond the energetics, structural information is important to understand the physical processes that occur during the solvation process. 
Initial neutron diffraction analysis led to results indicating minimal affects beyond the first ion hydration shell,\cite{khalideSoper2006} while later results showed that ion effects extend beyond the first solvation shell.\cite{perturbgood_Mancinelli2007}
Even further results suggested that cosmo/chaotropic effects have minimal bearing on molecular interactions between the solute and solvent, despite the changes in solvent structure.\cite{perturbbad_Mancinelli2007} 
X-ray absorption studies have indicated cationic destabilization of the hydrogen bond network indicative of extended cationic effects on the solvent, while anionic effects from Cl$^-$ indicated insignificant structural effects.\cite{nashell1_Waluyo2011, xraychaoWaluyo2014}
However, GGA-level \textit{ab initio} simulations have suggested that Na$^+$ affects the solvent only locally (not beyond the first solvation shell), while Cl$^-$ has longer-range effects,\cite{naclrdfs_Gaiduk2017} directly contrasting with interpretations of experimental results.
Moreover, nuclear quantum effects have been shown to be an important inclusion to yield more accurate \ce{Cl-} solvation structures,\cite{nqestructcl_Xu2021} and ionic polarizability also plays a significant role in whether solvation shells have the correct amount level of structure.\cite{polarzionstruct_DelloStritto2020}

In our study, we examine the effects of solution density on the solvation state stability, to better understand how the solvent thermodynamic properties, which
are strongly dependent on the underlying solvent model, affect the solvation process.
We address this question in a two-step approach. 
Firstly, we investigate the PMFs obtained by different classical models for water.
We evaluate separately the differences that can be associated to convergence and methodological approaches from those intrinsic to the model used to describe the water-water and water-ion interactions.
Then, using convergence criteria established within the first half of the work, we study how a first principles treatment of the interactions modifies what can be learned from purely classical models.
This is done within the formalism of density functional theory (DFT),
using two different approximations to the exchange and correlation (XC) functional.
Our choices are motivated by the need to compare to previously published results, aiming to make a fair assessment of similar, but not identical methodological approaches.
However our choices are fundamentally driven because these two different \textit{ab initio} models predict a rather distinct structural and dynamical description of liquid water.
Hence, we can evaluate whether a model which predicts
a low density liquid-type order of liquid water at normal conditions  like PBE\cite{pbe}
is more or less sensitive to ionic chaotropic/cosmotropic effects than a model which favors high density liquid-order type, like the van der Waals (vdW) XC functional vdW-cx.\cite{bh,marivi_dpps_vdwrhoeq_Fritz2016}

Finally in the last part of the manuscript we discuss the fundamental differences observed between classical and DFT models. In the case of classical models, we analyze the entropic contributions to the free energy of solvation.
We show that the contact ion pair potential depth is hardly dependent on the classical model for water or in small changes on the solvent density when using the same ion parameters.
However the solvent-separated ion pair is, mostly due to small entropic
differences between the models.

This is not the case for DFT simulations with different XC potentials.
Here both CIP and SSIP potential depths depend on the model and
solvent conditions. 
We argue that the reasons are both enthalpic and entropic and
fundamentally driven by changes in the polarizability and dielectric properties of the solvent water molecules.

\section{\label{methodology} Methodology}

\begin{figure}[!ht]
    \centering
    \includegraphics[scale=0.5]{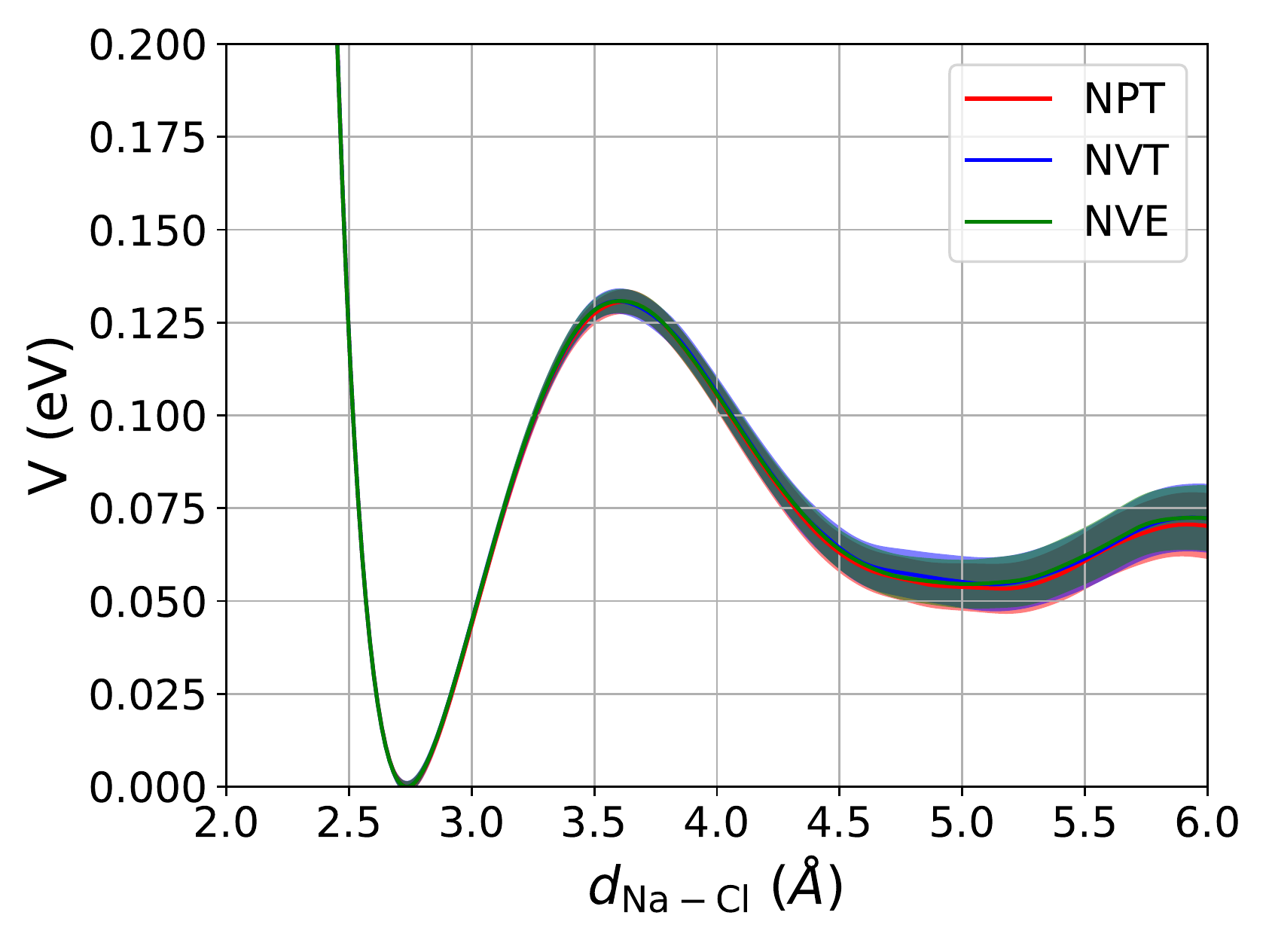}
    \caption{Testing the convergence of the potential of mean force in different simulation ensembles with TIP4P water and the default OPLS-AA ion parameters in \texttt{GROMACS}. The agreement between the different ensembles is almost exact, with a CIP depth of $\sim 125\ \si{meV}$. Error estimates come from the bootstrapping techniques implemented in \texttt{g-wham}. Temperatures in the Boltzmann factors for the NVE trajectories are the average temperatures of the simulation.}
    \label{fig:convnptnvtnve}
\end{figure}

\begin{figure}
    \centering
    \includegraphics[scale=0.5]{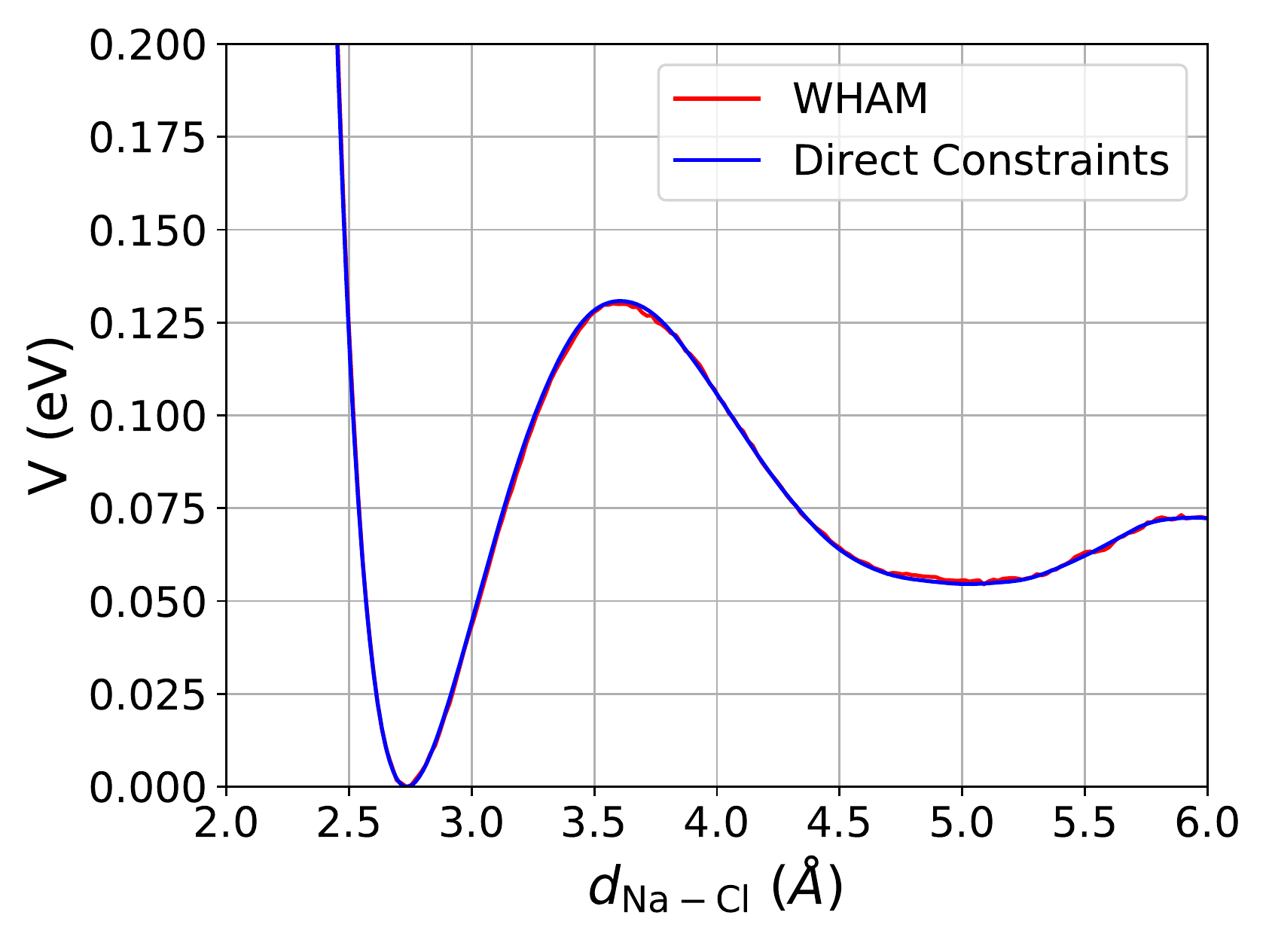}
    \caption{Testing the PMF convergence through different sampling methods, here shown with umbrella sampling and directly integrating holonomic constraints. The exact agreement allows for more controlled sampling of the configuration space in our \textit{ab initio} simulations. The solution used the TIP4P water model and default OPLS-AA ion parameters from \texttt{GROMACS} trajectories.}
    \label{fig:convwhamdir}
\end{figure}

\subsection{\label{methodology:pmf} Potential of Mean Force}
The potential of mean force (PMF) between two particles $U(r)$ is defined as:\cite{pmfdefKirkwood1935}
 \begin{equation}
    U(r) = -kT\ln g(r), \label{eq:pmfdef}
\end{equation}
where $g(r)$ is the inter-particle radial distribution function.
In simulations, the evaluation of $g(r)$ for charged particles in an implicit polar solvent like water is the computational bottleneck for calculating $U(r)$, because infinitely long trajectories are needed to completely explore the radial configuration space.
In particular, electrostatic interactions make the hydration environments of some ionic species long-lasting,\cite{dynamresLi2008} requiring biasing methods to force the exploration.
One popular method in literature is to use umbrella sampling and reconstruct the PMF by weighted-histogram analysis.\cite{whamKumar1992}
Another popular alternative to this procedure is to directly evaluate the PMF by integrating constraint forces between the two particles, along a discrete grid of interparticle distances.
Both methods map the infinitely long trajectory problem into a finite set of finite time molecular dynamic simulations of constrained particles.

When sampling with constraints, they must be accounted for in the derivation of the free energy of separation between the particles.
Drawing from the statistical mechanics identity relating the Helmholtz free energy $F$ to the canonical partition function $Z$ $$F=-kT\ln Z,$$ we can suppose a different free energy $W$ is related to a constrained probability distribution $P$ by
\begin{equation}
    W(r_o) = -kT\ln P_{\mr{r_o}}(r).
\end{equation}
Here, $P_{\mr{r_o}}(r)$ is the conditional statistical average given by \begin{align}
   P_{\mr{r_o}}(r) &= \left\langle \delta(|\bv{r}|-r_o) \right\rangle \nonumber\\
   &= \frac{1}{Z}\int d^{3N}r\ e^{-\beta \Phi(\{\bv{r}_i\})} \delta\left( |\bv{r}_1-\bv{r}_2|-r_o\right). \label{eq:constrprob}
\end{align}
This new definition is necessary because in order to use different methods of PMF calculation, we must force the system into thermodynamically unfavorable configurations through $P$.
We note that the definition in Eq. \eqref{eq:constrprob} implies that $P_{\mr{r_o}}(r)$ is a constrained partition function, measuring the likelihood of the interionic distance $r$ taking a value of $r_o$.
Constraints are assumed to only depend on particle positions, not momenta, hence the integration over only position-space.

In following through with relating the constrained free energy $W(r)$ to the PMF $U(r)$, the constraints bring about a correction factor $2kT/r$ that must be used to amend the average forces, yielding \begin{equation}
    \frac{dU(r)}{dr} = -f(r) + \frac{2kT}{r} \label{eq:dUdr},
\end{equation} which is integrated to yield the PMF.
This correction factor can be interpreted as an energy correction added back on to the free energy $W(r)$, arising from the constraints reducing the overall entropy of the available phase space of the system.\cite{pmfderivLi2007}
Of further note is that this correction factor need only be applied to our \textit{ab initio} simulations, where we sample the physical forces on the ions (i.e. -$\partial_r \Phi(r)$).
Since either alternative PMF construction methods are used in the classical simulations (e.g. the WHAM method), or holonomic constraint forces are used to integrate for the PMF, the correction factor is not needed, as these methods directly yield the free energy.\cite{jacobian_Ciccotti2018}

\section{\label{convtesting} Convergence Testing}
When initially determining production simulation parameters, the trade-off between efficiency and accuracy is key.
In particular, it is not currently feasible to simulate \textit{ab initio} systems at the same speed (and therefore, total trajectory time) as classical force fields would allow for the same system.
For this reason, we examined the simulation ensemble, sampling method, distance discretization, time step, and simulation length to try to find an appropriate balance between computationally efficient \textit{ab initio} simulations and accurate potential of mean force data.

Additionally, there is ambiguity in choice of simulation system physical parameters, most notably the number of waters and size of the simulation box.
In pure water simulations of constant-volume, one typically chooses a box length from a constant-pressure equilibration that reproduces the bulk characteristics of the water system, for example the density of water at which temperature you equilibrate.
For an infinite bulk system, introduction of ions into the solvent will have vanishing effect on this choice as a large bulk is overwhelmingly independent of any perturbing interactions from the solute.
This is not so for smaller systems. 

As \textit{ab initio} molecular dynamics are much more computationally intensive than classical force field dynamics, and since the reconstruction of a PMF often increases the expense by an at least an order of magnitude, we are forced to limit our system sizes to reasonable levels at the expense of bulk statistics.
Since we are limited in size, it is not clear what physical characteristic we should choose in determining box size.
Moreover, the density of liquid water and ice is known to be dependent on the concentration of dissolved salts,\cite{appdensDougherty2001, phenomdensNovotny1988} so choice of simulation parameters might then affect extrapolation to bulk characteristics.
For this study we chose to solvate a single \ce{NaCl} molecule in 96 \ce{H2O} molecules.
We use a box size of 14.373 \AA\ for the \textit{ab initio} simulations, chosen such that the number density of the system was close to that of liquid water, $n=\SI{0.033}{\angstrom^{-3}}$, corresponding to a solution concentration of 0.56 M.
With this choice, we can investigate the local solvent changes near the ion interface while having the relatively limited bulk still approximate the infinitely dilute limit. 
Additionally, to investigate the effects of solution density on solvation state stability, we choose another cubic box length of 14.725 \AA\, yielding a number density of $\SI{0.0306}{\angstrom^{-3}}$, slightly greater than that of ice, corresponding to a concentration of 0.52 M.

\textcolor{black}{It should be noted that when comparing PMF predictions with different methods, it's not straightforward what configuration should be used as the reference. 
Ideally, one would use the value of the PMF at very large interionic separations, where the bulk begins and the ions effectively have no knowledge of each other.
The typical size of \textit{ab initio} simulation boxes limits achieving convergence at such large distances.
We instead choose the zeroes of our PMFs to be the more stable of the CIP or SSIP minima.
This alignment choice does not affect the evaluation of stability between the two states, because the state stability ordering arises from the energy difference between the two minima. 
This value is a relative value, and can be used to compare different PMF predictions.
We present these values in Table~\ref{tbl:hdldaimd}.
We further present stabilities relative to the transition barrier, to better characterize what happens to the individual states as parameters change.
When describing affects of parameter changes it is with respect to these values that are independent of the zero-point, but for completeness whenever a change is noted we specify that it with respect to the transition barrier.}

\subsection{Ensemble}
For ensemble testing, a single \ce{NaCl} pair was solvated with 96 TIP4P\cite{tip4pJorgensen1983} \ce{H2O} molecules using the \texttt{GROMACS}\cite{gromacsAbraham2015} simulation suite.
With a time step of 0.5 femtoseconds, each NVE and NVT trajectory was underwent a 25 picosecond NVT equilibration, while the NPT trajectory equilibrated with NPT conditions.
Each trajectory then underwent a production run in the corresponding ensemble for 250 picoseconds.
For any given ionic distance window, there were eight initial random seeds given to different equilibrations, resulting in 2 nanoseconds of total trajectory time for each window.
Since we are less limited by computational expense when using classical force fields, we first use umbrella sampling to reconstruct the PMF.
Each restraining potential was centered on a reference ion distance starting at 2 \AA\ and ending at 6 \AA, in steps of 0.1 \AA.
With our harmonically constrained windows, the WHAM\cite{whamKumar1992} equations are used to reconstruct the free energy of solvation as the ions are separated.
In particular, both the the \texttt{g-wham} method\cite{gwhamHub2010} implemented in \texttt{GROMACS} and the external \texttt{wham} program\cite{whamGrossfield} were used and yielded consistent results.
Error estimates for WHAM generated PMFs come from the bootstrapping method built into \texttt{g-wham}, and are shown as windows around the average value lines in Fig.~\ref{fig:convnptnvtnve}.

We note the almost exact agreement of all the results.
With the agreement of the ensembles to within error, to optimize simulation efficiency in the \textit{ab initio} simulations we choose to use the NVE ensemble in our \textit{ab initio} simulations. This has further benefit by allowing us to sample the true dynamics of the systems we simulate.

\subsection{Sampling Method}

The WHAM method requires convergence of each sampled umbrella window histogram to properly converge the PMF.
This means that windows centered around inter-ion distances nearing the completely solvated range will require much more time to accurately converge, as the configuration states in these windows are more or less equally likely.
Rather than risk this computational bottleneck, we instead choose to directly constraint the distances in our \textit{ab initio} simulations.
The results comparing the WHAM and direct integration methods are plotted in Fig.~\ref{fig:convwhamdir}.
Ion distances are constrained with the SHAKE\cite{shake_Ryckaert1977} algorithm, and the holonomic constraint forces are directly integrated to yield the same result as generated with umbrella sampling.

\begin{figure}
    \centering
    \includegraphics[scale=0.5]{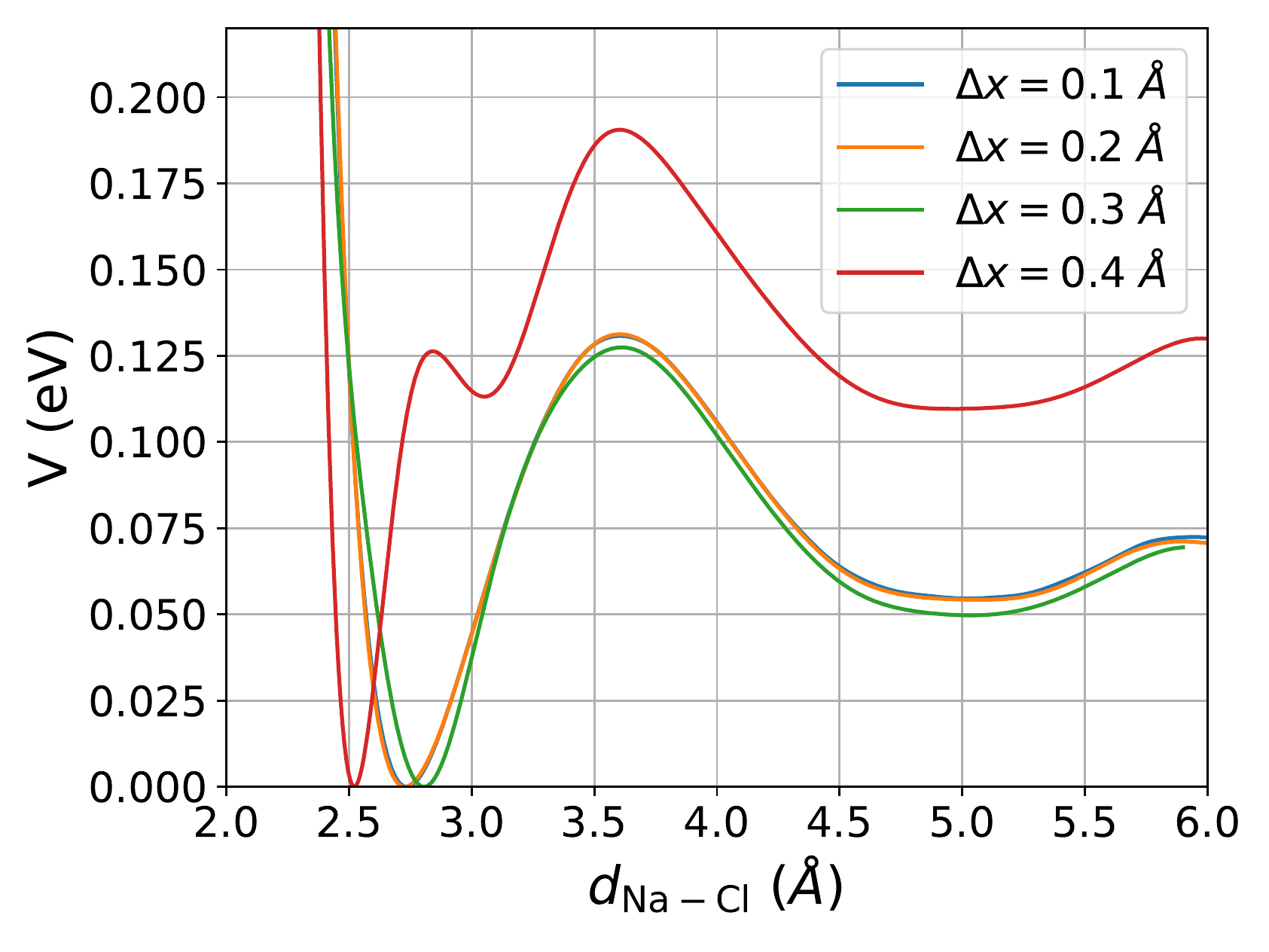}
    \caption{PMF convergence testing parameterized by changing step size in the discretization of distance between the ions, using a fixed-distance restraint in a TIP4P solution with default OPLS-AA ion parameters in \texttt{GROMACS}. Increments of 0.3 \AA\ and 0.4 \AA\ (green and red, respectively) do not yield converged results, while increments of 0.2 \AA\ and 0.1 \AA\ (yellow and blue, respectively) converge to consistency.
    }
    \label{fig:convdistance}
\end{figure}

\subsection{Distance Discretization}

For direct constraints PMF simulations,
me must choose a discretization of  the mesh of inter ionic distances on which the PMF is integrated.
A choice must be made between computational cost and accuracy, which can significantly affect the integrated potential results.
Fig. \ref{fig:convdistance} shows results for the directly constrained NVE PMF from 2 to 6 \AA\ as parameterized by the distance mesh separating the two points.

The results exactly agree whether using distance steps of 0.1 \AA\ or 0.2 \AA\ .
Some slight disagreement is seen when the step is increased to 0.3 \AA\, and increasing to 0.4 \AA\ completely ruins any convergence.
For the purposes of our \textit{ab initio} simulations, we chose a distance step of 0.2 \AA\, to have a more regular grid of sample points while keeping the number of necessary simulations at a reasonable level.

\subsection{Simulation Length}

\begin{figure}
    \centering
    \includegraphics[scale=0.5]{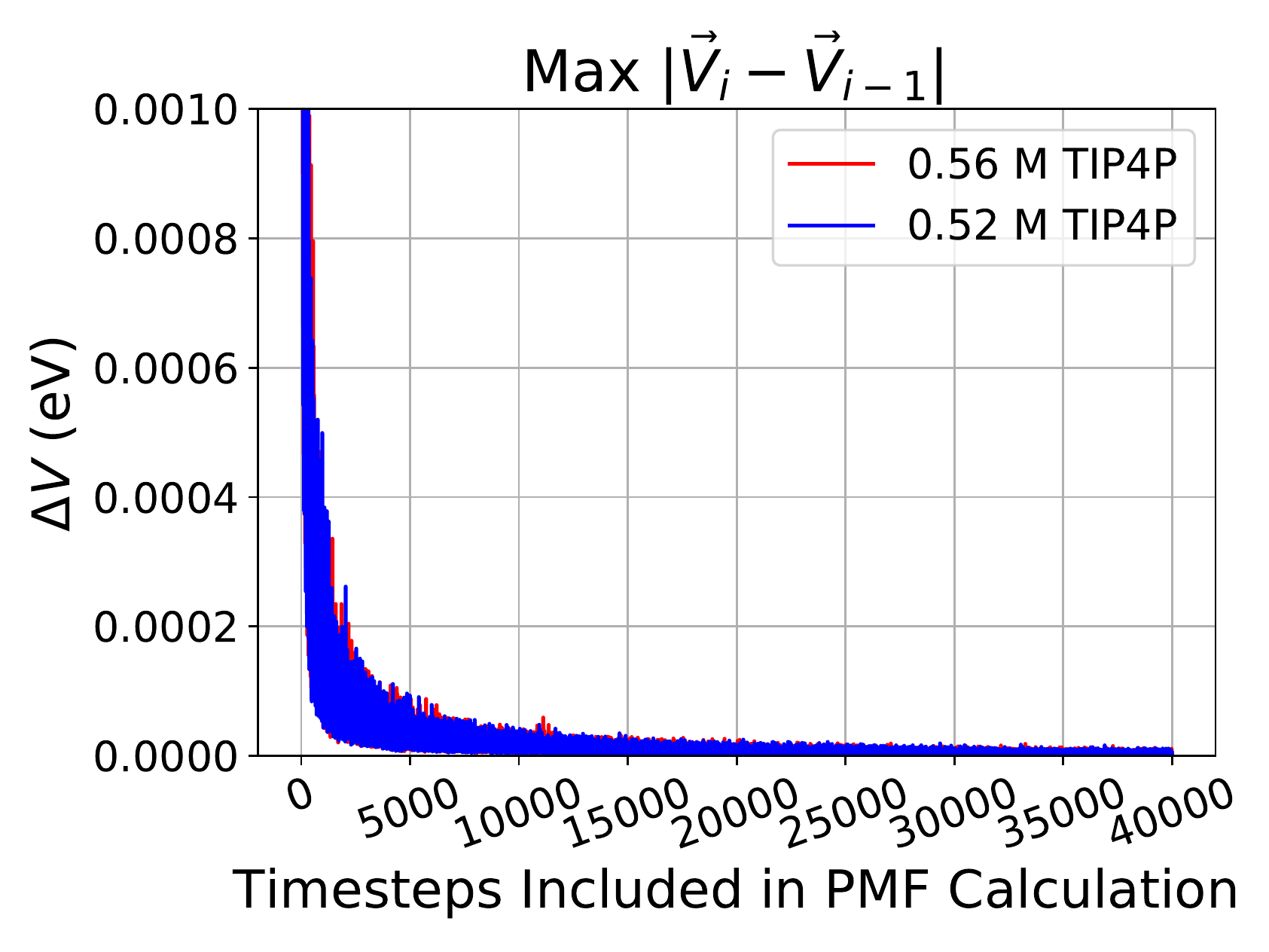}
    \caption{\textcolor{black}{Examining the convergence of the directly integrated NVE PMF through including progressively more of the trajectory of TIP4P water with OPLS-AA ion parameters in \texttt{GROMACS}. The maximum of the absolute error is chosen for convergence estimates. At each time step, the PMF is generated with all frames up to and including that step.}}
    \label{fig:convtimelength}
\end{figure}

\textcolor{black}{With above parameters determined, we compared the total length of necessary trajectory sampling needed to converge to a consistent potential.
For the given trajectories, increasing amounts of time steps were used to generate successive PMFs.
The PMF of any given step was subtracted from the previous estimate, and the absolute value is taken.
The maximum value of the current error between the two PMF estimates is taken as a convergence parameter.
The results are shown in Fig. \ref{fig:convtimelength}.
We find that at 15,000 time steps the error between successive PMF estimates converges to sufficient accuracy, and that the convergence slows down considerably past this amount of trajectory.
With less sampling, the results diverge more starkly and no longer agree within a reasonable tolerance.
For this reason we choose our \textit{ab initio} sample size to be 20,000 snapshots at each reference distance, which, with a time step of 0.5 femtoseconds, yields 10 picosecond trajectories across the 21 distances at which we constrain the ions.
We allow the system to equilibrate to \textit{ab initio} conditions and use the final 15,000 time steps in any analyses conducted.}

\subsection{\textit{Ab Initio} Simulation Convergence}

In Born-Oppenheimer DFT \textit{ab initio} molecular dynamics, the time step of the simulation is determined by the need to accurately sample the shortest vibrational modes of the system.
As classical models are usually rigid, classical molecular dynamic simulations can afford a longer time step than quantum mechanical simulations.
Having a time step too large has been shown to affect the vibrational structure of the bulk system, thus changing the effective hydrogen bond strength between water molecules.\cite{waterrigidAllesch2004, watervibPraprotnik2005}
In particular, increasing the simulation time step has been shown to redshift the \ce{O-H} stretching and, to a lesser extent, bending modes.

In the Supplementary Information, Fig.~\prettyref{supp-fig:convtimestep} shows the vibrational spectra of an \textit{ab initio} simulation of the same box of 96 water molecules using the vdW-DF-cx exchange and correlation (XC) functional\cite{bhBerland2014} with time steps of 0.5 and 1.5 fs, for the same total amount of steps, as well as the peak resonances of intermittent time steps from 1.0 to 1.5 fs.
The redshift of the vibrational modes is clearly evident as the time step increases, most notably for the \ce{O-H} stretching modes.
Redshifted stretching mode peaks correspond to water with a weaker hydrogen bond network, and thus less structure between water molecules.
This is shown in Fig.~\prettyref{supp-fig:dt_goo}, where we plot the \ce{O-O} radial distribution functions for pure water for both 0.5 and 1.5 fs simulation time steps against experimental results.\cite{Soper2013}

A weaker bulk system could have significant effects on any reconstructed PMF in that system.
In this case, as the ions are pulled apart from each other, there is less strength in the first hydration shell to resist the separation, which would effectively decrease the stability of the contact ion pair state.
For this reason, we choose an \textit{ab initio} time step of 0.5 femtoseconds, to accurately simulate the hydrogen bond network of the solvent.
We believe that longer time step sizes can significantly compromise the comparison of results\cite{pmf4_classical_scancpmd_Yao2018} obtained with the same model.

We further note that there are many convergence properties of \textit{ab initio} simulations which are basis-set dependent.
In particular, the \textit{ab initio} framework utilized in this work uses numerical atomic orbitals (NAOs), which, while more computationally efficient, lack clear systematic ways of improvement for a fixed basis size that plane-wave methods may employ.
Such convergence tests have been explored in great detail, in particular for the liquid water bulk in our current system of interest.\cite{marivi_waterbasisconvCorsetti2013}
We do not discuss such properties in this work.

\subsection{Single-/Multi-dimensional Reaction Coordinates}
Previous works have shown that multidimensional reaction axes are necessary to accurately probe the transition dynamics\cite{cn2_Geissler1999} and dynamic correction factors to rate constants.\cite{cn1_Mullen2014}
In particular, in addition to the ionic distance reaction coordinate, a coordinate involving the solvent structure nearby the ions might also be imposed.
One way to achieve an accurate sampling of solvent configurations is to apply a harmonic constraint to a coordination number around a given ion -- in our case, around the \ce{Na+} ion.
To compare effects multidimensional sampling have on our calculated PMF, we used the \texttt{coordNum} reaction coordinate from the \texttt{COLVAR}\cite{colvar_fiorin13} package implemented in \texttt{LAMMPS}.\cite{lammps_Plimpton1995}
This reaction coordinate is defined as \begin{equation}
    C(G_1, G_2)=\sum_{i\in G_1}\sum_{j\in G_2} \frac{1-\left(\frac{r_{ij}}{d_0}\right)^n}{1-\left(\frac{r_{ij}}{d_0}\right)^m} \label{eq:colvarcn}
\end{equation}
where $G_1$ and $G_2$ are atoms groups composed of the Na atom and O atoms in SPC/E water, respectively. 
Ion parameters were chosen to match that of the solvent model.\cite{naclt4pew_Joung2008}
The cutoff distance $d_0$ was chosen to be 3.0 \AA, just before the end of the first solvation shell of the \ce{Na-O} radial distribution function as seen in Fig.~\prettyref{supp-fig:gmx_aimd_gnao_ion_tdyn}.
The exponents $n$ and $m$ maintained their default values of 6 and 12, respectively.

\begin{figure}
    \centering
    \includegraphics[scale=0.5]{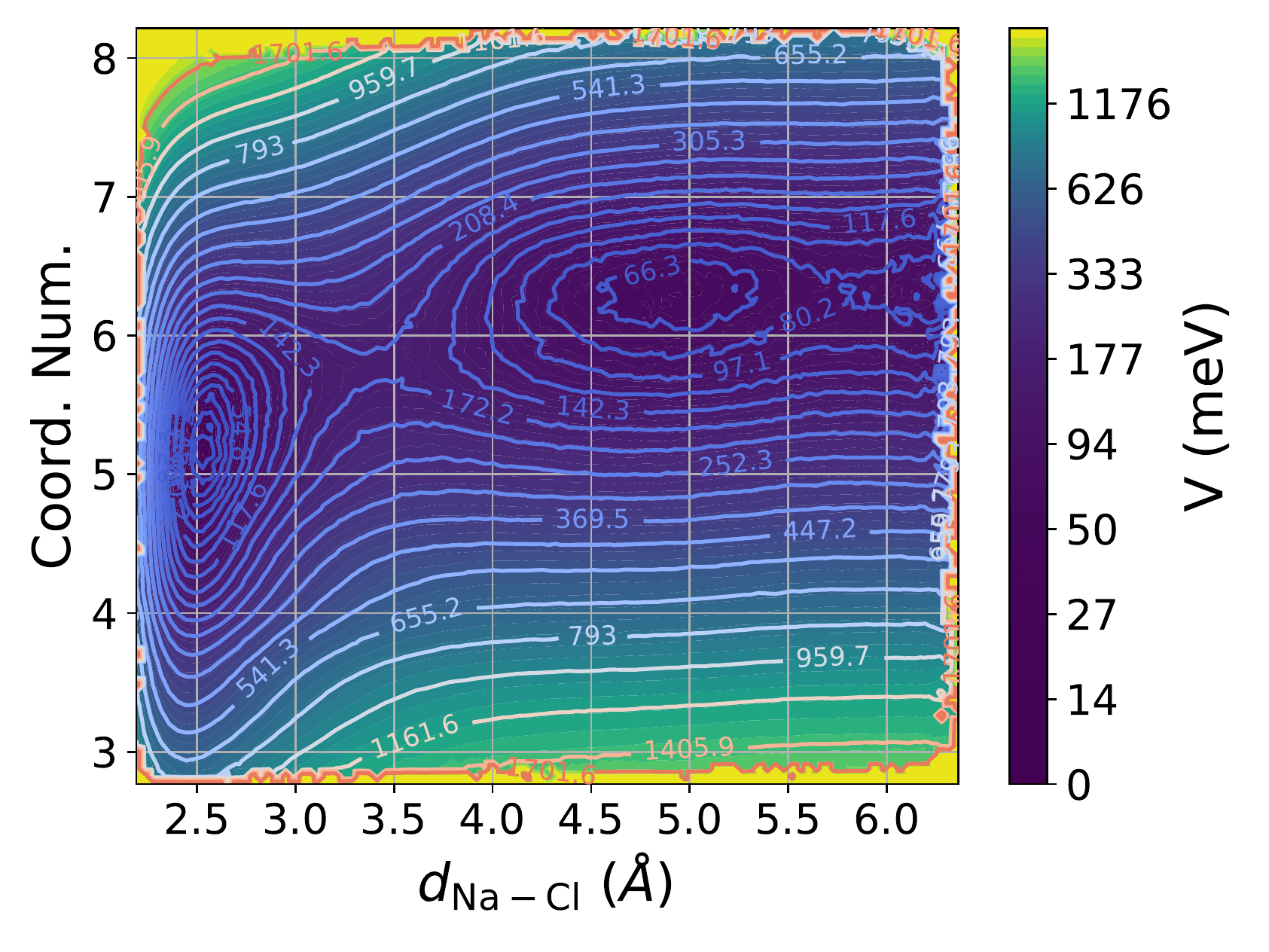}
    \caption{The two-dimensional free energy surface resulting from the umbrella sampling as mentioned in the main text. The surface is zeroed at the CIP, and the SSIP state energy is $\sim60$ meV.}
    \label{fig:2dpmf}
\end{figure}

For the two dimensional FES reconstructed with umbrella sampling, the distance windows had centers from 2.0 to 6.0 \AA, in steps of 0.2 \AA.
The coordination number reaction coordinate had window centers from 3.0 to 8.0 in increments of 0.25.
The full two-dimensional free-energy heatmap is shown in Fig.~\ref{fig:2dpmf}.
Of note is the much deeper CIP region than the SSIP, in agreement with our one-dimensional results.
The SSIP shows to be less stable than the CIP by a value on the order of 60 meV.

\begin{figure}
    \centering
    \includegraphics[scale=0.5]{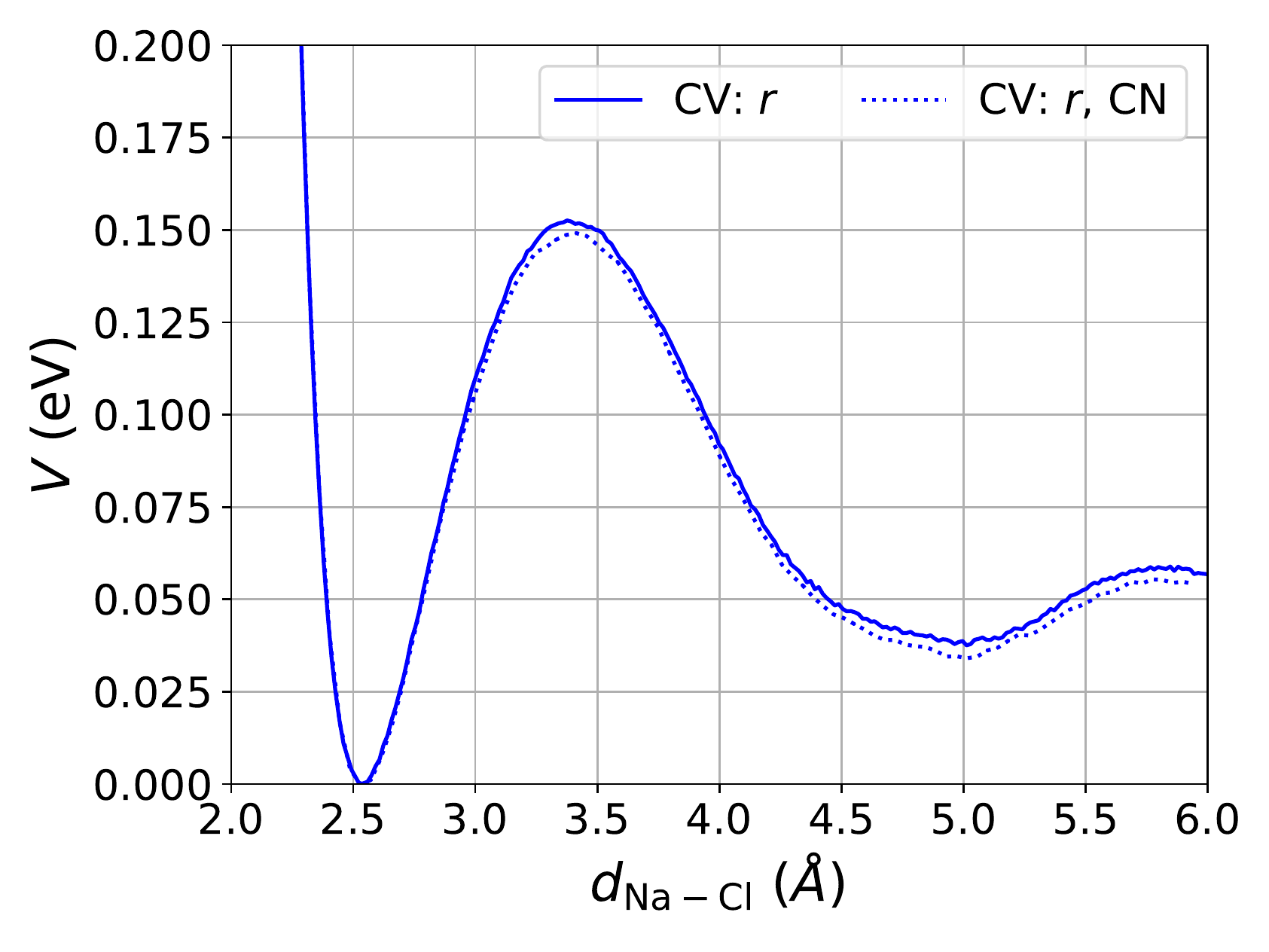}
    \caption{A comparison between the two-dimensional PMF after integrating out the coordination number reaction coordinate (dotted) and the one-dimensional PMF using just the interionic distance. The coordination number degree of freedom is integrated out using Boltzmann factor weighting across the sampled coordination numbers at a given distance.}
    \label{fig:cnrd_rd_comp}
\end{figure}

Moreover, integrating out the extra degree of freedom to compare one-dimensional and two-dimensional PMFs shows no significant difference in the free energy change during solvation, as seen in Fig.~\ref{fig:cnrd_rd_comp}.
Indeed, while forcing the solvent configuration sampling to occur through use of a harmonic potential might speed up the process, a sufficient number of snapshots at a given reference distance should sample the second reaction coordinate space reasonably well, and without introducing potentially unphysical configurations by forcing too many waters to be close to a given ion.
For this reasoning, we choose to focus on the one-dimensional PMFs generated through the interionic distance coordinate, to maximize efficiency in simulation expense while maintaining accuracy in our results.

\subsection{Parameter Convergence Summary}
So far we have evaluated how the PMF depends on primary simulation parameters,
independently of the chosen water-water and water-ion interaction model.
The previous classical and \textit{ab initio} convergence tests give us insights as to what parameters are appropriate for our studied system.
We use symplectic integrators for both classical and \textit{ab initio} simulations, with no temperature or pressure couplings, so that we can accurately simulate the true dynamics of the systems in an NVE ensemble.
We choose to directly constraint the \textit{ab initio} interionic distances from 2 to 6 \AA\, in steps of 0.2 \AA, to have better control over the mean force sampling convergence.
We use a 0.5 fs time step for our \textit{ab initio} simulations and a total time of 10 ps at each distance to be able to efficiently and accurately reconstruct the PMF.

\section{Results: Model and density-dependent potentials of mean force}

The goal of our study is to understand the physical origin for the 
observed disparity of relative stabilities of CIP and SSIP obtained
with different models.\cite{galli_Zhang2020,pmf4_classical_scancpmd_Yao2018}
In particular we want to disentangle the effects from water-water and ion-water interactions.
To do so we choose to evaluate separately classical and \textit{ab initio} models using two different molarities, both within the highly diluted limit.
A summary table with PMF properties for all our simulations is presented in Table~\ref{tbl:hdldaimd}.
For both classical and \textit{ab initio} setups, we use two cubic boxes of \ce{NaCl}+\ce{96H2O}, one with a side length of $L=14.373$ \AA\ and one with $L=14.725$ \AA, to reproduce the number densities previously mentioned.
These lengths yield solution concentrations of 0.56 M and 0.52 M, respectively.
For the different classical models we also obtain
PMFs with different sets of Lennard-Jones parameters and water models to describe the ion-water interactions.

\subsection{Classical PMFs}

At the classical level, we use OPLS-AA\cite{oplsJorgensen1996} force-field parameters included in \texttt{GROMACS}, with the ions solvated in TIP4P water.
%


\begin{table*}
	\caption{State transition barriers for classical and \textit{ab initio} PMFs.}
	\label{tbl:hdldaimd}
	\begin{ruledtabular}
		\begin{tabular}{cc|ccccc} 
			\makecell{Model} & \makecell{Concentration (M)} &  \makecell{$V_{TS}-V_{CIP}$\\ (meV)} & \makecell{$V_{TS}-V_{SSIP}$ \\ (meV)} & \makecell{$V_{SSIP}-V_{CIP}$ \\ (meV)} & $\log_{10}K_a$ & $\log_{10} K_e$\\ \hline
			\multirow{2}{*}{TIP4P+OPLS-AA} & 0.56& $130.7$ & $76.2$ & $54.5$ & 3.13 & 0.37 \\
			& 0.52 & $133.2$ & $69.3 $ & $63.9$ & 3.21 & 0.23\\ \hline    
			TIP4P/2005+Madrid & 0.56 & $78.4$ & $76.3$ & $2.1$ & 2.51 & 1.17 \\ \hline
			SPC(/E)+OPLS-AA & 0.56 & $130.9$ & $95.5 $ & $35.4$ & 3.02 & 0.58 \\ \hline
			SPC/E+JC & 0.56& $128.7$ & $122.6 $ & $6.1$ & 2.89 & 1.17 \\ \hline
			\multirow{2}{*}{PBE} & 0.56& $180.7$ & $181.3$ & $-0.6$ & 3.47 & 0.90\\
			& 0.52 & $77.2$ & $213$  & $-135$ & 3.51 & 2.64\\ \hline
			\multirow{2}{*}{vdW-DF-cx} & 0.56& $116.1$ & $129.0$ & $-13$ & 3.21 & 1.40 \\
			& 0.52 & $106.3$ & $120$ & $-14$ & 3.36 & 0.95\\ \hline
			\multirow{2}{*}{PBE\cite{galli_Zhang2020}} & 0.68\footnote{NPT, 400 K, 1 atm} & $49.8$ & -- & $-30.3$ & -- & --\\
			& 0.68\footnote{NPT, 1000 K, 11 GPa} & $79.8$ & -- & $-28.1$ & -- & --\\  \end{tabular}
	\end{ruledtabular}
\end{table*}

\begin{figure}
    \centering
    \includegraphics[scale=0.5]{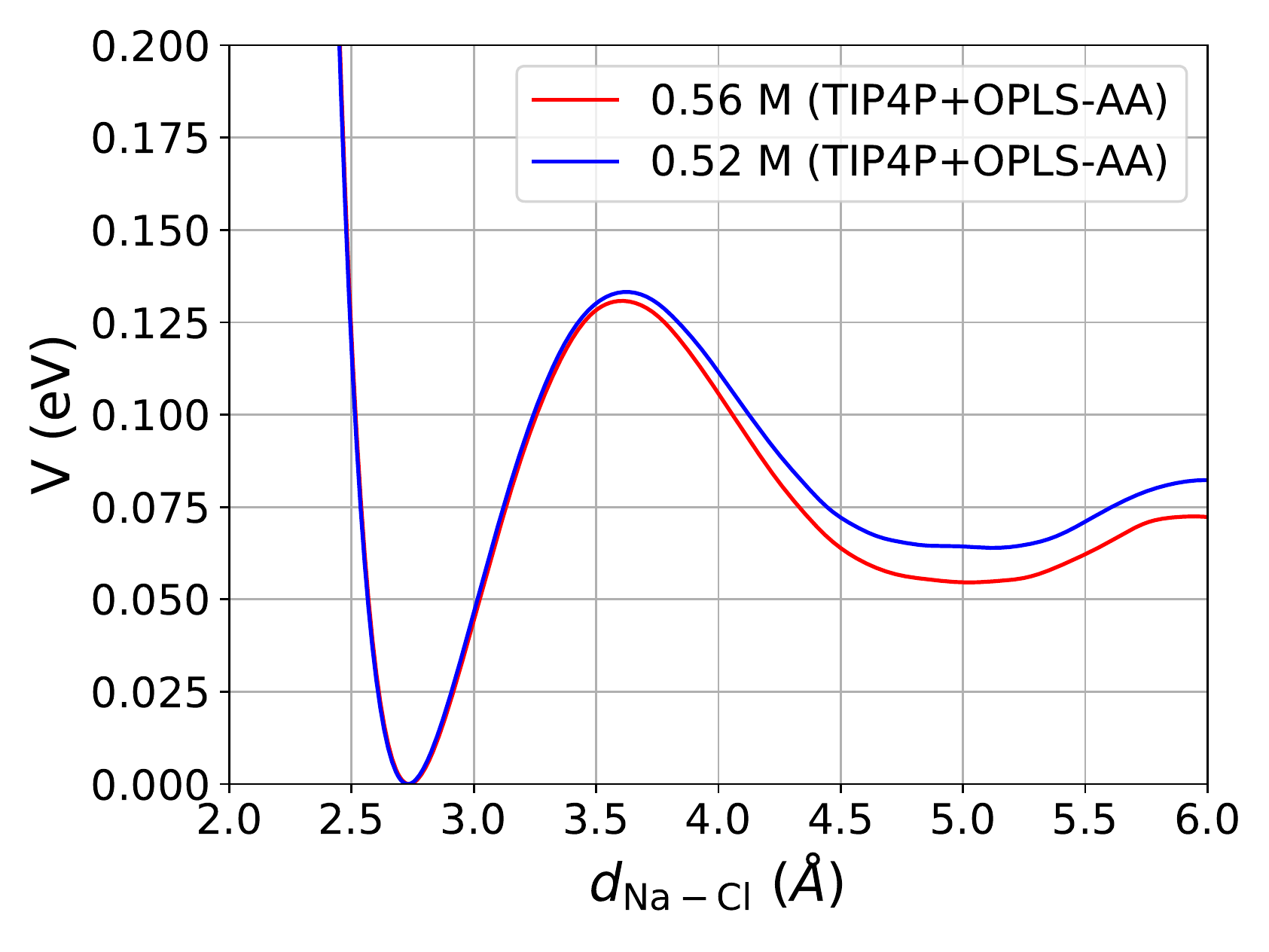}
    \caption{The PMFs from the trajectories generated from directly constraining the ion distances in \texttt{GROMACS}. There is seemingly minimal effect of molarity on the classical CIP, as evidenced by the 0.56 M (red) and 0.52 M (blue) having the same overall structure, with a minima of comparable depths. The SSIP region is slightly affected, however, as the higher density predicts a marginally more stable state than the lower density simulation. 
    }
    \label{fig:pmfclassical}
\end{figure}

With the previously mentioned system compositions, we reconstruct the NVE ensemble PMF of NaCl as shown in Fig.~\ref{fig:pmfclassical}.
We note the relatively equal depths \textcolor{black}{(relative to the transition barrier)} in both conditions for the CIP state, on the order of 125 meV, both of which are much more stable than the SSIP state as we would expect from the classical force field.
We further note that the SSIP state in the 0.56 M solution is slightly more stable than that of the 0.52 M solution.
\textcolor{black}{Equivalently, the CIP state is slightly destabilized in the 0.56 M solution when compared to the 0.52 M solution.}
As noted by a previous study,\cite{galli_Zhang2020} increased pressure has the effect of stabilizing the SSIP through entropic contributions.
Therefore it is to be expected that the smaller system has a more stabilized SSIP.

\begin{figure}
    \centering
    \includegraphics[scale=0.5]{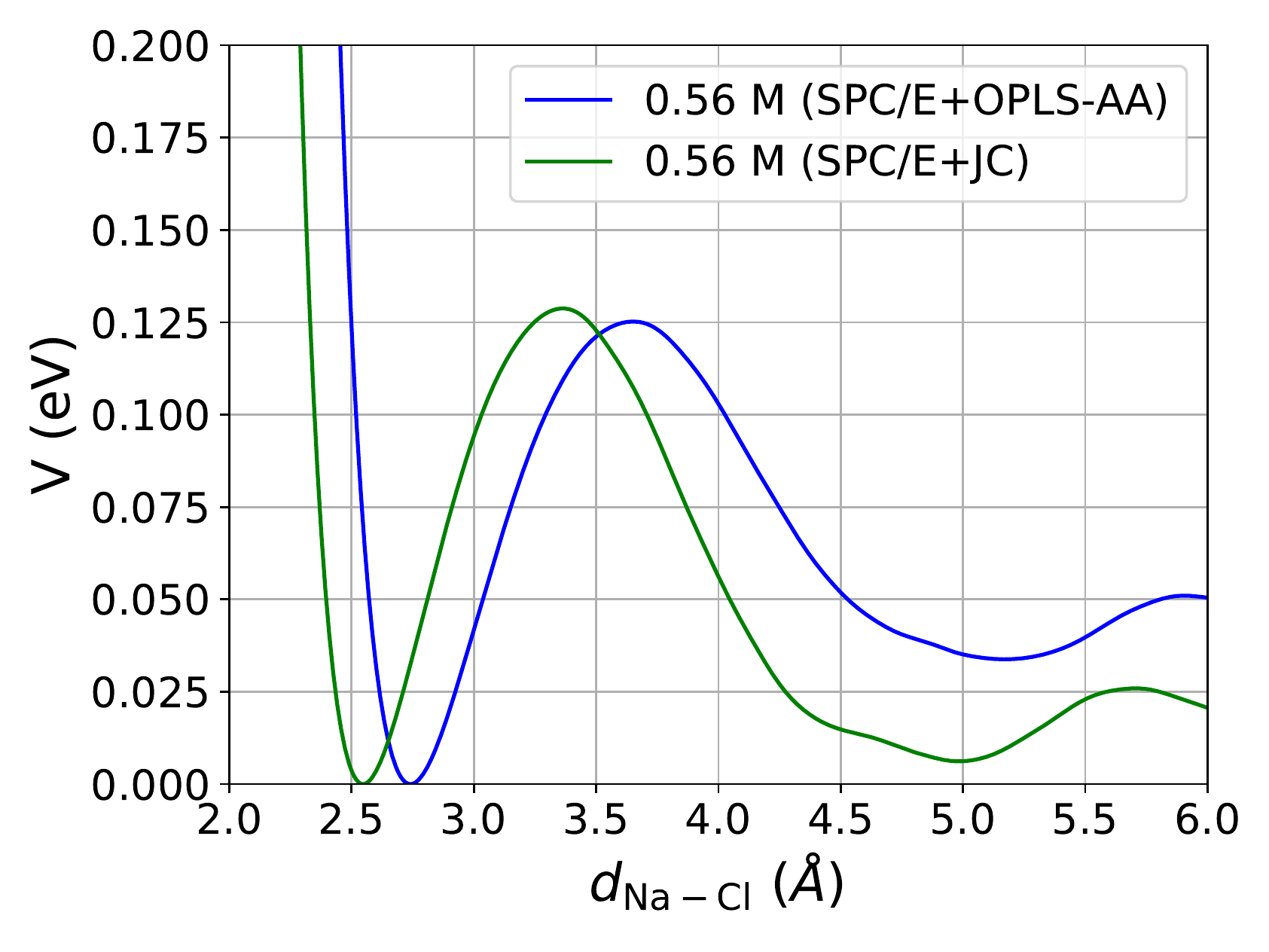}
    \caption{PMFs generated with different parameter choices in 0.56 M solution. Note the enhanced stability of the SPC/E (blue) SSIP states when compared to those in Fig.~\ref{fig:pmfclassical}. Moreover, changing from the OPLS-AA parameters in the SPC/E solvent to the JC parameters (green) further enhances the stability of the SSIP, and affects the CIP location as well.}
    \label{fig:gmxdcspc}
\end{figure}

\begin{figure*}
    \centering
    \begin{subfigure}[b]{0.49\textwidth}
    \includegraphics[scale=0.5]{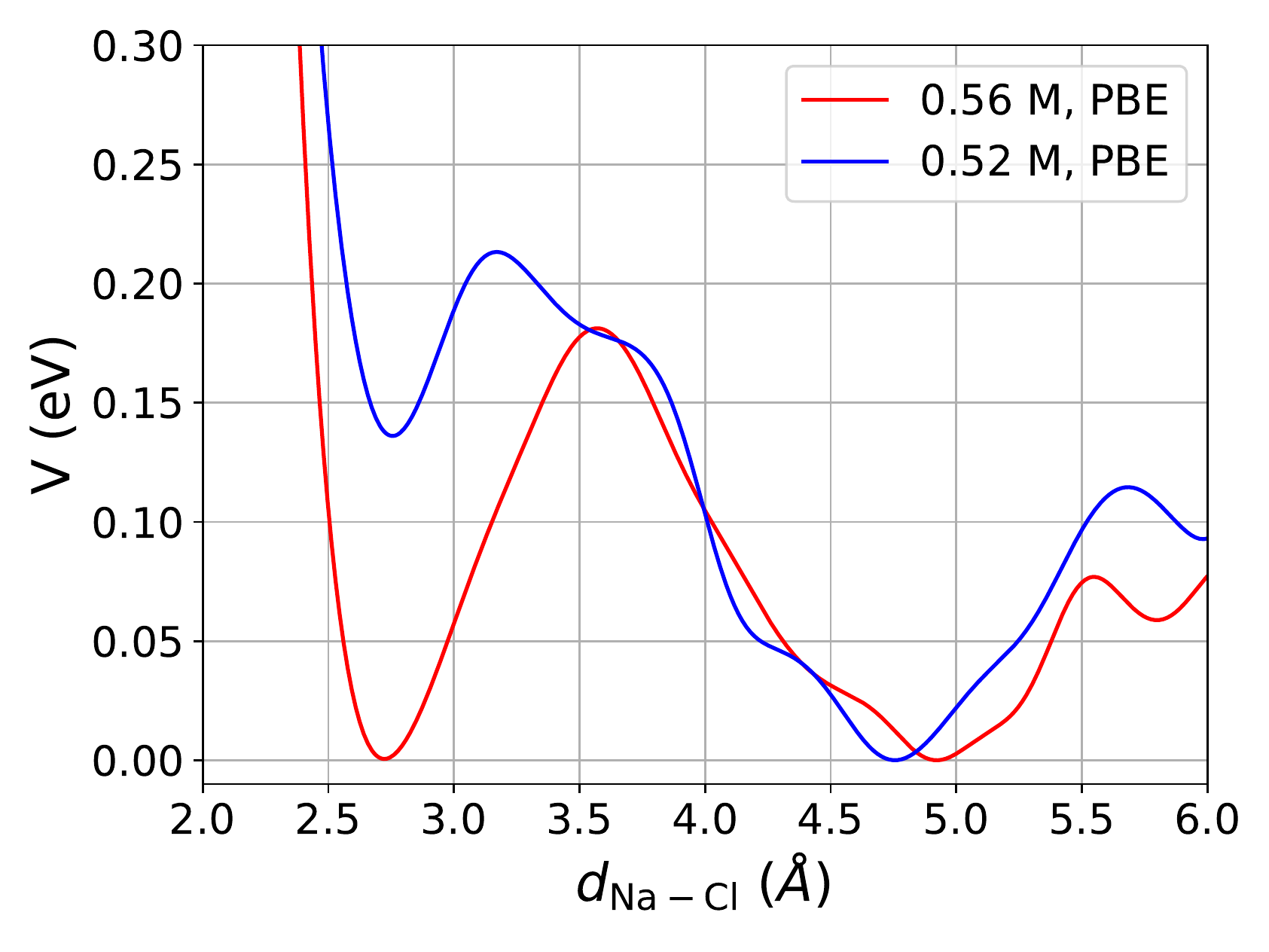}
        \caption{PBE}
        \label{fig:pmfaimdhd}
    \end{subfigure}
    ~ 
    \begin{subfigure}[b]{0.49\textwidth}
    \includegraphics[scale=0.5]{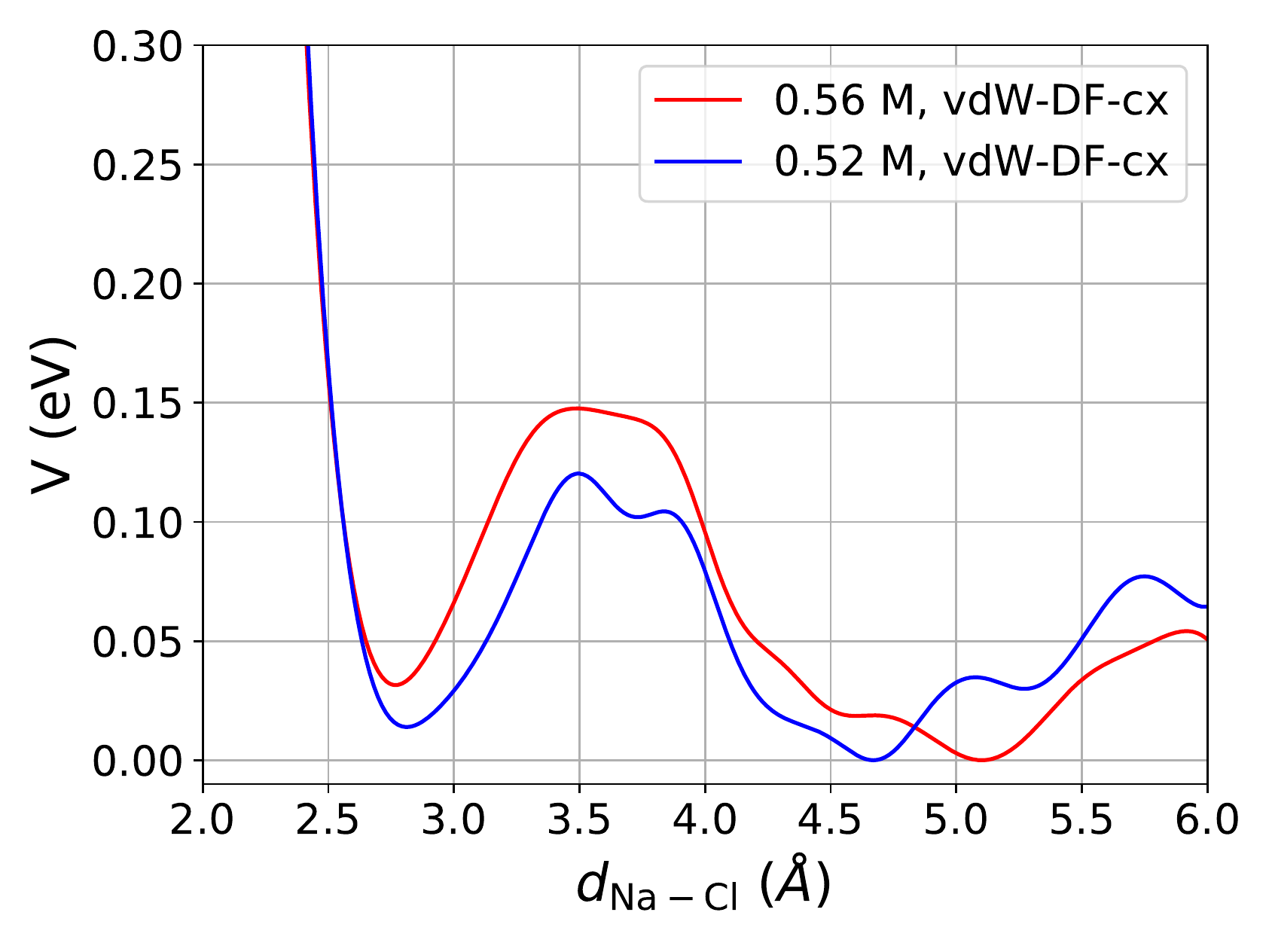}
        \caption{vdW-DF-cx}
        \label{fig:pmfaimdld}
    \end{subfigure}
    \caption{\textbf{(a)} The potentials of mean force from the fixed distance constraint method implemented in SIESTA on the PBE trajectories. In this case, the lower concentration PMF (blue) is the only one that predicts a markedly more stable SSIP than CIP. \textbf{(b)} The potentials of mean force from the fixed distance constraint method implemented in SIESTA on the vdW-DF-cx trajectories. The inclusion of van der Waals corrections makes the simulation less sensitive to density perturbations. 
    }\label{fig:pmfaimdhdld}
\end{figure*}

Fig.~\ref{fig:gmxdcspc} shows the PMFs for the 0.56 M simulation box but with a combination of different water models and ion parameters.
The OPLS-AA force field parameters are used to compare the effects of changing the water model from TIP4P to SPC/E.\cite{spce_Berendsen1987}
As shown in Table~\ref{tbl:hdldaimd}, while for both models the CIP is the favored minimum, the SSIP is stabilized by 20 meV \textcolor{black}{(relative to the barrier)} in SPC/E.
This stabilization must be due to the changes in the water model.
We also evaluate how the change of ion parameters modify the stabilities.
For the SPC/E simulations two different sets of ion parameters are used -- the OPLS-AA parameters and parameters optimized by Joung and Cheatham (JC).\cite{naclt4pew_Joung2008}
The JC parameters further stabilize the SSIP \textcolor{black}{(with respect to the barrier)}, while also shifting the location of the CIP.
Both changes are expected because the JC parameters strengthen the water-ion interactions and change the sizes of the ions.
These results for the classical transition barriers are summarized in Table~\ref{tbl:hdldaimd}.
We also simulated a system with the SPC water model and OPLS-AA ion parameters, shown in Fig.~\prettyref{supp-fig:gmxdcspc}.
The lack of drastic difference between the SPC and SPC/E simulations with the OPLS-AA ion parameters can be attributed to the minimal difference between the water models, as opposed to the more substantial change in going from TIP4P to either of the SPC models.

\textcolor{black}{Finally, we note that scaled charge models have produced various results that agree well with experimental values.\cite{madrid17_Benavides2017,madrid19_Zeron2019}
In an effort to cover all feasible parameter changes, we simulate the 0.56 M system using scaled charge ion parameters\cite{madrid19_Zeron2019} in a TIP4P/2005\cite{t4p05Abascal2005} solvent.
For brevity, the resulting PMF is shown in the Supplementary Information section as Fig.~\prettyref{supp-fig:naclmadrid}.
We note, however, that this model decreases the transition barrier between the two states relative to the other classical simulations presented here.
In addition to this barrier weakening, the CIP and SSIP states are of relatively equal stability with respect to the transition barrier.
}

\subsection{\textit{Ab initio} PMFs}

The DFT-based PMFs are computed using the \texttt{SIESTA} code,
which uses numerical atomic orbitals as basis sets for valence electrons and norm-conserving pseudopotentials.\cite{siestaordernSoler2002}
For all atoms, we used double-$\zeta$+P basis sets.\cite{corsetti2013optimal, marivi_dpps_vdwrhoeq_Fritz2016}
In all of the \textit{ab initio} simulations, we do 10 picosend NVE-ensemble runs with a time step of 0.5 femtoseconds, seeded with coordinates from the corresponding box's classical simulations.
For each box size, we do a set of simulations using the PBE\cite{pbePerdew1996} exchange-correlation functional in the generalized-gradient approximation, and another set including van der Waals interactions through the vdW-DF-cx\cite{bhBerland2014} XC parameterization.
We make this choice because we want to evaluate the differences in PMF using two models which clearly produce two very different types of liquid water structure.
PBE is well known to produce an overstructured,  low-density-type liquid,\cite{marivi_vdwrhoeq_Corsetti2013,wang2011density} and its equilibrium density
is $\approx 0.9 g/cm^3$.
vdW-DF-cx on the other hand produces a high density liquid,\cite{marivi_dpps_vdwrhoeq_Fritz2016}
with an equilibrium density of $\approx 1.1 g/cm^3$.
The distance between the Na-Cl ions was constrained from 2 to 6 \AA, incremented in steps of 0.2 \AA, and the PMF was reconstructed by integrating the mean force projection onto the vector separating the ions.
Due to the fact that we sample the physical forces on the \textit{ab initio} ions as opposed to holonomic constraint forces, the forces integrated in these \texttt{SIESTA} simulations include a Jacobian correction factor.\cite{jacobian_Ciccotti2018} 
After integrating Eq.~\ref{eq:dUdr}, the Jacobian factor in the case of a two-atom distance constraint takes the form $2k_\mathrm{B}T\ln r$, with $r$ the distance between ions, and can be interpreted as an entropic correction.\cite{pmfderivLi2007}

To seed the \textit{ab initio} simulations, coordinates from the classical trajectory were chosen as initial configurations for the dynamics at each reference distance.
Only the last 15,000 time steps are used in the analysis, to allow the systems time to equilibrate.
\textcolor{black}{\textit{Ab initio} convergence figures analogous to that shown in Fig.~\ref{fig:convtimelength} are presented in the Supplementary Information, in Fig.~\prettyref{supp-fig:aimdconv}.}
The resulting data from the normal and low number density simulations are shown in Figs. \ref{fig:pmfaimdhd} and \ref{fig:pmfaimdld}, respectively.
The transition barrier energies are summarized alongside the classical results in Table \ref{tbl:hdldaimd}.

\begin{figure}[ht]
    \centering
    \begin{subfigure}[b]{0.49\textwidth}
    \includegraphics[scale=0.45]{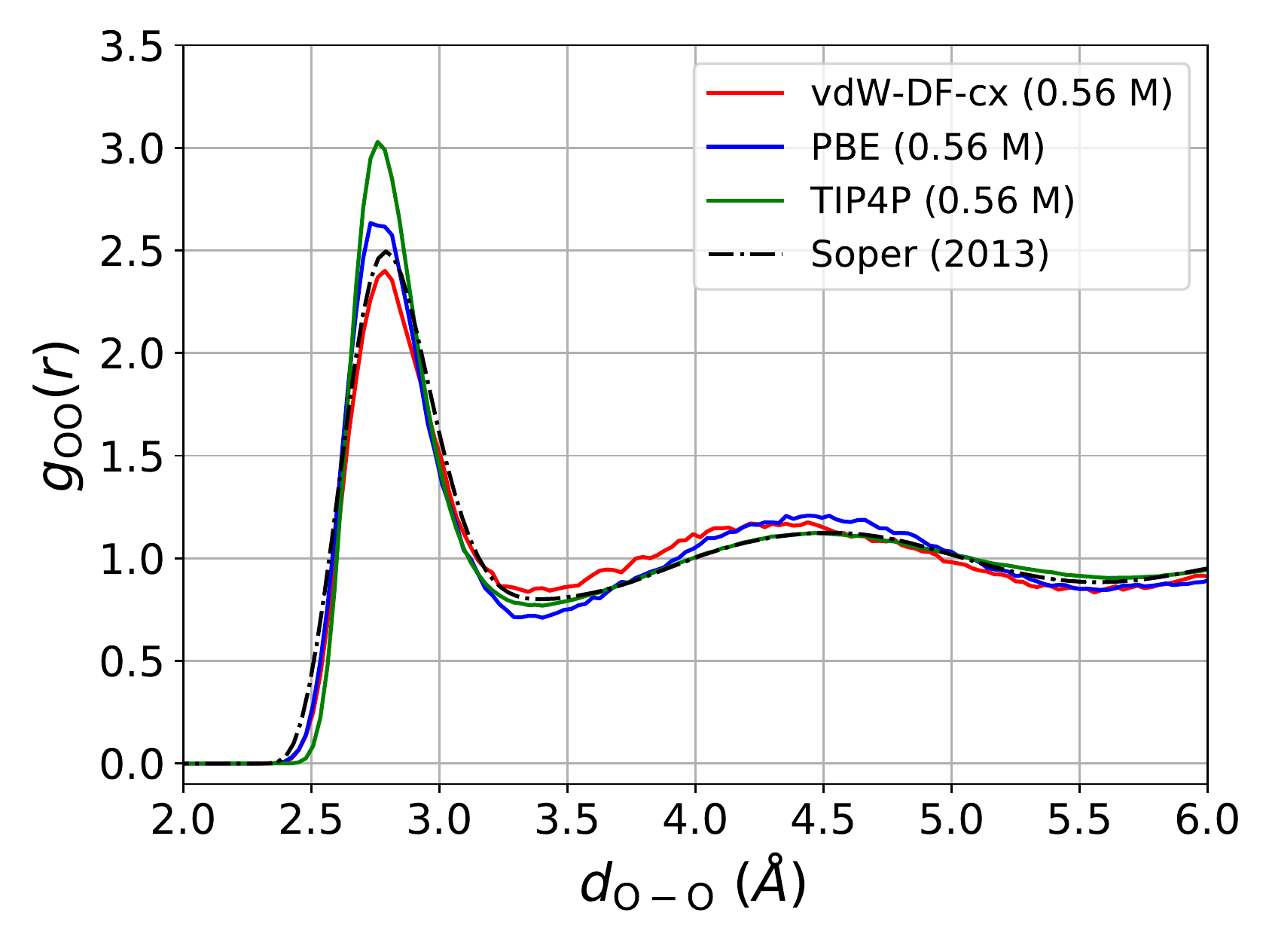}
    \caption{Pure Water Solutions}
    \label{fig:hdld_goo}
    \end{subfigure}
    \begin{subfigure}[b]{0.49\textwidth}
    \includegraphics[scale=0.45]{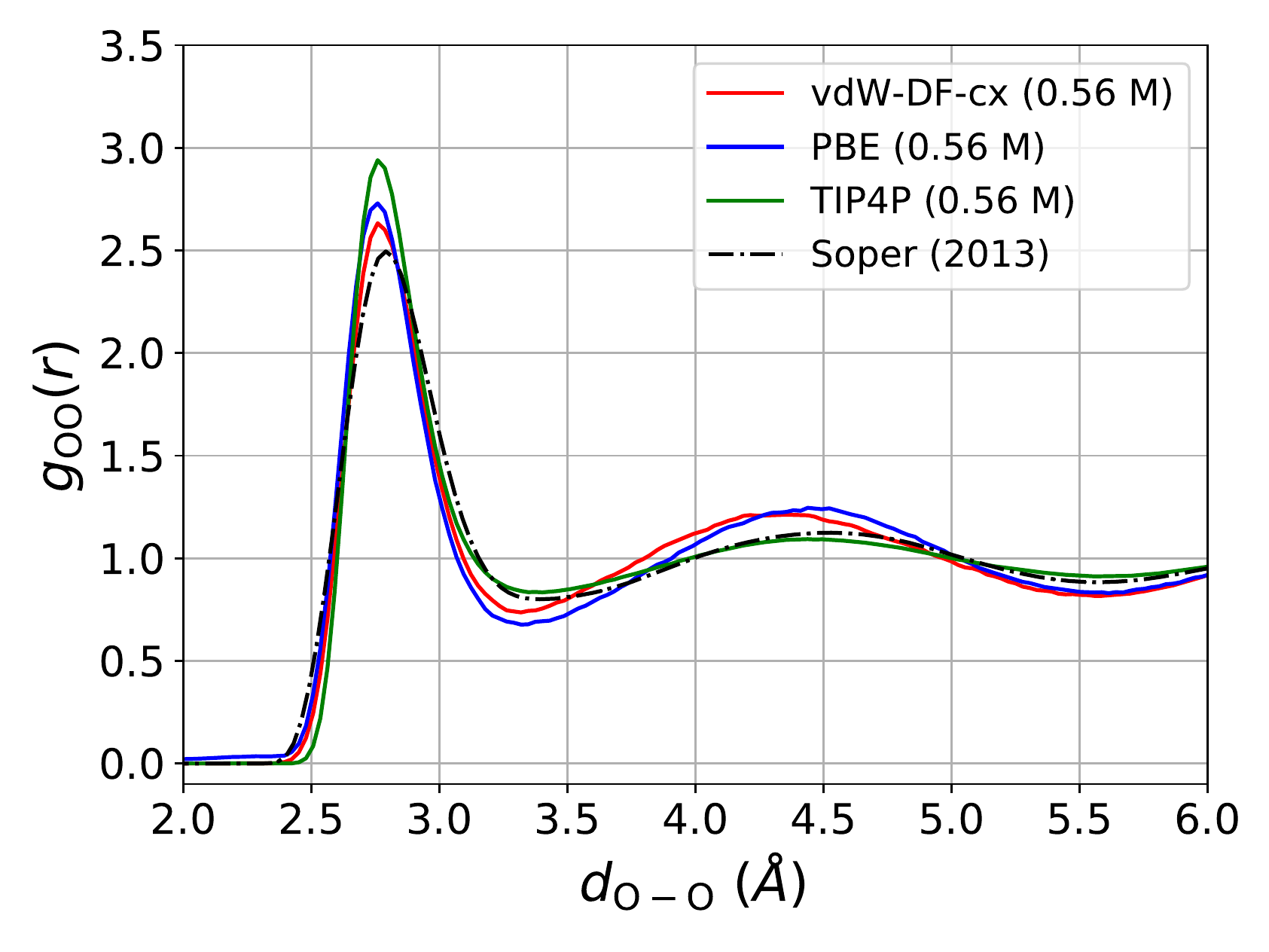}
    \caption{Ionic Solutions}
    \label{fig:hdld_goo_ion_tdyn}
    \end{subfigure}
    \caption{\textbf{(a)} The \ce{O-O} radial distributions functions of pure water using TIP4P, vdW-DF-cx, and PBE (green, red, and blue, respectively) with the box size that corresponds to the 0.56 M solutions with the ions removed. \textbf{(b)} The therymodynamically averaged \ce{O-O} radial distributions functions of the ion solutions for the TIP4P, vdW-DF-cx, and PBE simulations with concentrations of 0.56 M. Both figures show experimental results at ambient conditions for comparison.\cite{Soper2013} The inclusion of the ions serves to increase the structure in the \textit{ab initio} cases.}
\end{figure}

Contrary to the classical simulations, the DFT PMF stabilities show a much stronger
dependence on the density of liquid water.
In particular, the 0.52 M PBE solution predicts a much more stable SSIP state than CIP, whereas the higher concentration PBE solution has equally favorable state stabilities.
The 0.52 M PBE solution is the only result here to have the favored SSIP, as predicted by others,\cite{pmf4_classical_scancpmd_Yao2018, galli_Zhang2020} \textcolor{black}{and is indeed qualitatively similar to the behavior of a PMF generated using flexible SPC water molecules.\cite{response1_Guardia1991}}
However, one of these studies\cite{pmf4_classical_scancpmd_Yao2018} uses a time step of 1.5 fs in their simulations, which we have shown weakens the hydrogen-bond network, thereby enhancing SSIP stability.
All other DFT simulation results here predict relatively equal stability for the solvation states, with the vdW-DF-cx simulations predicting a slightly more stable SSIP.
Moreover, the vdW inclusive functional yields PMFs that seem to be more resistant than PBE to changes in molarity.
It is, however, striking how much more sensitive the DFT PMFs are to small changes in the simulation density than classical results.
This can originate from the very different compressibilities of the DFT pure liquids.\cite{pbecompress_MVFSWang2011,marivi_dpps_vdwrhoeq_Fritz2016}
To explicitly show the network behavior in the different densities and XC functionals, we plot the \ce{O-O} radial distribution functions for pure water in Fig. \ref{fig:hdld_goo}.
Average thermodynamic values for the systems plotted are given in Table~\prettyref{supp-tbl:tdynavgs} for the ionic solutions, and in Table~\prettyref{supp-tbl:pureh2oavgs} for the pure solutions.
We immediately see that the inclusion of the ions serves to structure both sets of system.
However, the effect is similar for both XC functionals.
This indicates that the reason behind the high sensitivity of the PBE PMF to small
changes in density of water must be rooted in energetic and/or entropic reasons.
We will explore this in the following section.

Finally, under the approximation that we are near the infinitely-dilute regime, we can reconstruct the ion-ion radial distribution function by inverting Eq.~\ref{eq:pmfdef} for $g(r)$, which we will denote as $g^\infty(r)$ to emphasize the approximation.
With this, we can find the ion-ion association constant and solvation state equilibrium constant.\cite{kamChialvo1995}
We adapt the referenced work's equations for the radial cutoffs given in our simulations, yielding \begin{align*}
    K_a &= 4\pi \int_{0}^{R_D} g^\infty(r)\cdot r^2dr\\
    K_e &= \frac{\int_{R_{TS}}^{R_D} g^\infty(r)\cdot r^2dr}{\int_{0}^{R_{TS}} g^\infty(r)\cdot r^2dr},
\end{align*} where $K_a$ and $K_e$ are the association and equilibrium constants, respectively.
$R_{TS}$ and $R_D$ are the transition state barrier radius and maximum simulation constraint radius, respectively.
Formally, the lower integration limit of 0 to cover the CIP state must be replaced by an artificial cutoff of 2.0 \AA, which is the smallest constraint distance we utilize.
In our simulations $R_D$ is 6.0 \AA, while $R_{TS}$ varies depending on the XC functional and density.
The results for the constants generated from our simulations are shown in Table \ref{tbl:hdldaimd}.
In all cases, the $K_e$ values predict a higher population of SSIP pairs, even when the CIP is more energetically favored as is the case in the simulations with OPLS-AA ion parameters.
This is likely a result of the extended region over which the SSIP is stable.
Through the larger volumetric probability, the ions are more likely to be found in the SSIP regime.

\begin{figure*}[!ht]
     \centering
     \begin{subfigure}[b]{0.32\textwidth}
     \includegraphics[scale=0.32]{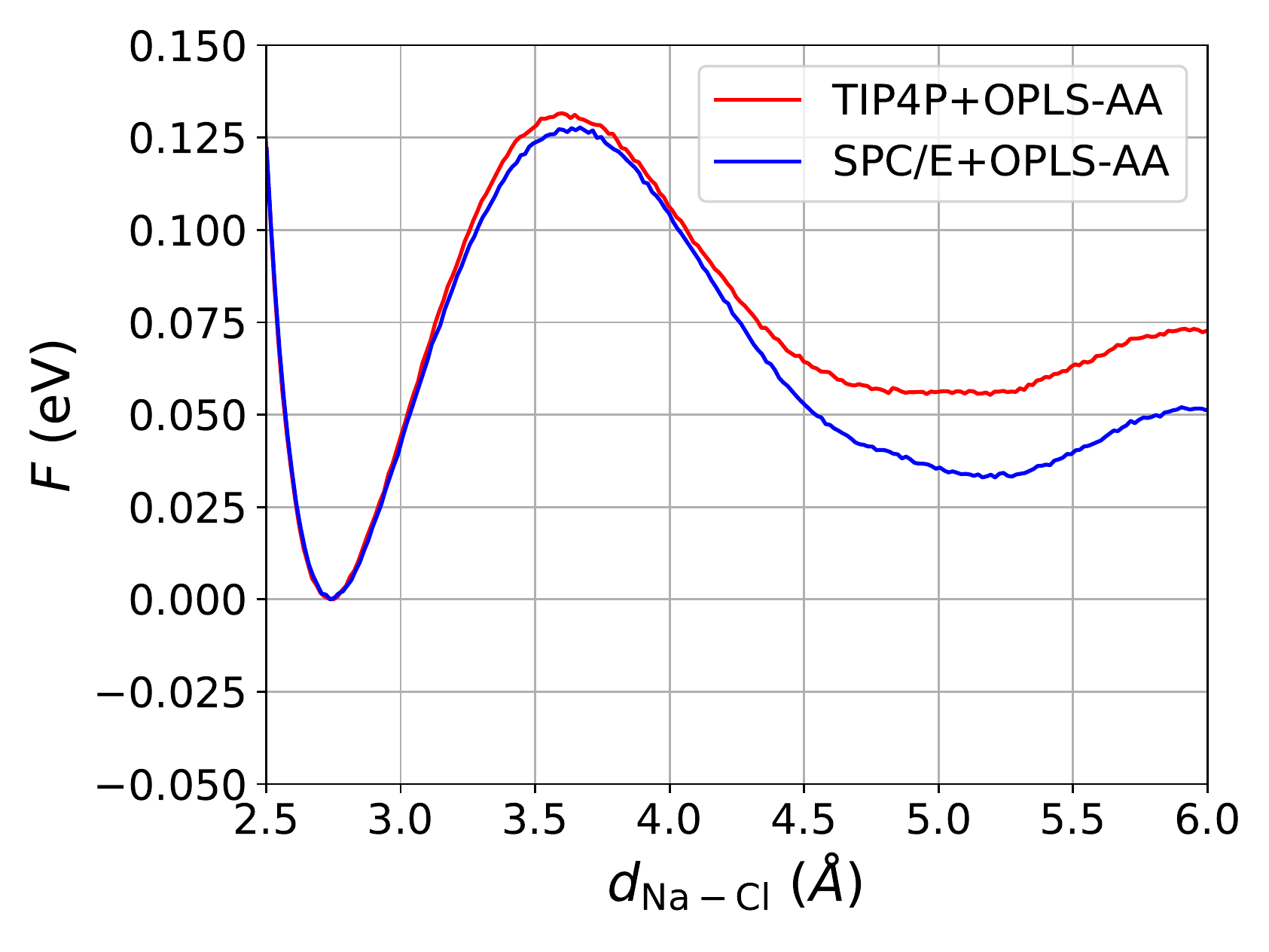}
         \caption{}
         \label{fig:t4pspcepmf}
     \end{subfigure}
     \begin{subfigure}[b]{0.32\textwidth}
     \includegraphics[scale=0.32]{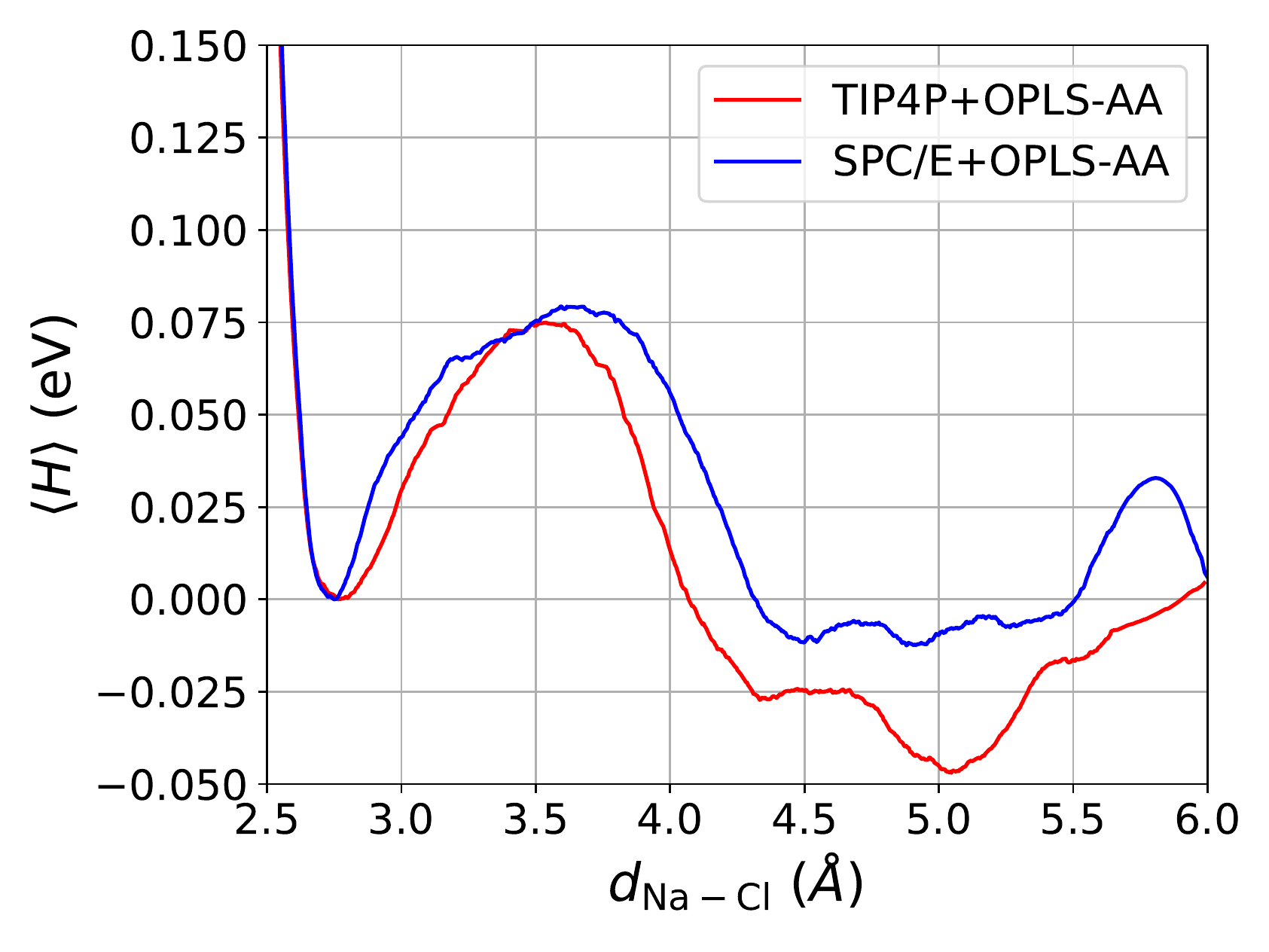}
         \caption{}
         \label{fig:t4pspcepoten}
     \end{subfigure}
     ~ 
     \begin{subfigure}[b]{0.32\textwidth}
     \includegraphics[scale=0.32]{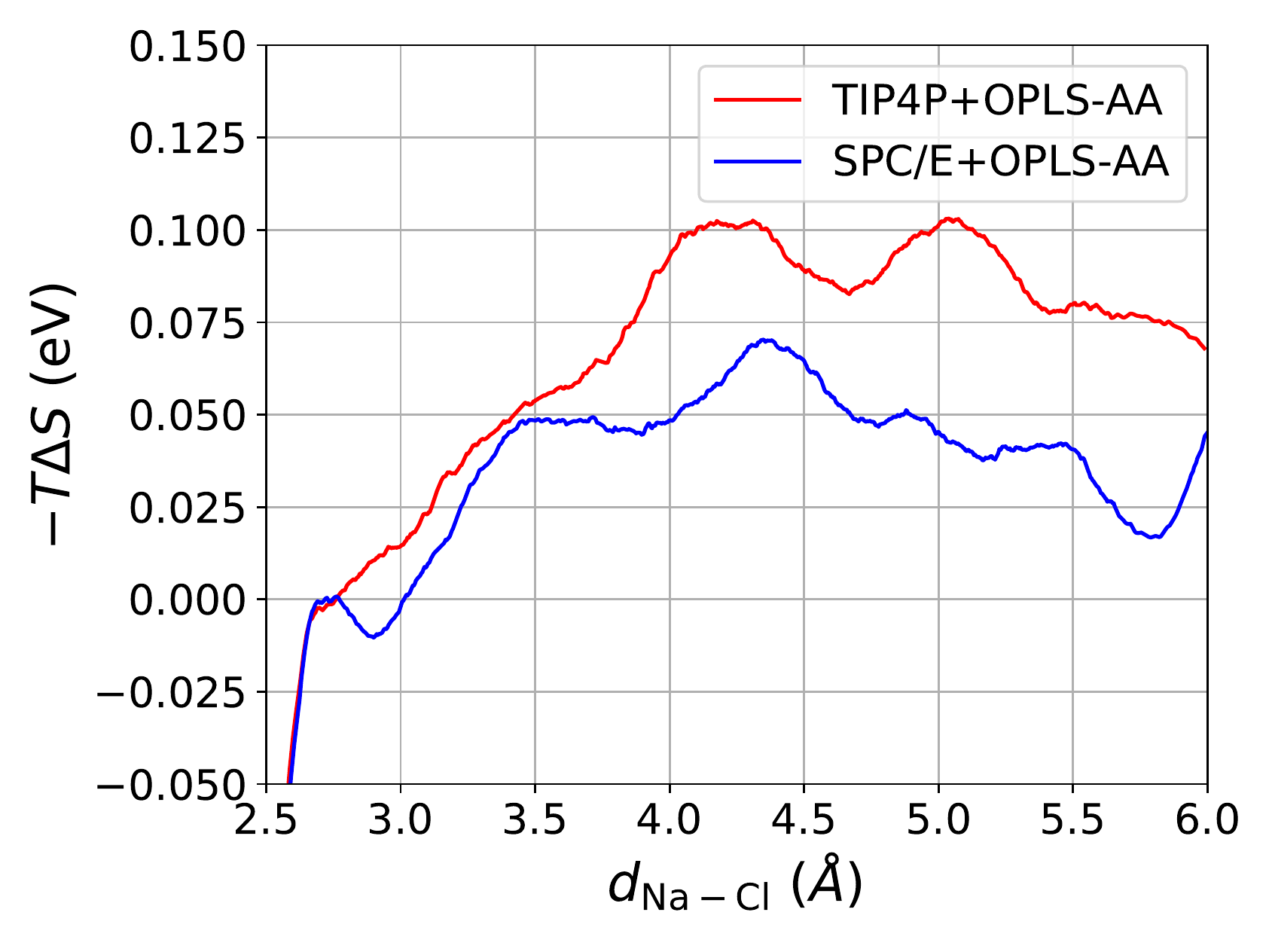}
         \caption{}
         \label{fig:t4pspcetent}
     \end{subfigure}
     \caption{\textbf{(a)} The WHAM generated NVT PMFs for OPLS-AA ions solvated in TIP4P (red) and SPC/E (blue) water models. \textbf{(b)} The histogram-binned total potential energies along the reaction coordinate, giving the enthalpic contribution to the free energy. \textbf{(c)} The entropic contribution to the free energy, taken as $-T\Delta S = F - H$.}
\end{figure*}

The $K_a$ values we calculate differ significantly from empirical fits to experimental values at lower solvent densities and higher temperatures,\cite{kake_expZimmerman2012} but behave consistently in that the association constant is expected to increase as the temperature decreases and the ions can more readily overcome thermal fluctuations.
We further note that our \textit{ab initio} $K_a$ values are less than those predicted by a QM/EFP study of the same system,\cite{pmf3_classical_ihs_qmefp_Ghosh2013} but again are consistent in that the QM/EFP study thermostatted their system to 300 K, while ours equilibrated themselves to a higher temperature, thereby causing increased thermal fluctuation strength.
We note that the $K_e$ value of 2.64 for the low density PBE simulation predicts an approximate ratio of 1:400 of CIP to SSIP solvation state population, in agreement with the steep decrease in CIP stability.

As structural changes induced by the ions are of considerable important in understanding the solvation process, we dicsuss the relevant results here but relegate a large amount of figures to the SI. Notably, we see in Fig.~\prettyref{supp-fig:tdynrdfs} the effects the ions have on the neighboring solution through the thermodynamically averaged radial distribution functions of the ion-oxygen pairs.
The presence of a clear second shell in the \ce{Na-O} distributions (Fig.~\prettyref{supp-fig:gmx_aimd_gnao_ion_tdyn}) is indicative of slightly extended perturbations on the surrounding solvent molecules, what one might see referred to as a ``cosmotropic'' effect in that it introduces structure into the surrounding solvent.
An analogous effect is present in the \ce{Cl-O} distributions (Fig.~\prettyref{supp-fig:gmx_aimd_gclo_ion_tdyn}), although not as strongly as is the case with the \ce{Na-O} pair.

Additionally, indications of local cosmotropic effects are present in Table~\prettyref{supp-tbl:rdfs}, where the first hydration shell radius ($r_1$) for the \ce{Na-O} RDF is notably pulled in to lower distances.
Moreover, the PBE \ce{Na-O} second coordination shell is markedly more contracted inwards than the vdW-DF-cx simulations, for both molarities.
This indicates that the PBE solution is more prone to nonlocal cosmotropic perturbations induced by the \ce{Na+} cation.
Table~\prettyref{supp-tbl:rdfs} further shows that the \ce{Cl-O} RDFs show extended perturbation in solvation structure, in agreement with previous results.\cite{naclrdfs_Gaiduk2017}
The solvation radii for both the first and second solvation shells ($r_1$ and $r_2$, respectively) are notably shifted to higher distances, indicative of a weakening of the hydrogen bond network in the vicinity of the \ce{Cl-} ion -- a ``chaotropic'' response.

\section{Discussion}

From all the previous results two main observations can be made:

\textbf{(i)}  Classical PMFs using the same ion parameters reliably 
predict the same potential depth for the contact ion pair.
The solvent-separated ion pair minimum and depth, on the other hand, is more dependent on the water model.
Small changes in the density of the solution do not alter these results.
These two outcomes are not totally unexpected.
The CIP potential depth and position is mostly dependent on the ion sizes and their charges.
For rigid, non-polarizable ion models one would expect them to not change, independently of the water force field.
The change of the SSIP when the water model is changed is more complex.
The reasons could be entropic, enthalpic or a combination of the two.

\textbf{(ii)} DFT results show that both CIP and SSIP are strongly dependent
both on the exchange and correlation approximation and the density
of the solution.
The latter effect is particularly large for PBE.
The fact that the CIP is so sensitive to the XC model indicates that the ion size and polarizability are strongly sensitive to the XC approximation, a conclusion supported by previous studies.\cite{polarzionstruct_DelloStritto2020}
Furthermore, the solvent dielectric properties will also be heavily dependent on the XC approximations employed, as well as the density and other thermodynamic conditions of the solution.\cite{dielwaterextreme_Pan2013,aimddielicewater_Lu2008}
The dependence with the density indicates that there is an additional contribution to
this polarizability on the density of the surrounding water molecules.
Hence these results point to the important role that the dielectric screening plays in
shaping the PMF.
Although in this study we do not aim to make an exhaustive study of how the dielectric 
screening of charges influences the PMF, we can gain some information of this by
partitioning the free energy into entropic and enthalpic contributions.

In order to do this we can sample the potential energy of the configurations that
produced the PMF.
Because most of the system is composed of liquid water, large fluctuations dominate the
potential energy and it is necessary to sample long trajectories to observe the ion-ion potential energy contribution.
This limits our potential energy study to classical trajectories.
Results are shown in Fig.~\ref{fig:t4pspcepoten} for TIP4P and SPC/E, both using the same ion parameters.
As the simulations are done at constant volume, the enthalpy is obtained using the average potential energy $U$.
The free energy $F$ is nothing else but the potential of mean force, from which we can extract the entropic component through $F-U=-T\Delta S$, in a manner similar to previous studies.\cite{galli_Zhang2020}
Thermostatting is required for this decomposition, otherwise we don't get realistic energetic partitioning as the ions are separated, thus the partitioned simulations were done in the NVT ensemble.

Two interesting observations can me made. 
Firstly, the SSIP is more stable than the CIP when only the total potential energy (enthalpic) contribution is accounted for.
This indicates the CIP stabilization at room temperature is dominated by the entropic contribution.
Moreover, the entropy is relatively constant along larger interionic separation distances, up to the transition barrier region between the SSIP and the CIP states.
Here, the two systems' entropic contributions increase with similar slope. 
Note we plot $-T\Delta S$ in the figure, thus the entropy is \textit{increasing} in the regime between transition barrier peak inwards to the CIP.
This entropic stabilization can be associated to the entropy increase of the liquid by the release of the water molecule(s) involved in the
SSIP shared solvation shell.
These results, obtained with models that only differ on their electrostatic description of water,
indicate that the model-dependent dielectric screening of the ions significantly contributes to the PMF's differences. 
We note that the dielectric constant of TIP4P water is $\sim$ 50 while that of SPC/E water is $\sim$ 70.\cite{spct4pdielconst_Elton2014}
Larger screening effects will soften the enthalpic contribution to the SSIP.
At the same time, a larger dielectric constant will also result in a larger entropy increase, as seen in Fig.~\ref{fig:t4pspcetent}, which shows that the SPC/E entropy is larger than that of TIP4P.
Hence, this decomposition unveils that the stability of the two minima is a balance of opposite interactions, and small changes in the model can result in big differences in the PMF.
This might be even more significant in polarizable
models, explaining why DFT-based simulations
are much more sensitive to small changes in simulation conditions and XC functionals.

It is not feasible to obtain an equivalent figure for the DFT simulations, given that one would need nanosecond length simulations to obtain a potential energy plot discernible from its fluctuations.
It is possible to attempt to obtain an estimate of the entropic contribution using local connectivity fluctuations through Voronoi tesselations. 
We present such a study in the Supplementary Information, but notions of entropy in such characterizations are not readily connected to the physical entropies we produce above.
Nevertheless, the behavior of local ionic volumes reveal markedly different behavior in classical and \textit{ab initio} systems, as shown in Fig.~\prettyref{supp-fig:vorvols}.
Overall, these results indicate that the entropic effects are very similar for all models.
The leading conclusion for this points to the importance of polarizability and overall dielectric properties of water as the source of the large variations of PMF energies observed across the models.

\section{Conclusion}

We have reconstructed potentials of mean force for \ce{NaCl} solutions of different molarities using solvent number densities corresponding to normal and almost ice-like water, in both the classical and \textit{ab initio} regime.
The goal of this study was to understand the differences 
observed in PMFs computed using different models, sampling methods,
and simulation parameters.
Overall we find that when simulation-based parameters are converged, classical, non-polarizable models produce similar PMFs when
using the same ion parameters.
In all classical simulations, the CIP solvation state is more stable than the SSIP state under the conditions chosen.
Moreover, the depth of the CIP free energy minimum is independent of the specific water model.
The choice of water model does, however, have a greater effect on the SSIP stability.
By decomposing the free energy into entropic and enthalpic contributions we show that differences between models 
are likely rooted in the dielectric screening of the ions, which for non-polarizable models mostly affects the SSIP.

On the other hand, \textit{ab initio} simulations have much more variability in the solvation state stabilities.
In particular, the solvent density has a much more profound impact on the energetically preferred state.
These results point to a much larger sensitivity of the polarizability and dielectric response in these models to the thermodynamic conditions.

A detailed study of how the dielectric screening influences the ion pair association and stability will be the 
subject of a future research study.

\section{Supplementary Material} 
In the supplementary material, various analyses are presented. We analyze the bulk thermodynamic characteristics of the simulations in consideration and evaluate effects on the solution equilibrium caused by the ions. These results are presented in Table~\prettyref{supp-tbl:tdynavgs}. In Tables~\prettyref{supp-tbl:tdynavgs} and \prettyref{supp-tbl:vorvol}, we present data from Voronoi tesselations that characterizes volumetric differences between the ions in different conditions, as well as analyze a type of entropy arising from the connectivity of the tesselation. Lastly, we discuss structural effects the ions have on the solvent beyond what is presented in the main body, with a numerical summary provided in Table~\prettyref{supp-tbl:rdfs}. Figures contained within the supplementary material include those related to the above mentioned analyses and those referenced in the main body of the text that were moved for formatting concerns. Template input files with all used parameters are available upon request.

\begin{acknowledgments}
This material is based upon work supported by the Department of Energy under award number DE-SC0001137.
In addition, this research used resources of the National Energy Research Scientific Computing Center, a DOE Office of Science User Facility supported by the Office of Science of the U.S. Department of Energy under Contract No. DE-AC02-05CH11231.
We also thank Stony Brook Research Computing and Cyberinfrastructure, and the Institute for Advanced Computational Science at Stony Brook University for access to the high-performance SeaWulf computing system, which was made possible by a National Science Foundation grant No. 1531492.
\end{acknowledgments}

\section{Data Availability Statement}
The data that support the findings of this study are available from the corresponding author upon reasonable request and within the article's supplementary material.

\bibliography{__refs}

\begin{thebibliography}{61}%
\makeatletter
\providecommand \@ifxundefined [1]{%
 \@ifx{#1\undefined}
}%
\providecommand \@ifnum [1]{%
 \ifnum #1\expandafter \@firstoftwo
 \else \expandafter \@secondoftwo
 \fi
}%
\providecommand \@ifx [1]{%
 \ifx #1\expandafter \@firstoftwo
 \else \expandafter \@secondoftwo
 \fi
}%
\providecommand \natexlab [1]{#1}%
\providecommand \enquote  [1]{``#1''}%
\providecommand \bibnamefont  [1]{#1}%
\providecommand \bibfnamefont [1]{#1}%
\providecommand \citenamefont [1]{#1}%
\providecommand \href@noop [0]{\@secondoftwo}%
\providecommand \href [0]{\begingroup \@sanitize@url \@href}%
\providecommand \@href[1]{\@@startlink{#1}\@@href}%
\providecommand \@@href[1]{\endgroup#1\@@endlink}%
\providecommand \@sanitize@url [0]{\catcode `\\12\catcode `\$12\catcode
  `\&12\catcode `\#12\catcode `\^12\catcode `\_12\catcode `\%12\relax}%
\providecommand \@@startlink[1]{}%
\providecommand \@@endlink[0]{}%
\providecommand \url  [0]{\begingroup\@sanitize@url \@url }%
\providecommand \@url [1]{\endgroup\@href {#1}{\urlprefix }}%
\providecommand \urlprefix  [0]{URL }%
\providecommand \Eprint [0]{\href }%
\providecommand \doibase [0]{http://dx.doi.org/}%
\providecommand \selectlanguage [0]{\@gobble}%
\providecommand \bibinfo  [0]{\@secondoftwo}%
\providecommand \bibfield  [0]{\@secondoftwo}%
\providecommand \translation [1]{[#1]}%
\providecommand \BibitemOpen [0]{}%
\providecommand \bibitemStop [0]{}%
\providecommand \bibitemNoStop [0]{.\EOS\space}%
\providecommand \EOS [0]{\spacefactor3000\relax}%
\providecommand \BibitemShut  [1]{\csname bibitem#1\endcsname}%
\let\auto@bib@innerbib\@empty
\bibitem [{\citenamefont {Waluyo}\ \emph {et~al.}(2014)\citenamefont {Waluyo},
  \citenamefont {Nordlund}, \citenamefont {Bergmann}, \citenamefont
  {Schlesinger}, \citenamefont {Pettersson},\ and\ \citenamefont
  {Nilsson}}]{xraychaoWaluyo2014}%
  \BibitemOpen
  \bibfield  {author} {\bibinfo {author} {\bibfnamefont {I.}~\bibnamefont
  {Waluyo}}, \bibinfo {author} {\bibfnamefont {D.}~\bibnamefont {Nordlund}},
  \bibinfo {author} {\bibfnamefont {U.}~\bibnamefont {Bergmann}}, \bibinfo
  {author} {\bibfnamefont {D.}~\bibnamefont {Schlesinger}}, \bibinfo {author}
  {\bibfnamefont {L.~G.~M.}\ \bibnamefont {Pettersson}}, \ and\ \bibinfo
  {author} {\bibfnamefont {A.}~\bibnamefont {Nilsson}},\ }\bibfield  {title}
  {\enquote {\bibinfo {title} {{A different view of structure-making and
  structure-breaking in alkali halide aqueous solutions through x-ray
  absorption spectroscopy}},}\ }\href {\doibase 10.1063/1.4881600} {\bibfield
  {journal} {\bibinfo  {journal} {The Journal of Chemical Physics}\ }\textbf
  {\bibinfo {volume} {140}},\ \bibinfo {pages} {244506} (\bibinfo {year}
  {2014})}\BibitemShut {NoStop}%
\bibitem [{\citenamefont {Soper}\ and\ \citenamefont
  {Weckstr{\"{o}}m}(2006)}]{khalideSoper2006}%
  \BibitemOpen
  \bibfield  {author} {\bibinfo {author} {\bibfnamefont {A.~K.}\ \bibnamefont
  {Soper}}\ and\ \bibinfo {author} {\bibfnamefont {K.}~\bibnamefont
  {Weckstr{\"{o}}m}},\ }\bibfield  {title} {\enquote {\bibinfo {title} {{Ion
  solvation and water structure in potassium halide aqueous solutions}},}\
  }\href {\doibase 10.1016/j.bpc.2006.04.009} {\bibfield  {journal} {\bibinfo
  {journal} {Biophysical Chemistry}\ }\textbf {\bibinfo {volume} {124}},\
  \bibinfo {pages} {180--191} (\bibinfo {year} {2006})}\BibitemShut {NoStop}%
\bibitem [{\citenamefont {Mancinelli}\ \emph
  {et~al.}(2007{\natexlab{a}})\citenamefont {Mancinelli}, \citenamefont
  {Botti}, \citenamefont {Bruni}, \citenamefont {Ricci},\ and\ \citenamefont
  {Soper}}]{na_k_cl_Mancinelli2007}%
  \BibitemOpen
  \bibfield  {author} {\bibinfo {author} {\bibfnamefont {R.}~\bibnamefont
  {Mancinelli}}, \bibinfo {author} {\bibfnamefont {A.}~\bibnamefont {Botti}},
  \bibinfo {author} {\bibfnamefont {F.}~\bibnamefont {Bruni}}, \bibinfo
  {author} {\bibfnamefont {M.~A.}\ \bibnamefont {Ricci}}, \ and\ \bibinfo
  {author} {\bibfnamefont {A.~K.}\ \bibnamefont {Soper}},\ }\bibfield  {title}
  {\enquote {\bibinfo {title} {{Perturbation of water structure due to
  monovalent ions in solution}},}\ }\href {\doibase 10.1039/b701855j}
  {\bibfield  {journal} {\bibinfo  {journal} {Physical Chemistry Chemical
  Physics}\ }\textbf {\bibinfo {volume} {9}},\ \bibinfo {pages} {2959--2967}
  (\bibinfo {year} {2007}{\natexlab{a}})}\BibitemShut {NoStop}%
\bibitem [{\citenamefont {Kropman}\ and\ \citenamefont
  {Bakker}(2004)}]{nacosmoKropman2004}%
  \BibitemOpen
  \bibfield  {author} {\bibinfo {author} {\bibfnamefont {M.~F.}\ \bibnamefont
  {Kropman}}\ and\ \bibinfo {author} {\bibfnamefont {H.~J.}\ \bibnamefont
  {Bakker}},\ }\bibfield  {title} {\enquote {\bibinfo {title} {{Effect of ions
  on the vibrational relaxation of liquid water}},}\ }\href {\doibase
  10.1021/ja039147r} {\bibfield  {journal} {\bibinfo  {journal} {Journal of the
  American Chemical Society}\ }\textbf {\bibinfo {volume} {126}},\ \bibinfo
  {pages} {9135--9141} (\bibinfo {year} {2004})}\BibitemShut {NoStop}%
\bibitem [{\citenamefont {Remsing}\ \emph {et~al.}(2018)\citenamefont
  {Remsing}, \citenamefont {Duignan}, \citenamefont {Baer}, \citenamefont
  {Schenter}, \citenamefont {Mundy},\ and\ \citenamefont
  {Weeks}}]{lonepairRemsing2018}%
  \BibitemOpen
  \bibfield  {author} {\bibinfo {author} {\bibfnamefont {R.~C.}\ \bibnamefont
  {Remsing}}, \bibinfo {author} {\bibfnamefont {T.~T.}\ \bibnamefont
  {Duignan}}, \bibinfo {author} {\bibfnamefont {M.~D.}\ \bibnamefont {Baer}},
  \bibinfo {author} {\bibfnamefont {G.~K.}\ \bibnamefont {Schenter}}, \bibinfo
  {author} {\bibfnamefont {C.~J.}\ \bibnamefont {Mundy}}, \ and\ \bibinfo
  {author} {\bibfnamefont {J.~D.}\ \bibnamefont {Weeks}},\ }\bibfield  {title}
  {\enquote {\bibinfo {title} {{Water Lone Pair Delocalization in Classical and
  Quantum Descriptions of the Hydration of Model Ions}},}\ }\href {\doibase
  10.1021/acs.jpcb.7b10722} {\bibfield  {journal} {\bibinfo  {journal} {Journal
  of Physical Chemistry B}\ }\textbf {\bibinfo {volume} {122}},\ \bibinfo
  {pages} {3519--3527} (\bibinfo {year} {2018})}\BibitemShut {NoStop}%
\bibitem [{\citenamefont {Fifen}\ and\ \citenamefont
  {Agmon}(2019)}]{naradFifen2019}%
  \BibitemOpen
  \bibfield  {author} {\bibinfo {author} {\bibfnamefont {J.~J.}\ \bibnamefont
  {Fifen}}\ and\ \bibinfo {author} {\bibfnamefont {N.}~\bibnamefont {Agmon}},\
  }\bibfield  {title} {\enquote {\bibinfo {title} {{Ionic radii of hydrated
  sodium cation from QTAIM}},}\ }\href {\doibase 10.1063/1.5020150} {\bibfield
  {journal} {\bibinfo  {journal} {The Journal of Chemical Physics}\ }\textbf
  {\bibinfo {volume} {150}},\ \bibinfo {pages} {034304} (\bibinfo {year}
  {2019})}\BibitemShut {NoStop}%
\bibitem [{\citenamefont {Khavrutskii}, \citenamefont {Dzubiella},\ and\
  \citenamefont {McCammon}(2008)}]{pmf1charmmggahfbKhavrutskii2008}%
  \BibitemOpen
  \bibfield  {author} {\bibinfo {author} {\bibfnamefont {I.~V.}\ \bibnamefont
  {Khavrutskii}}, \bibinfo {author} {\bibfnamefont {J.}~\bibnamefont
  {Dzubiella}}, \ and\ \bibinfo {author} {\bibfnamefont {J.~A.}\ \bibnamefont
  {McCammon}},\ }\bibfield  {title} {\enquote {\bibinfo {title} {{Computing
  accurate potentials of mean force in electrolyte solutions with the
  generalized gradient-augmented harmonic Fourier beads method}},}\ }\href
  {\doibase 10.1063/1.2825620} {\bibfield  {journal} {\bibinfo  {journal}
  {Journal of Chemical Physics}\ }\textbf {\bibinfo {volume} {128}} (\bibinfo
  {year} {2008}),\ 10.1063/1.2825620}\BibitemShut {NoStop}%
\bibitem [{\citenamefont {Timko}, \citenamefont {Bucher},\ and\ \citenamefont
  {Kuyucak}(2010)}]{pmf2_classical_cpmd_Timko2010}%
  \BibitemOpen
  \bibfield  {author} {\bibinfo {author} {\bibfnamefont {J.}~\bibnamefont
  {Timko}}, \bibinfo {author} {\bibfnamefont {D.}~\bibnamefont {Bucher}}, \
  and\ \bibinfo {author} {\bibfnamefont {S.}~\bibnamefont {Kuyucak}},\
  }\bibfield  {title} {\enquote {\bibinfo {title} {{Dissociation of NaCl in
  water from ab initio molecular dynamics simulations}},}\ }\href {\doibase
  10.1063/1.3360310} {\bibfield  {journal} {\bibinfo  {journal} {Journal of
  Chemical Physics}\ }\textbf {\bibinfo {volume} {132}} (\bibinfo {year}
  {2010}),\ 10.1063/1.3360310}\BibitemShut {NoStop}%
\bibitem [{\citenamefont {Ghosh}\ \emph {et~al.}(2013)\citenamefont {Ghosh},
  \citenamefont {Re}, \citenamefont {Feig}, \citenamefont {Sugita},\ and\
  \citenamefont {Choi}}]{pmf3_classical_ihs_qmefp_Ghosh2013}%
  \BibitemOpen
  \bibfield  {author} {\bibinfo {author} {\bibfnamefont {M.~K.}\ \bibnamefont
  {Ghosh}}, \bibinfo {author} {\bibfnamefont {S.}~\bibnamefont {Re}}, \bibinfo
  {author} {\bibfnamefont {M.}~\bibnamefont {Feig}}, \bibinfo {author}
  {\bibfnamefont {Y.}~\bibnamefont {Sugita}}, \ and\ \bibinfo {author}
  {\bibfnamefont {C.~H.}\ \bibnamefont {Choi}},\ }\bibfield  {title} {\enquote
  {\bibinfo {title} {{Interionic Hydration Structures of NaCl in Aqueous
  Solution: A Combined Study of Quantum Mechanical Cluster Calculations and
  QM/EFP-MD Simulations}},}\ }\href {\doibase 10.1021/jp308731z} {\bibfield
  {journal} {\bibinfo  {journal} {The Journal of Physical Chemistry B}\
  }\textbf {\bibinfo {volume} {117}},\ \bibinfo {pages} {289--295} (\bibinfo
  {year} {2013})}\BibitemShut {NoStop}%
\bibitem [{\citenamefont {Yao}\ and\ \citenamefont
  {Kanai}(2018)}]{pmf4_classical_scancpmd_Yao2018}%
  \BibitemOpen
  \bibfield  {author} {\bibinfo {author} {\bibfnamefont {Y.}~\bibnamefont
  {Yao}}\ and\ \bibinfo {author} {\bibfnamefont {Y.}~\bibnamefont {Kanai}},\
  }\bibfield  {title} {\enquote {\bibinfo {title} {{Free Energy Profile of NaCl
  in Water: First-Principles Molecular Dynamics with SCAN and $\omega$B97X-V
  Exchange–Correlation Functionals}},}\ }\href {\doibase
  10.1021/acs.jctc.7b00846} {\bibfield  {journal} {\bibinfo  {journal} {Journal
  of Chemical Theory and Computation}\ }\textbf {\bibinfo {volume} {14}},\
  \bibinfo {pages} {884--893} (\bibinfo {year} {2018})}\BibitemShut {NoStop}%
\bibitem [{\citenamefont {Roy}\ \emph {et~al.}(2017)\citenamefont {Roy},
  \citenamefont {Baer}, \citenamefont {Mundy},\ and\ \citenamefont
  {Schenter}}]{pmf5_classical_marcusthry_Roy2017}%
  \BibitemOpen
  \bibfield  {author} {\bibinfo {author} {\bibfnamefont {S.}~\bibnamefont
  {Roy}}, \bibinfo {author} {\bibfnamefont {M.~D.}\ \bibnamefont {Baer}},
  \bibinfo {author} {\bibfnamefont {C.~J.}\ \bibnamefont {Mundy}}, \ and\
  \bibinfo {author} {\bibfnamefont {G.~K.}\ \bibnamefont {Schenter}},\
  }\bibfield  {title} {\enquote {\bibinfo {title} {{Marcus Theory of
  Ion-Pairing}},}\ }\href {\doibase 10.1021/acs.jctc.7b00332} {\bibfield
  {journal} {\bibinfo  {journal} {Journal of Chemical Theory and Computation}\
  }\textbf {\bibinfo {volume} {13}},\ \bibinfo {pages} {3470--3477} (\bibinfo
  {year} {2017})}\BibitemShut {NoStop}%
\bibitem [{\citenamefont {Car}\ and\ \citenamefont
  {Parrinello}(1985)}]{cpmdCar1985}%
  \BibitemOpen
  \bibfield  {author} {\bibinfo {author} {\bibfnamefont {R.}~\bibnamefont
  {Car}}\ and\ \bibinfo {author} {\bibfnamefont {M.}~\bibnamefont
  {Parrinello}},\ }\bibfield  {title} {\enquote {\bibinfo {title} {{Unified
  Approach for Molecular Dynamics and Density-Functional Theory}},}\ }\href
  {\doibase 10.1103/PhysRevLett.55.2471} {\bibfield  {journal} {\bibinfo
  {journal} {Physical Review Letters}\ }\textbf {\bibinfo {volume} {55}},\
  \bibinfo {pages} {2471--2474} (\bibinfo {year} {1985})}\BibitemShut {NoStop}%
\bibitem [{\citenamefont {Zhang}\ \emph {et~al.}(2020)\citenamefont {Zhang},
  \citenamefont {Giberti}, \citenamefont {Sevgen}, \citenamefont {de~Pablo},
  \citenamefont {Gygi},\ and\ \citenamefont {Galli}}]{galli_Zhang2020}%
  \BibitemOpen
  \bibfield  {author} {\bibinfo {author} {\bibfnamefont {C.}~\bibnamefont
  {Zhang}}, \bibinfo {author} {\bibfnamefont {F.}~\bibnamefont {Giberti}},
  \bibinfo {author} {\bibfnamefont {E.}~\bibnamefont {Sevgen}}, \bibinfo
  {author} {\bibfnamefont {J.~J.}\ \bibnamefont {de~Pablo}}, \bibinfo {author}
  {\bibfnamefont {F.}~\bibnamefont {Gygi}}, \ and\ \bibinfo {author}
  {\bibfnamefont {G.}~\bibnamefont {Galli}},\ }\bibfield  {title} {\enquote
  {\bibinfo {title} {{Dissociation of salts in water under pressure}},}\ }\href
  {\doibase 10.1038/s41467-020-16704-9} {\bibfield  {journal} {\bibinfo
  {journal} {Nature Communications}\ }\textbf {\bibinfo {volume} {11}},\
  \bibinfo {pages} {3037} (\bibinfo {year} {2020})}\BibitemShut {NoStop}%
\bibitem [{\citenamefont {Gu{\`{a}}rdia}, \citenamefont {Rey},\ and\
  \citenamefont {Padr{\'{o}}}(1991)}]{response1_Guardia1991}%
  \BibitemOpen
  \bibfield  {author} {\bibinfo {author} {\bibfnamefont {E.}~\bibnamefont
  {Gu{\`{a}}rdia}}, \bibinfo {author} {\bibfnamefont {R.}~\bibnamefont {Rey}},
  \ and\ \bibinfo {author} {\bibfnamefont {J.}~\bibnamefont {Padr{\'{o}}}},\
  }\bibfield  {title} {\enquote {\bibinfo {title} {{Potential of mean force by
  constrained molecular dynamics: A sodium chloride ion-pair in water}},}\
  }\href {\doibase 10.1016/0301-0104(91)87019-R} {\bibfield  {journal}
  {\bibinfo  {journal} {Chemical Physics}\ }\textbf {\bibinfo {volume} {155}},\
  \bibinfo {pages} {187--195} (\bibinfo {year} {1991})}\BibitemShut {NoStop}%
\bibitem [{\citenamefont {Mancinelli}\ \emph
  {et~al.}(2007{\natexlab{b}})\citenamefont {Mancinelli}, \citenamefont
  {Botti}, \citenamefont {Bruni}, \citenamefont {Ricci},\ and\ \citenamefont
  {Soper}}]{perturbgood_Mancinelli2007}%
  \BibitemOpen
  \bibfield  {author} {\bibinfo {author} {\bibfnamefont {R.}~\bibnamefont
  {Mancinelli}}, \bibinfo {author} {\bibfnamefont {A.}~\bibnamefont {Botti}},
  \bibinfo {author} {\bibfnamefont {F.}~\bibnamefont {Bruni}}, \bibinfo
  {author} {\bibfnamefont {M.~A.}\ \bibnamefont {Ricci}}, \ and\ \bibinfo
  {author} {\bibfnamefont {A.~K.}\ \bibnamefont {Soper}},\ }\bibfield  {title}
  {\enquote {\bibinfo {title} {{Perturbation of water structure due to
  monovalent ions in solution}},}\ }\href {\doibase 10.1039/b701855j}
  {\bibfield  {journal} {\bibinfo  {journal} {Physical Chemistry Chemical
  Physics}\ }\textbf {\bibinfo {volume} {9}},\ \bibinfo {pages} {2959--2967}
  (\bibinfo {year} {2007}{\natexlab{b}})}\BibitemShut {NoStop}%
\bibitem [{\citenamefont {Mancinelli}\ \emph
  {et~al.}(2007{\natexlab{c}})\citenamefont {Mancinelli}, \citenamefont
  {Botti}, \citenamefont {Bruni}, \citenamefont {Ricci},\ and\ \citenamefont
  {Soper}}]{perturbbad_Mancinelli2007}%
  \BibitemOpen
  \bibfield  {author} {\bibinfo {author} {\bibfnamefont {R.}~\bibnamefont
  {Mancinelli}}, \bibinfo {author} {\bibfnamefont {A.}~\bibnamefont {Botti}},
  \bibinfo {author} {\bibfnamefont {F.}~\bibnamefont {Bruni}}, \bibinfo
  {author} {\bibfnamefont {M.~A.}\ \bibnamefont {Ricci}}, \ and\ \bibinfo
  {author} {\bibfnamefont {A.~K.}\ \bibnamefont {Soper}},\ }\bibfield  {title}
  {\enquote {\bibinfo {title} {{Hydration of sodium, potassium, and chloride
  ions in solution and the concept of structure maker/breaker}},}\ }\href
  {\doibase 10.1021/jp075913v} {\bibfield  {journal} {\bibinfo  {journal}
  {Journal of Physical Chemistry B}\ }\textbf {\bibinfo {volume} {111}},\
  \bibinfo {pages} {13570--13577} (\bibinfo {year}
  {2007}{\natexlab{c}})}\BibitemShut {NoStop}%
\bibitem [{\citenamefont {Waluyo}\ \emph {et~al.}(2011)\citenamefont {Waluyo},
  \citenamefont {Huang}, \citenamefont {Nordlund}, \citenamefont {Bergmann},
  \citenamefont {Weiss}, \citenamefont {Pettersson},\ and\ \citenamefont
  {Nilsson}}]{nashell1_Waluyo2011}%
  \BibitemOpen
  \bibfield  {author} {\bibinfo {author} {\bibfnamefont {I.}~\bibnamefont
  {Waluyo}}, \bibinfo {author} {\bibfnamefont {C.}~\bibnamefont {Huang}},
  \bibinfo {author} {\bibfnamefont {D.}~\bibnamefont {Nordlund}}, \bibinfo
  {author} {\bibfnamefont {U.}~\bibnamefont {Bergmann}}, \bibinfo {author}
  {\bibfnamefont {T.~M.}\ \bibnamefont {Weiss}}, \bibinfo {author}
  {\bibfnamefont {L.~G.~M.}\ \bibnamefont {Pettersson}}, \ and\ \bibinfo
  {author} {\bibfnamefont {A.}~\bibnamefont {Nilsson}},\ }\bibfield  {title}
  {\enquote {\bibinfo {title} {{The structure of water in the hydration shell
  of cations from x-ray Raman and small angle x-ray scattering
  measurements}},}\ }\href {\doibase 10.1063/1.3533958} {\bibfield  {journal}
  {\bibinfo  {journal} {The Journal of Chemical Physics}\ }\textbf {\bibinfo
  {volume} {134}},\ \bibinfo {pages} {064513} (\bibinfo {year}
  {2011})}\BibitemShut {NoStop}%
\bibitem [{\citenamefont {Gaiduk}\ and\ \citenamefont
  {Galli}(2017)}]{naclrdfs_Gaiduk2017}%
  \BibitemOpen
  \bibfield  {author} {\bibinfo {author} {\bibfnamefont {A.~P.}\ \bibnamefont
  {Gaiduk}}\ and\ \bibinfo {author} {\bibfnamefont {G.}~\bibnamefont {Galli}},\
  }\bibfield  {title} {\enquote {\bibinfo {title} {{Local and Global Effects of
  Dissolved Sodium Chloride on the Structure of Water}},}\ }\href {\doibase
  10.1021/acs.jpclett.7b00239} {\bibfield  {journal} {\bibinfo  {journal} {The
  Journal of Physical Chemistry Letters}\ }\textbf {\bibinfo {volume} {8}},\
  \bibinfo {pages} {1496--1502} (\bibinfo {year} {2017})}\BibitemShut {NoStop}%
\bibitem [{\citenamefont {Xu}\ \emph {et~al.}(2021)\citenamefont {Xu},
  \citenamefont {Sun}, \citenamefont {Zhang}, \citenamefont {DelloStritto},
  \citenamefont {Lu}, \citenamefont {Klein},\ and\ \citenamefont
  {Wu}}]{nqestructcl_Xu2021}%
  \BibitemOpen
  \bibfield  {author} {\bibinfo {author} {\bibfnamefont {J.}~\bibnamefont
  {Xu}}, \bibinfo {author} {\bibfnamefont {Z.}~\bibnamefont {Sun}}, \bibinfo
  {author} {\bibfnamefont {C.}~\bibnamefont {Zhang}}, \bibinfo {author}
  {\bibfnamefont {M.}~\bibnamefont {DelloStritto}}, \bibinfo {author}
  {\bibfnamefont {D.}~\bibnamefont {Lu}}, \bibinfo {author} {\bibfnamefont
  {M.~L.}\ \bibnamefont {Klein}}, \ and\ \bibinfo {author} {\bibfnamefont
  {X.}~\bibnamefont {Wu}},\ }\bibfield  {title} {\enquote {\bibinfo {title}
  {{Importance of nuclear quantum effects on the hydration of chloride ion}},}\
  }\href {\doibase 10.1103/PhysRevMaterials.5.L012801} {\bibfield  {journal}
  {\bibinfo  {journal} {Physical Review Materials}\ }\textbf {\bibinfo {volume}
  {5}},\ \bibinfo {pages} {L012801} (\bibinfo {year} {2021})},\ \Eprint
  {http://arxiv.org/abs/2009.07304} {arXiv:2009.07304} \BibitemShut {NoStop}%
\bibitem [{\citenamefont {DelloStritto}\ \emph {et~al.}(2020)\citenamefont
  {DelloStritto}, \citenamefont {Xu}, \citenamefont {Wu},\ and\ \citenamefont
  {Klein}}]{polarzionstruct_DelloStritto2020}%
  \BibitemOpen
  \bibfield  {author} {\bibinfo {author} {\bibfnamefont {M.}~\bibnamefont
  {DelloStritto}}, \bibinfo {author} {\bibfnamefont {J.}~\bibnamefont {Xu}},
  \bibinfo {author} {\bibfnamefont {X.}~\bibnamefont {Wu}}, \ and\ \bibinfo
  {author} {\bibfnamefont {M.~L.}\ \bibnamefont {Klein}},\ }\bibfield  {title}
  {\enquote {\bibinfo {title} {{Aqueous solvation of the chloride ion revisited
  with density functional theory: impact of correlation and exchange
  approximations}},}\ }\href {\doibase 10.1039/C9CP06821J} {\bibfield
  {journal} {\bibinfo  {journal} {Physical Chemistry Chemical Physics}\
  }\textbf {\bibinfo {volume} {22}},\ \bibinfo {pages} {10666--10675} (\bibinfo
  {year} {2020})}\BibitemShut {NoStop}%
\bibitem [{\citenamefont {{J. Perdew, K. Burke, M. Ernzerhof}}(1996)}]{pbe}%
  \BibitemOpen
  \bibfield  {author} {\bibinfo {author} {\bibnamefont {{J. Perdew, K. Burke,
  M. Ernzerhof}}},\ }\bibfield  {title} {\enquote {\bibinfo {title}
  {{Generalized gradient approximation made simple}},}\ }\href@noop {}
  {\bibfield  {journal} {\bibinfo  {journal} {Phys. Rev. Lett.}\ }\textbf
  {\bibinfo {volume} {77}},\ \bibinfo {pages} {3865--3868} (\bibinfo {year}
  {1996})}\BibitemShut {NoStop}%
\bibitem [{\citenamefont {{K. Berland, P. Hyldgaard}}(2014)}]{bh}%
  \BibitemOpen
  \bibfield  {author} {\bibinfo {author} {\bibnamefont {{K. Berland, P.
  Hyldgaard}}},\ }\bibfield  {title} {\enquote {\bibinfo {title} {{Exchange
  functional that tests the robustness of the plasmon description of the van
  der Waals density functional}},}\ }\href@noop {} {\bibfield  {journal}
  {\bibinfo  {journal} {Phys. Rev. B}\ }\textbf {\bibinfo {volume} {89}},\
  \bibinfo {pages} {1--8} (\bibinfo {year} {2014})}\BibitemShut {NoStop}%
\bibitem [{\citenamefont {Fritz}, \citenamefont {Fern{\'{a}}ndez-Serra},\ and\
  \citenamefont {Soler}(2016)}]{marivi_dpps_vdwrhoeq_Fritz2016}%
  \BibitemOpen
  \bibfield  {author} {\bibinfo {author} {\bibfnamefont {M.}~\bibnamefont
  {Fritz}}, \bibinfo {author} {\bibfnamefont {M.}~\bibnamefont
  {Fern{\'{a}}ndez-Serra}}, \ and\ \bibinfo {author} {\bibfnamefont {J.~M.}\
  \bibnamefont {Soler}},\ }\bibfield  {title} {\enquote {\bibinfo {title}
  {{Optimization of an exchange-correlation density functional for water}},}\
  }\href {\doibase 10.1063/1.4953081} {\bibfield  {journal} {\bibinfo
  {journal} {Journal of Chemical Physics}\ }\textbf {\bibinfo {volume} {144}}
  (\bibinfo {year} {2016}),\ 10.1063/1.4953081},\ \Eprint
  {http://arxiv.org/abs/1603.07302} {arXiv:1603.07302} \BibitemShut {NoStop}%
\bibitem [{\citenamefont {Kirkwood}(1935)}]{pmfdefKirkwood1935}%
  \BibitemOpen
  \bibfield  {author} {\bibinfo {author} {\bibfnamefont {J.~G.}\ \bibnamefont
  {Kirkwood}},\ }\bibfield  {title} {\enquote {\bibinfo {title} {{Statistical
  mechanics of fluid mixtures}},}\ }\href {\doibase 10.1063/1.1749657}
  {\bibfield  {journal} {\bibinfo  {journal} {The Journal of Chemical Physics}\
  }\textbf {\bibinfo {volume} {3}},\ \bibinfo {pages} {300--313} (\bibinfo
  {year} {1935})}\BibitemShut {NoStop}%
\bibitem [{\citenamefont {Li}\ \emph {et~al.}(2008)\citenamefont {Li},
  \citenamefont {Duan}, \citenamefont {Zhang}, \citenamefont {Zhang},\ and\
  \citenamefont {Weare}}]{dynamresLi2008}%
  \BibitemOpen
  \bibfield  {author} {\bibinfo {author} {\bibfnamefont {M.}~\bibnamefont
  {Li}}, \bibinfo {author} {\bibfnamefont {Z.}~\bibnamefont {Duan}}, \bibinfo
  {author} {\bibfnamefont {Z.}~\bibnamefont {Zhang}}, \bibinfo {author}
  {\bibfnamefont {C.}~\bibnamefont {Zhang}}, \ and\ \bibinfo {author}
  {\bibfnamefont {J.}~\bibnamefont {Weare}},\ }\bibfield  {title} {\enquote
  {\bibinfo {title} {{The structure, dynamics and solvation mechanisms of ions
  in water from long time molecular dynamics simulations: a case study of CaCl
  2 (aq) aqueous solutions}},}\ }\href {\doibase 10.1080/00268970802634981}
  {\bibfield  {journal} {\bibinfo  {journal} {Molecular Physics}\ }\textbf
  {\bibinfo {volume} {106}},\ \bibinfo {pages} {2685--2697} (\bibinfo {year}
  {2008})}\BibitemShut {NoStop}%
\bibitem [{\citenamefont {Kumar}\ \emph {et~al.}(1992)\citenamefont {Kumar},
  \citenamefont {Rosenberg}, \citenamefont {Bouzida}, \citenamefont
  {Swendsen},\ and\ \citenamefont {Kollman}}]{whamKumar1992}%
  \BibitemOpen
  \bibfield  {author} {\bibinfo {author} {\bibfnamefont {S.}~\bibnamefont
  {Kumar}}, \bibinfo {author} {\bibfnamefont {J.~M.}\ \bibnamefont
  {Rosenberg}}, \bibinfo {author} {\bibfnamefont {D.}~\bibnamefont {Bouzida}},
  \bibinfo {author} {\bibfnamefont {R.~H.}\ \bibnamefont {Swendsen}}, \ and\
  \bibinfo {author} {\bibfnamefont {P.~A.}\ \bibnamefont {Kollman}},\
  }\bibfield  {title} {\enquote {\bibinfo {title} {{THE weighted histogram
  analysis method for free‐energy calculations on biomolecules. I. The
  method}},}\ }\href {\doibase 10.1002/jcc.540130812} {\bibfield  {journal}
  {\bibinfo  {journal} {Journal of Computational Chemistry}\ }\textbf {\bibinfo
  {volume} {13}},\ \bibinfo {pages} {1011--1021} (\bibinfo {year}
  {1992})}\BibitemShut {NoStop}%
\bibitem [{\citenamefont {Li}\ \emph {et~al.}(2007)\citenamefont {Li},
  \citenamefont {Car}, \citenamefont {Tang},\ and\ \citenamefont
  {Wingreen}}]{pmfderivLi2007}%
  \BibitemOpen
  \bibfield  {author} {\bibinfo {author} {\bibfnamefont {J.-L.}\ \bibnamefont
  {Li}}, \bibinfo {author} {\bibfnamefont {R.}~\bibnamefont {Car}}, \bibinfo
  {author} {\bibfnamefont {C.}~\bibnamefont {Tang}}, \ and\ \bibinfo {author}
  {\bibfnamefont {N.~S.}\ \bibnamefont {Wingreen}},\ }\bibfield  {title}
  {\enquote {\bibinfo {title} {{Hydrophobic interaction and hydrogen-bond
  network for a methane pair in liquid water}},}\ }\href {\doibase
  10.1073/pnas.0610945104} {\bibfield  {journal} {\bibinfo  {journal}
  {Proceedings of the National Academy of Sciences}\ }\textbf {\bibinfo
  {volume} {104}},\ \bibinfo {pages} {2626--2630} (\bibinfo {year}
  {2007})}\BibitemShut {NoStop}%
\bibitem [{\citenamefont {Ciccotti}\ and\ \citenamefont
  {Ferrario}(2018)}]{jacobian_Ciccotti2018}%
  \BibitemOpen
  \bibfield  {author} {\bibinfo {author} {\bibfnamefont {G.}~\bibnamefont
  {Ciccotti}}\ and\ \bibinfo {author} {\bibfnamefont {M.}~\bibnamefont
  {Ferrario}},\ }\bibfield  {title} {\enquote {\bibinfo {title} {{Holonomic
  Constraints: A Case for Statistical Mechanics of Non-Hamiltonian Systems}},}\
  }\href {\doibase 10.3390/computation6010011} {\bibfield  {journal} {\bibinfo
  {journal} {Computation}\ }\textbf {\bibinfo {volume} {6}},\ \bibinfo {pages}
  {11} (\bibinfo {year} {2018})}\BibitemShut {NoStop}%
\bibitem [{\citenamefont {Dougherty}(2001)}]{appdensDougherty2001}%
  \BibitemOpen
  \bibfield  {author} {\bibinfo {author} {\bibfnamefont {R.~C.}\ \bibnamefont
  {Dougherty}},\ }\bibfield  {title} {\enquote {\bibinfo {title} {{Density of
  Salt Solutions: Effect of Ions on the Apparent Density of Water}},}\ }\href
  {\doibase 10.1021/jp010097r} {\bibfield  {journal} {\bibinfo  {journal} {The
  Journal of Physical Chemistry B}\ }\textbf {\bibinfo {volume} {105}},\
  \bibinfo {pages} {4514--4519} (\bibinfo {year} {2001})}\BibitemShut {NoStop}%
\bibitem [{\citenamefont {Novotn{\'{y}}}\ and\ \citenamefont
  {S{\"{o}}hnel}(1988)}]{phenomdensNovotny1988}%
  \BibitemOpen
  \bibfield  {author} {\bibinfo {author} {\bibfnamefont {P.}~\bibnamefont
  {Novotn{\'{y}}}}\ and\ \bibinfo {author} {\bibfnamefont {O.}~\bibnamefont
  {S{\"{o}}hnel}},\ }\bibfield  {title} {\enquote {\bibinfo {title} {{Densities
  of Binary Aqueous Solutions of 306 Inorganic Substances}},}\ }\href {\doibase
  10.1021/je00051a018} {\bibfield  {journal} {\bibinfo  {journal} {Journal of
  Chemical and Engineering Data}\ }\textbf {\bibinfo {volume} {33}},\ \bibinfo
  {pages} {49--55} (\bibinfo {year} {1988})}\BibitemShut {NoStop}%
\bibitem [{\citenamefont {Jorgensen}\ \emph {et~al.}(1983)\citenamefont
  {Jorgensen}, \citenamefont {Chandrasekhar}, \citenamefont {Madura},
  \citenamefont {Impey},\ and\ \citenamefont {Klein}}]{tip4pJorgensen1983}%
  \BibitemOpen
  \bibfield  {author} {\bibinfo {author} {\bibfnamefont {W.~L.}\ \bibnamefont
  {Jorgensen}}, \bibinfo {author} {\bibfnamefont {J.}~\bibnamefont
  {Chandrasekhar}}, \bibinfo {author} {\bibfnamefont {J.~D.}\ \bibnamefont
  {Madura}}, \bibinfo {author} {\bibfnamefont {R.~W.}\ \bibnamefont {Impey}}, \
  and\ \bibinfo {author} {\bibfnamefont {M.~L.}\ \bibnamefont {Klein}},\
  }\bibfield  {title} {\enquote {\bibinfo {title} {{Comparison of simple
  potential functions for simulating liquid water}},}\ }\href {\doibase
  10.1063/1.445869} {\bibfield  {journal} {\bibinfo  {journal} {The Journal of
  Chemical Physics}\ }\textbf {\bibinfo {volume} {79}},\ \bibinfo {pages}
  {926--935} (\bibinfo {year} {1983})}\BibitemShut {NoStop}%
\bibitem [{\citenamefont {Abraham}\ \emph {et~al.}(2015)\citenamefont
  {Abraham}, \citenamefont {Murtola}, \citenamefont {Schulz}, \citenamefont
  {P{\'{a}}ll}, \citenamefont {Smith}, \citenamefont {Hess},\ and\
  \citenamefont {Lindahl}}]{gromacsAbraham2015}%
  \BibitemOpen
  \bibfield  {author} {\bibinfo {author} {\bibfnamefont {M.~J.}\ \bibnamefont
  {Abraham}}, \bibinfo {author} {\bibfnamefont {T.}~\bibnamefont {Murtola}},
  \bibinfo {author} {\bibfnamefont {R.}~\bibnamefont {Schulz}}, \bibinfo
  {author} {\bibfnamefont {S.}~\bibnamefont {P{\'{a}}ll}}, \bibinfo {author}
  {\bibfnamefont {J.~C.}\ \bibnamefont {Smith}}, \bibinfo {author}
  {\bibfnamefont {B.}~\bibnamefont {Hess}}, \ and\ \bibinfo {author}
  {\bibfnamefont {E.}~\bibnamefont {Lindahl}},\ }\bibfield  {title} {\enquote
  {\bibinfo {title} {{GROMACS: High performance molecular simulations through
  multi-level parallelism from laptops to supercomputers}},}\ }\href {\doibase
  10.1016/j.softx.2015.06.001} {\bibfield  {journal} {\bibinfo  {journal}
  {SoftwareX}\ }\textbf {\bibinfo {volume} {1-2}},\ \bibinfo {pages} {19--25}
  (\bibinfo {year} {2015})}\BibitemShut {NoStop}%
\bibitem [{\citenamefont {Hub}, \citenamefont {de~Groot},\ and\ \citenamefont
  {van~der Spoel}(2010)}]{gwhamHub2010}%
  \BibitemOpen
  \bibfield  {author} {\bibinfo {author} {\bibfnamefont {J.~S.}\ \bibnamefont
  {Hub}}, \bibinfo {author} {\bibfnamefont {B.~L.}\ \bibnamefont {de~Groot}}, \
  and\ \bibinfo {author} {\bibfnamefont {D.}~\bibnamefont {van~der Spoel}},\
  }\bibfield  {title} {\enquote {\bibinfo {title} {{g{\_}wham—A Free Weighted
  Histogram Analysis Implementation Including Robust Error and Autocorrelation
  Estimates}},}\ }\href {\doibase 10.1021/ct100494z} {\bibfield  {journal}
  {\bibinfo  {journal} {Journal of Chemical Theory and Computation}\ }\textbf
  {\bibinfo {volume} {6}},\ \bibinfo {pages} {3713--3720} (\bibinfo {year}
  {2010})}\BibitemShut {NoStop}%
\bibitem [{\citenamefont {Grossfield}()}]{whamGrossfield}%
  \BibitemOpen
  \bibfield  {author} {\bibinfo {author} {\bibfnamefont {A.}~\bibnamefont
  {Grossfield}},\ }\href@noop {} {\enquote {\bibinfo {title} {Wham: an
  implementation of the weighted histogram analysis method},}\ }\bibinfo
  {howpublished} {http://membrane.urmc.rochester.edu/content/wham/},\ \bibinfo
  {note} {version 2.0.9}\BibitemShut {NoStop}%
\bibitem [{\citenamefont {Ryckaert}, \citenamefont {Ciccotti},\ and\
  \citenamefont {Berendsen}(1977)}]{shake_Ryckaert1977}%
  \BibitemOpen
  \bibfield  {author} {\bibinfo {author} {\bibfnamefont {J.-P.}\ \bibnamefont
  {Ryckaert}}, \bibinfo {author} {\bibfnamefont {G.}~\bibnamefont {Ciccotti}},
  \ and\ \bibinfo {author} {\bibfnamefont {H.~J.}\ \bibnamefont {Berendsen}},\
  }\bibfield  {title} {\enquote {\bibinfo {title} {{Numerical integration of
  the cartesian equations of motion of a system with constraints: molecular
  dynamics of n-alkanes}},}\ }\href {\doibase 10.1016/0021-9991(77)90098-5}
  {\bibfield  {journal} {\bibinfo  {journal} {Journal of Computational
  Physics}\ }\textbf {\bibinfo {volume} {23}},\ \bibinfo {pages} {327--341}
  (\bibinfo {year} {1977})}\BibitemShut {NoStop}%
\bibitem [{\citenamefont {Allesch}\ \emph {et~al.}(2004)\citenamefont
  {Allesch}, \citenamefont {Schwegler}, \citenamefont {Gygi},\ and\
  \citenamefont {Galli}}]{waterrigidAllesch2004}%
  \BibitemOpen
  \bibfield  {author} {\bibinfo {author} {\bibfnamefont {M.}~\bibnamefont
  {Allesch}}, \bibinfo {author} {\bibfnamefont {E.}~\bibnamefont {Schwegler}},
  \bibinfo {author} {\bibfnamefont {F.}~\bibnamefont {Gygi}}, \ and\ \bibinfo
  {author} {\bibfnamefont {G.}~\bibnamefont {Galli}},\ }\bibfield  {title}
  {\enquote {\bibinfo {title} {{A first principles simulation of rigid
  water}},}\ }\href {\doibase 10.1063/1.1647529} {\bibfield  {journal}
  {\bibinfo  {journal} {Journal of Chemical Physics}\ }\textbf {\bibinfo
  {volume} {120}},\ \bibinfo {pages} {5192--5198} (\bibinfo {year} {2004})},\
  \Eprint {http://arxiv.org/abs/0401267} {arXiv:0401267 [cond-mat]}
  \BibitemShut {NoStop}%
\bibitem [{\citenamefont {Praprotnik}\ and\ \citenamefont
  {Jane{\v{z}}i{\v{c}}}(2005)}]{watervibPraprotnik2005}%
  \BibitemOpen
  \bibfield  {author} {\bibinfo {author} {\bibfnamefont {M.}~\bibnamefont
  {Praprotnik}}\ and\ \bibinfo {author} {\bibfnamefont {D.}~\bibnamefont
  {Jane{\v{z}}i{\v{c}}}},\ }\bibfield  {title} {\enquote {\bibinfo {title}
  {{Molecular dynamics integration and molecular vibrational theory. III. The
  infrared spectrum of water}},}\ }\href {\doibase 10.1063/1.1884609}
  {\bibfield  {journal} {\bibinfo  {journal} {The Journal of Chemical Physics}\
  }\textbf {\bibinfo {volume} {122}},\ \bibinfo {pages} {174103} (\bibinfo
  {year} {2005})}\BibitemShut {NoStop}%
\bibitem [{\citenamefont {Berland}\ and\ \citenamefont
  {Hyldgaard}(2014)}]{bhBerland2014}%
  \BibitemOpen
  \bibfield  {author} {\bibinfo {author} {\bibfnamefont {K.}~\bibnamefont
  {Berland}}\ and\ \bibinfo {author} {\bibfnamefont {P.}~\bibnamefont
  {Hyldgaard}},\ }\bibfield  {title} {\enquote {\bibinfo {title} {{Exchange
  functional that tests the robustness of the plasmon description of the van
  der Waals density functional}},}\ }\href {\doibase
  10.1103/PhysRevB.89.035412} {\bibfield  {journal} {\bibinfo  {journal}
  {Physical Review B - Condensed Matter and Materials Physics}\ }\textbf
  {\bibinfo {volume} {89}},\ \bibinfo {pages} {1--8} (\bibinfo {year}
  {2014})},\ \Eprint {http://arxiv.org/abs/arXiv:1309.1756v1}
  {arXiv:arXiv:1309.1756v1} \BibitemShut {NoStop}%
\bibitem [{\citenamefont {Soper}(2013)}]{Soper2013}%
  \BibitemOpen
  \bibfield  {author} {\bibinfo {author} {\bibfnamefont {A.~K.}\ \bibnamefont
  {Soper}},\ }\bibfield  {title} {\enquote {\bibinfo {title} {{The Radial
  Distribution Functions of Water as Derived from Radiation Total Scattering
  Experiments: Is There Anything We Can Say for Sure?}}}\ }\href {\doibase
  10.1155/2013/279463} {\bibfield  {journal} {\bibinfo  {journal} {ISRN
  Physical Chemistry}\ }\textbf {\bibinfo {volume} {2013}},\ \bibinfo {pages}
  {1--67} (\bibinfo {year} {2013})}\BibitemShut {NoStop}%
\bibitem [{\citenamefont {Corsetti}\ \emph
  {et~al.}(2013{\natexlab{a}})\citenamefont {Corsetti}, \citenamefont
  {Fern{\'{a}}ndez-Serra}, \citenamefont {Soler},\ and\ \citenamefont
  {Artacho}}]{marivi_waterbasisconvCorsetti2013}%
  \BibitemOpen
  \bibfield  {author} {\bibinfo {author} {\bibfnamefont {F.}~\bibnamefont
  {Corsetti}}, \bibinfo {author} {\bibfnamefont {M.-V.}\ \bibnamefont
  {Fern{\'{a}}ndez-Serra}}, \bibinfo {author} {\bibfnamefont {J.~M.}\
  \bibnamefont {Soler}}, \ and\ \bibinfo {author} {\bibfnamefont
  {E.}~\bibnamefont {Artacho}},\ }\bibfield  {title} {\enquote {\bibinfo
  {title} {{Optimal finite-range atomic basis sets for liquid water and
  ice}},}\ }\href {\doibase 10.1088/0953-8984/25/43/435504} {\bibfield
  {journal} {\bibinfo  {journal} {Journal of Physics: Condensed Matter}\
  }\textbf {\bibinfo {volume} {25}},\ \bibinfo {pages} {435504} (\bibinfo
  {year} {2013}{\natexlab{a}})}\BibitemShut {NoStop}%
\bibitem [{\citenamefont {Geissler}, \citenamefont {Dellago},\ and\
  \citenamefont {Chandler}(1999)}]{cn2_Geissler1999}%
  \BibitemOpen
  \bibfield  {author} {\bibinfo {author} {\bibfnamefont {P.~L.}\ \bibnamefont
  {Geissler}}, \bibinfo {author} {\bibfnamefont {C.}~\bibnamefont {Dellago}}, \
  and\ \bibinfo {author} {\bibfnamefont {D.}~\bibnamefont {Chandler}},\
  }\bibfield  {title} {\enquote {\bibinfo {title} {{Kinetic Pathways of Ion
  Pair Dissociation in Water}},}\ }\href {\doibase 10.1021/jp984837g}
  {\bibfield  {journal} {\bibinfo  {journal} {The Journal of Physical Chemistry
  B}\ }\textbf {\bibinfo {volume} {103}},\ \bibinfo {pages} {3706--3710}
  (\bibinfo {year} {1999})}\BibitemShut {NoStop}%
\bibitem [{\citenamefont {Mullen}, \citenamefont {Shea},\ and\ \citenamefont
  {Peters}(2014)}]{cn1_Mullen2014}%
  \BibitemOpen
  \bibfield  {author} {\bibinfo {author} {\bibfnamefont {R.~G.}\ \bibnamefont
  {Mullen}}, \bibinfo {author} {\bibfnamefont {J.-E.}\ \bibnamefont {Shea}}, \
  and\ \bibinfo {author} {\bibfnamefont {B.}~\bibnamefont {Peters}},\
  }\bibfield  {title} {\enquote {\bibinfo {title} {{Transmission Coefficients,
  Committors, and Solvent Coordinates in Ion-Pair Dissociation}},}\ }\href
  {\doibase 10.1021/ct4009798} {\bibfield  {journal} {\bibinfo  {journal}
  {Journal of Chemical Theory and Computation}\ }\textbf {\bibinfo {volume}
  {10}},\ \bibinfo {pages} {659--667} (\bibinfo {year} {2014})}\BibitemShut
  {NoStop}%
\bibitem [{\citenamefont {Fiorin}, \citenamefont {Klein},\ and\ \citenamefont
  {H{\'{e}}nin}(2013)}]{colvar_fiorin13}%
  \BibitemOpen
  \bibfield  {author} {\bibinfo {author} {\bibfnamefont {G.}~\bibnamefont
  {Fiorin}}, \bibinfo {author} {\bibfnamefont {M.~L.}\ \bibnamefont {Klein}}, \
  and\ \bibinfo {author} {\bibfnamefont {J.}~\bibnamefont {H{\'{e}}nin}},\
  }\bibfield  {title} {\enquote {\bibinfo {title} {{Using collective variables
  to drive molecular dynamics simulations}},}\ }\href {\doibase
  10.1080/00268976.2013.813594} {\bibfield  {journal} {\bibinfo  {journal}
  {Molecular Physics}\ }\textbf {\bibinfo {volume} {111}},\ \bibinfo {pages}
  {3345--3362} (\bibinfo {year} {2013})}\BibitemShut {NoStop}%
\bibitem [{\citenamefont {Plimpton}(1995)}]{lammps_Plimpton1995}%
  \BibitemOpen
  \bibfield  {author} {\bibinfo {author} {\bibfnamefont {S.}~\bibnamefont
  {Plimpton}},\ }\bibfield  {title} {\enquote {\bibinfo {title} {{Fast Parallel
  Algorithms for Short-Range Molecular Dynamics}},}\ }\href {\doibase
  10.1006/jcph.1995.1039} {\bibfield  {journal} {\bibinfo  {journal} {Journal
  of Computational Physics}\ }\textbf {\bibinfo {volume} {117}},\ \bibinfo
  {pages} {1--19} (\bibinfo {year} {1995})},\ \Eprint
  {http://arxiv.org/abs/nag.2347} {arXiv:nag.2347 [10.1002]} \BibitemShut
  {NoStop}%
\bibitem [{\citenamefont {Joung}\ and\ \citenamefont
  {Cheatham}(2008)}]{naclt4pew_Joung2008}%
  \BibitemOpen
  \bibfield  {author} {\bibinfo {author} {\bibfnamefont {I.~S.}\ \bibnamefont
  {Joung}}\ and\ \bibinfo {author} {\bibfnamefont {T.~E.}\ \bibnamefont
  {Cheatham}},\ }\bibfield  {title} {\enquote {\bibinfo {title} {{Determination
  of Alkali and Halide Monovalent Ion Parameters for Use in Explicitly Solvated
  Biomolecular Simulations}},}\ }\href {\doibase 10.1021/jp8001614} {\bibfield
  {journal} {\bibinfo  {journal} {The Journal of Physical Chemistry B}\
  }\textbf {\bibinfo {volume} {112}},\ \bibinfo {pages} {9020--9041} (\bibinfo
  {year} {2008})}\BibitemShut {NoStop}%
\bibitem [{\citenamefont {Jorgensen}, \citenamefont {Maxwell},\ and\
  \citenamefont {Tirado-Rives}(1996)}]{oplsJorgensen1996}%
  \BibitemOpen
  \bibfield  {author} {\bibinfo {author} {\bibfnamefont {W.~L.}\ \bibnamefont
  {Jorgensen}}, \bibinfo {author} {\bibfnamefont {D.~S.}\ \bibnamefont
  {Maxwell}}, \ and\ \bibinfo {author} {\bibfnamefont {J.}~\bibnamefont
  {Tirado-Rives}},\ }\bibfield  {title} {\enquote {\bibinfo {title}
  {{Development and Testing of the OLPS All-Atom Force Field on Conformational
  Energetics and Properties of Organic Liquids}},}\ }\href {\doibase
  10.1021/ja9621760} {\bibfield  {journal} {\bibinfo  {journal} {J. Am. Chem.
  Soc.}\ }\textbf {\bibinfo {volume} {118}},\ \bibinfo {pages} {11225--11236}
  (\bibinfo {year} {1996})},\ \Eprint {http://arxiv.org/abs/{\_}barata
  Materials and Techniques of polychrome wooden sculpture} {arXiv:{\_}barata
  Materials and Techniques of polychrome wooden sculpture} \BibitemShut
  {NoStop}%
\bibitem [{\citenamefont {Berendsen}, \citenamefont {Grigera},\ and\
  \citenamefont {Straatsma}(1987)}]{spce_Berendsen1987}%
  \BibitemOpen
  \bibfield  {author} {\bibinfo {author} {\bibfnamefont {H.~J.~C.}\
  \bibnamefont {Berendsen}}, \bibinfo {author} {\bibfnamefont {J.~R.}\
  \bibnamefont {Grigera}}, \ and\ \bibinfo {author} {\bibfnamefont {T.~P.}\
  \bibnamefont {Straatsma}},\ }\bibfield  {title} {\enquote {\bibinfo {title}
  {{The missing term in effective pair potentials}},}\ }\href {\doibase
  10.1021/j100308a038} {\bibfield  {journal} {\bibinfo  {journal} {The Journal
  of Physical Chemistry}\ }\textbf {\bibinfo {volume} {91}},\ \bibinfo {pages}
  {6269--6271} (\bibinfo {year} {1987})}\BibitemShut {NoStop}%
\bibitem [{\citenamefont {Benavides}\ \emph {et~al.}(2017)\citenamefont
  {Benavides}, \citenamefont {Portillo}, \citenamefont {Chamorro},
  \citenamefont {Espinosa}, \citenamefont {Abascal},\ and\ \citenamefont
  {Vega}}]{madrid17_Benavides2017}%
  \BibitemOpen
  \bibfield  {author} {\bibinfo {author} {\bibfnamefont {A.~L.}\ \bibnamefont
  {Benavides}}, \bibinfo {author} {\bibfnamefont {M.~A.}\ \bibnamefont
  {Portillo}}, \bibinfo {author} {\bibfnamefont {V.~C.}\ \bibnamefont
  {Chamorro}}, \bibinfo {author} {\bibfnamefont {J.~R.}\ \bibnamefont
  {Espinosa}}, \bibinfo {author} {\bibfnamefont {J.~L.~F.}\ \bibnamefont
  {Abascal}}, \ and\ \bibinfo {author} {\bibfnamefont {C.}~\bibnamefont
  {Vega}},\ }\bibfield  {title} {\enquote {\bibinfo {title} {{A potential model
  for sodium chloride solutions based on the TIP4P/2005 water model}},}\ }\href
  {\doibase 10.1063/1.5001190} {\bibfield  {journal} {\bibinfo  {journal} {The
  Journal of Chemical Physics}\ }\textbf {\bibinfo {volume} {147}},\ \bibinfo
  {pages} {104501} (\bibinfo {year} {2017})}\BibitemShut {NoStop}%
\bibitem [{\citenamefont {Zeron}, \citenamefont {Abascal},\ and\ \citenamefont
  {Vega}(2019)}]{madrid19_Zeron2019}%
  \BibitemOpen
  \bibfield  {author} {\bibinfo {author} {\bibfnamefont {I.~M.}\ \bibnamefont
  {Zeron}}, \bibinfo {author} {\bibfnamefont {J.~L.~F.}\ \bibnamefont
  {Abascal}}, \ and\ \bibinfo {author} {\bibfnamefont {C.}~\bibnamefont
  {Vega}},\ }\bibfield  {title} {\enquote {\bibinfo {title} {{A force field of
  Li + , Na + , K + , Mg 2+ , Ca 2+ , Cl − , and SO42− in aqueous solution
  based on the TIP4P/2005 water model and scaled charges for the ions}},}\
  }\href {\doibase 10.1063/1.5121392} {\bibfield  {journal} {\bibinfo
  {journal} {The Journal of Chemical Physics}\ }\textbf {\bibinfo {volume}
  {151}},\ \bibinfo {pages} {134504} (\bibinfo {year} {2019})}\BibitemShut
  {NoStop}%
\bibitem [{\citenamefont {Abascal}\ and\ \citenamefont
  {Vega}(2005)}]{t4p05Abascal2005}%
  \BibitemOpen
  \bibfield  {author} {\bibinfo {author} {\bibfnamefont {J.~L.~F.}\
  \bibnamefont {Abascal}}\ and\ \bibinfo {author} {\bibfnamefont
  {C.}~\bibnamefont {Vega}},\ }\bibfield  {title} {\enquote {\bibinfo {title}
  {{A general purpose model for the condensed phases of water: TIP4P/2005}},}\
  }\href {\doibase 10.1063/1.2121687} {\bibfield  {journal} {\bibinfo
  {journal} {The Journal of Chemical Physics}\ }\textbf {\bibinfo {volume}
  {123}},\ \bibinfo {pages} {234505} (\bibinfo {year} {2005})}\BibitemShut
  {NoStop}%
\bibitem [{\citenamefont {Soler}\ \emph {et~al.}(2002)\citenamefont {Soler},
  \citenamefont {Artacho}, \citenamefont {Gale}, \citenamefont {Garc\'{i}a},
  \citenamefont {Junquera}, \citenamefont {Ordej\'{o}n},\ and\ \citenamefont
  {Daniel}}]{siestaordernSoler2002}%
  \BibitemOpen
  \bibfield  {author} {\bibinfo {author} {\bibfnamefont {M.}~\bibnamefont
  {Soler}}, \bibinfo {author} {\bibfnamefont {E.}~\bibnamefont {Artacho}},
  \bibinfo {author} {\bibfnamefont {J.~D.}\ \bibnamefont {Gale}}, \bibinfo
  {author} {\bibfnamefont {A.}~\bibnamefont {Garc\'{i}a}}, \bibinfo {author}
  {\bibfnamefont {J.}~\bibnamefont {Junquera}}, \bibinfo {author}
  {\bibfnamefont {P.}~\bibnamefont {Ordej\'{o}n}}, \ and\ \bibinfo {author}
  {\bibfnamefont {S.-P.}\ \bibnamefont {Daniel}},\ }\bibfield  {title}
  {\enquote {\bibinfo {title} {{The SIESTA method for ab initio order-N
  materials}},}\ }\href {\doibase 10.1088/0953- 8984/14/11/302} {\bibfield
  {journal} {\bibinfo  {journal} {J. Phys. Cond. Mat.}\ }\textbf {\bibinfo
  {volume} {14}},\ \bibinfo {pages} {2745--2779} (\bibinfo {year}
  {2002})}\BibitemShut {NoStop}%
\bibitem [{\citenamefont {Corsetti}\ \emph
  {et~al.}(2013{\natexlab{b}})\citenamefont {Corsetti}, \citenamefont
  {Fern{\'a}ndez-Serra}, \citenamefont {Soler},\ and\ \citenamefont
  {Artacho}}]{corsetti2013optimal}%
  \BibitemOpen
  \bibfield  {author} {\bibinfo {author} {\bibfnamefont {F.}~\bibnamefont
  {Corsetti}}, \bibinfo {author} {\bibfnamefont {M.}~\bibnamefont
  {Fern{\'a}ndez-Serra}}, \bibinfo {author} {\bibfnamefont {J.~M.}\
  \bibnamefont {Soler}}, \ and\ \bibinfo {author} {\bibfnamefont
  {E.}~\bibnamefont {Artacho}},\ }\bibfield  {title} {\enquote {\bibinfo
  {title} {Optimal finite-range atomic basis sets for liquid water and ice},}\
  }\href@noop {} {\bibfield  {journal} {\bibinfo  {journal} {Journal of
  Physics: Condensed Matter}\ }\textbf {\bibinfo {volume} {25}},\ \bibinfo
  {pages} {435504} (\bibinfo {year} {2013}{\natexlab{b}})}\BibitemShut
  {NoStop}%
\bibitem [{\citenamefont {Perdew}, \citenamefont {Burke},\ and\ \citenamefont
  {Ernzerhof}(1996)}]{pbePerdew1996}%
  \BibitemOpen
  \bibfield  {author} {\bibinfo {author} {\bibfnamefont {J.~P.}\ \bibnamefont
  {Perdew}}, \bibinfo {author} {\bibfnamefont {K.}~\bibnamefont {Burke}}, \
  and\ \bibinfo {author} {\bibfnamefont {M.}~\bibnamefont {Ernzerhof}},\
  }\bibfield  {title} {\enquote {\bibinfo {title} {{Generalized gradient
  approximation made simple}},}\ }\href {\doibase 10.1103/PhysRevLett.77.3865}
  {\bibfield  {journal} {\bibinfo  {journal} {Physical Review Letters}\
  }\textbf {\bibinfo {volume} {77}},\ \bibinfo {pages} {3865--3868} (\bibinfo
  {year} {1996})},\ \Eprint {http://arxiv.org/abs/0927-0256(96)00008}
  {arXiv:0927-0256(96)00008 [10.1016]} \BibitemShut {NoStop}%
\bibitem [{\citenamefont {Corsetti}\ \emph
  {et~al.}(2013{\natexlab{c}})\citenamefont {Corsetti}, \citenamefont
  {Artacho}, \citenamefont {Soler}, \citenamefont {Alexandre},\ and\
  \citenamefont {Fern{\'{a}}ndez-Serra}}]{marivi_vdwrhoeq_Corsetti2013}%
  \BibitemOpen
  \bibfield  {author} {\bibinfo {author} {\bibfnamefont {F.}~\bibnamefont
  {Corsetti}}, \bibinfo {author} {\bibfnamefont {E.}~\bibnamefont {Artacho}},
  \bibinfo {author} {\bibfnamefont {J.~M.}\ \bibnamefont {Soler}}, \bibinfo
  {author} {\bibfnamefont {S.~S.}\ \bibnamefont {Alexandre}}, \ and\ \bibinfo
  {author} {\bibfnamefont {M.-V.}\ \bibnamefont {Fern{\'{a}}ndez-Serra}},\
  }\bibfield  {title} {\enquote {\bibinfo {title} {{Room temperature
  compressibility and diffusivity of liquid water from first principles}},}\
  }\href {\doibase 10.1063/1.4832141} {\bibfield  {journal} {\bibinfo
  {journal} {The Journal of Chemical Physics}\ }\textbf {\bibinfo {volume}
  {139}},\ \bibinfo {pages} {194502} (\bibinfo {year} {2013}{\natexlab{c}})},\
  \Eprint {http://arxiv.org/abs/1307.1645} {arXiv:1307.1645} \BibitemShut
  {NoStop}%
\bibitem [{\citenamefont {Wang}\ \emph
  {et~al.}(2011{\natexlab{a}})\citenamefont {Wang}, \citenamefont
  {Rom{\'a}n-P{\'e}rez}, \citenamefont {Soler}, \citenamefont {Artacho},\ and\
  \citenamefont {Fern{\'a}ndez-Serra}}]{wang2011density}%
  \BibitemOpen
  \bibfield  {author} {\bibinfo {author} {\bibfnamefont {J.}~\bibnamefont
  {Wang}}, \bibinfo {author} {\bibfnamefont {G.}~\bibnamefont
  {Rom{\'a}n-P{\'e}rez}}, \bibinfo {author} {\bibfnamefont {J.~M.}\
  \bibnamefont {Soler}}, \bibinfo {author} {\bibfnamefont {E.}~\bibnamefont
  {Artacho}}, \ and\ \bibinfo {author} {\bibfnamefont {M.-V.}\ \bibnamefont
  {Fern{\'a}ndez-Serra}},\ }\bibfield  {title} {\enquote {\bibinfo {title}
  {Density, structure, and dynamics of water: The effect of van der waals
  interactions},}\ }\href@noop {} {\bibfield  {journal} {\bibinfo  {journal}
  {The Journal of chemical physics}\ }\textbf {\bibinfo {volume} {134}},\
  \bibinfo {pages} {024516} (\bibinfo {year} {2011}{\natexlab{a}})}\BibitemShut
  {NoStop}%
\bibitem [{\citenamefont {Wang}\ \emph
  {et~al.}(2011{\natexlab{b}})\citenamefont {Wang}, \citenamefont
  {Rom{\'{a}}n-P{\'{e}}rez}, \citenamefont {Soler}, \citenamefont {Artacho},\
  and\ \citenamefont {Fern{\'{a}}ndez-Serra}}]{pbecompress_MVFSWang2011}%
  \BibitemOpen
  \bibfield  {author} {\bibinfo {author} {\bibfnamefont {J.}~\bibnamefont
  {Wang}}, \bibinfo {author} {\bibfnamefont {G.}~\bibnamefont
  {Rom{\'{a}}n-P{\'{e}}rez}}, \bibinfo {author} {\bibfnamefont {J.~M.}\
  \bibnamefont {Soler}}, \bibinfo {author} {\bibfnamefont {E.}~\bibnamefont
  {Artacho}}, \ and\ \bibinfo {author} {\bibfnamefont {M.-V.}\ \bibnamefont
  {Fern{\'{a}}ndez-Serra}},\ }\bibfield  {title} {\enquote {\bibinfo {title}
  {{Density, structure, and dynamics of water: The effect of van der Waals
  interactions}},}\ }\href {\doibase 10.1063/1.3521268} {\bibfield  {journal}
  {\bibinfo  {journal} {The Journal of Chemical Physics}\ }\textbf {\bibinfo
  {volume} {134}},\ \bibinfo {pages} {024516} (\bibinfo {year}
  {2011}{\natexlab{b}})}\BibitemShut {NoStop}%
\bibitem [{\citenamefont {Chialvo}\ \emph {et~al.}(1995)\citenamefont
  {Chialvo}, \citenamefont {Cummings}, \citenamefont {Cochran}, \citenamefont
  {Simonson},\ and\ \citenamefont {Mesmer}}]{kamChialvo1995}%
  \BibitemOpen
  \bibfield  {author} {\bibinfo {author} {\bibfnamefont {A.~A.}\ \bibnamefont
  {Chialvo}}, \bibinfo {author} {\bibfnamefont {P.~T.}\ \bibnamefont
  {Cummings}}, \bibinfo {author} {\bibfnamefont {H.~D.}\ \bibnamefont
  {Cochran}}, \bibinfo {author} {\bibfnamefont {J.~M.}\ \bibnamefont
  {Simonson}}, \ and\ \bibinfo {author} {\bibfnamefont {R.~E.}\ \bibnamefont
  {Mesmer}},\ }\bibfield  {title} {\enquote {\bibinfo {title} {{Na + Cl -- ion
  pair association in supercritical water}},}\ }\href {\doibase
  10.1063/1.470707} {\bibfield  {journal} {\bibinfo  {journal} {The Journal of
  Chemical Physics}\ }\textbf {\bibinfo {volume} {103}},\ \bibinfo {pages}
  {9379--9387} (\bibinfo {year} {1995})}\BibitemShut {NoStop}%
\bibitem [{\citenamefont {Zimmerman}, \citenamefont {Arcis},\ and\
  \citenamefont {Tremaine}(2012)}]{kake_expZimmerman2012}%
  \BibitemOpen
  \bibfield  {author} {\bibinfo {author} {\bibfnamefont {G.~H.}\ \bibnamefont
  {Zimmerman}}, \bibinfo {author} {\bibfnamefont {H.}~\bibnamefont {Arcis}}, \
  and\ \bibinfo {author} {\bibfnamefont {P.~R.}\ \bibnamefont {Tremaine}},\
  }\bibfield  {title} {\enquote {\bibinfo {title} {{Limiting conductivities and
  ion association constants of aqueous NaCl under hydrothermal conditions:
  Experimental data and correlations}},}\ }\href {\doibase 10.1021/je300361j}
  {\bibfield  {journal} {\bibinfo  {journal} {Journal of Chemical and
  Engineering Data}\ }\textbf {\bibinfo {volume} {57}},\ \bibinfo {pages}
  {2415--2429} (\bibinfo {year} {2012})}\BibitemShut {NoStop}%
\bibitem [{\citenamefont {Pan}\ \emph {et~al.}(2013)\citenamefont {Pan},
  \citenamefont {Spanu}, \citenamefont {Harrison}, \citenamefont {Sverjensky},\
  and\ \citenamefont {Galli}}]{dielwaterextreme_Pan2013}%
  \BibitemOpen
  \bibfield  {author} {\bibinfo {author} {\bibfnamefont {D.}~\bibnamefont
  {Pan}}, \bibinfo {author} {\bibfnamefont {L.}~\bibnamefont {Spanu}}, \bibinfo
  {author} {\bibfnamefont {B.}~\bibnamefont {Harrison}}, \bibinfo {author}
  {\bibfnamefont {D.~A.}\ \bibnamefont {Sverjensky}}, \ and\ \bibinfo {author}
  {\bibfnamefont {G.}~\bibnamefont {Galli}},\ }\bibfield  {title} {\enquote
  {\bibinfo {title} {{Dielectric properties of water under extreme conditions
  and transport of carbonates in the deep Earth}},}\ }\href {\doibase
  10.1073/pnas.1221581110} {\bibfield  {journal} {\bibinfo  {journal}
  {Proceedings of the National Academy of Sciences}\ }\textbf {\bibinfo
  {volume} {110}},\ \bibinfo {pages} {6646--6650} (\bibinfo {year}
  {2013})}\BibitemShut {NoStop}%
\bibitem [{\citenamefont {Lu}, \citenamefont {Gygi},\ and\ \citenamefont
  {Galli}(2008)}]{aimddielicewater_Lu2008}%
  \BibitemOpen
  \bibfield  {author} {\bibinfo {author} {\bibfnamefont {D.}~\bibnamefont
  {Lu}}, \bibinfo {author} {\bibfnamefont {F.}~\bibnamefont {Gygi}}, \ and\
  \bibinfo {author} {\bibfnamefont {G.}~\bibnamefont {Galli}},\ }\bibfield
  {title} {\enquote {\bibinfo {title} {{Dielectric Properties of Ice and Liquid
  Water from First-Principles Calculations}},}\ }\href {\doibase
  10.1103/PhysRevLett.100.147601} {\bibfield  {journal} {\bibinfo  {journal}
  {Physical Review Letters}\ }\textbf {\bibinfo {volume} {100}},\ \bibinfo
  {pages} {147601} (\bibinfo {year} {2008})}\BibitemShut {NoStop}%
\bibitem [{\citenamefont {Elton}\ and\ \citenamefont
  {Fern{\'{a}}ndez-Serra}(2014)}]{spct4pdielconst_Elton2014}%
  \BibitemOpen
  \bibfield  {author} {\bibinfo {author} {\bibfnamefont {D.~C.}\ \bibnamefont
  {Elton}}\ and\ \bibinfo {author} {\bibfnamefont {M.~V.}\ \bibnamefont
  {Fern{\'{a}}ndez-Serra}},\ }\bibfield  {title} {\enquote {\bibinfo {title}
  {{Polar nanoregions in water: A study of the dielectric properties of
  TIP4P/2005, TIP4P/2005f and TTM3F}},}\ }\href {\doibase 10.1063/1.4869110}
  {\bibfield  {journal} {\bibinfo  {journal} {Journal of Chemical Physics}\
  }\textbf {\bibinfo {volume} {140}} (\bibinfo {year} {2014}),\
  10.1063/1.4869110},\ \Eprint {http://arxiv.org/abs/1401.5090}
  {arXiv:1401.5090} \BibitemShut {NoStop}%
\end{thebibliography}%


\begin{thebibliography}{15}%
\makeatletter
\providecommand \@ifxundefined [1]{%
 \@ifx{#1\undefined}
}%
\providecommand \@ifnum [1]{%
 \ifnum #1\expandafter \@firstoftwo
 \else \expandafter \@secondoftwo
 \fi
}%
\providecommand \@ifx [1]{%
 \ifx #1\expandafter \@firstoftwo
 \else \expandafter \@secondoftwo
 \fi
}%
\providecommand \natexlab [1]{#1}%
\providecommand \enquote  [1]{``#1''}%
\providecommand \bibnamefont  [1]{#1}%
\providecommand \bibfnamefont [1]{#1}%
\providecommand \citenamefont [1]{#1}%
\providecommand \href@noop [0]{\@secondoftwo}%
\providecommand \href [0]{\begingroup \@sanitize@url \@href}%
\providecommand \@href[1]{\@@startlink{#1}\@@href}%
\providecommand \@@href[1]{\endgroup#1\@@endlink}%
\providecommand \@sanitize@url [0]{\catcode `\\12\catcode `\$12\catcode
  `\&12\catcode `\#12\catcode `\^12\catcode `\_12\catcode `\%12\relax}%
\providecommand \@@startlink[1]{}%
\providecommand \@@endlink[0]{}%
\providecommand \url  [0]{\begingroup\@sanitize@url \@url }%
\providecommand \@url [1]{\endgroup\@href {#1}{\urlprefix }}%
\providecommand \urlprefix  [0]{URL }%
\providecommand \Eprint [0]{\href }%
\providecommand \doibase [0]{http://dx.doi.org/}%
\providecommand \selectlanguage [0]{\@gobble}%
\providecommand \bibinfo  [0]{\@secondoftwo}%
\providecommand \bibfield  [0]{\@secondoftwo}%
\providecommand \translation [1]{[#1]}%
\providecommand \BibitemOpen [0]{}%
\providecommand \bibitemStop [0]{}%
\providecommand \bibitemNoStop [0]{.\EOS\space}%
\providecommand \EOS [0]{\spacefactor3000\relax}%
\providecommand \BibitemShut  [1]{\csname bibitem#1\endcsname}%
\let\auto@bib@innerbib\@empty
\bibitem [{\citenamefont {Vega}\ \emph {et~al.}(2005)\citenamefont {Vega},
  \citenamefont {McBride}, \citenamefont {Sanz},\ and\ \citenamefont
  {Abascal}}]{t4prhoeq_Vega2005}%
  \BibitemOpen
  \bibfield  {author} {\bibinfo {author} {\bibfnamefont {C.}~\bibnamefont
  {Vega}}, \bibinfo {author} {\bibfnamefont {C.}~\bibnamefont {McBride}},
  \bibinfo {author} {\bibfnamefont {E.}~\bibnamefont {Sanz}}, \ and\ \bibinfo
  {author} {\bibfnamefont {J.~L.~F.}\ \bibnamefont {Abascal}},\ }\bibfield
  {title} {\enquote {\bibinfo {title} {{Radial distribution functions and
  densities for the SPC/E, TIP4P and TIP5P models for liquid water and ices Ih,
  Ic, II, III, IV, V, VI, VII, VIII, IX, XI and XII}},}\ }\href {\doibase
  10.1039/b418934e} {\bibfield  {journal} {\bibinfo  {journal} {Physical
  Chemistry Chemical Physics}\ }\textbf {\bibinfo {volume} {7}},\ \bibinfo
  {pages} {1450} (\bibinfo {year} {2005})}\BibitemShut {NoStop}%
\bibitem [{\citenamefont {Gillan}, \citenamefont {Alf{\`{e}}},\ and\
  \citenamefont {Michaelides}(2016)}]{pberhoeq_Gillan2016}%
  \BibitemOpen
  \bibfield  {author} {\bibinfo {author} {\bibfnamefont {M.~J.}\ \bibnamefont
  {Gillan}}, \bibinfo {author} {\bibfnamefont {D.}~\bibnamefont {Alf{\`{e}}}},
  \ and\ \bibinfo {author} {\bibfnamefont {A.}~\bibnamefont {Michaelides}},\
  }\bibfield  {title} {\enquote {\bibinfo {title} {{Perspective: How good is
  DFT for water?}}}\ }\href {\doibase 10.1063/1.4944633} {\bibfield  {journal}
  {\bibinfo  {journal} {Journal of Chemical Physics}\ }\textbf {\bibinfo
  {volume} {144}} (\bibinfo {year} {2016}),\ 10.1063/1.4944633}\BibitemShut
  {NoStop}%
\bibitem [{\citenamefont {Corsetti}\ \emph {et~al.}(2013)\citenamefont
  {Corsetti}, \citenamefont {Artacho}, \citenamefont {Soler}, \citenamefont
  {Alexandre},\ and\ \citenamefont
  {Fern{\'{a}}ndez-Serra}}]{marivi_vdwrhoeq_Corsetti2013}%
  \BibitemOpen
  \bibfield  {author} {\bibinfo {author} {\bibfnamefont {F.}~\bibnamefont
  {Corsetti}}, \bibinfo {author} {\bibfnamefont {E.}~\bibnamefont {Artacho}},
  \bibinfo {author} {\bibfnamefont {J.~M.}\ \bibnamefont {Soler}}, \bibinfo
  {author} {\bibfnamefont {S.~S.}\ \bibnamefont {Alexandre}}, \ and\ \bibinfo
  {author} {\bibfnamefont {M.-V.}\ \bibnamefont {Fern{\'{a}}ndez-Serra}},\
  }\bibfield  {title} {\enquote {\bibinfo {title} {{Room temperature
  compressibility and diffusivity of liquid water from first principles}},}\
  }\href {\doibase 10.1063/1.4832141} {\bibfield  {journal} {\bibinfo
  {journal} {The Journal of Chemical Physics}\ }\textbf {\bibinfo {volume}
  {139}},\ \bibinfo {pages} {194502} (\bibinfo {year} {2013})},\ \Eprint
  {http://arxiv.org/abs/1307.1645} {arXiv:1307.1645} \BibitemShut {NoStop}%
\bibitem [{\citenamefont {Fritz}, \citenamefont {Fern{\'{a}}ndez-Serra},\ and\
  \citenamefont {Soler}(2016)}]{marivi_dpps_vdwrhoeq_Fritz2016}%
  \BibitemOpen
  \bibfield  {author} {\bibinfo {author} {\bibfnamefont {M.}~\bibnamefont
  {Fritz}}, \bibinfo {author} {\bibfnamefont {M.}~\bibnamefont
  {Fern{\'{a}}ndez-Serra}}, \ and\ \bibinfo {author} {\bibfnamefont {J.~M.}\
  \bibnamefont {Soler}},\ }\bibfield  {title} {\enquote {\bibinfo {title}
  {{Optimization of an exchange-correlation density functional for water}},}\
  }\href {\doibase 10.1063/1.4953081} {\bibfield  {journal} {\bibinfo
  {journal} {Journal of Chemical Physics}\ }\textbf {\bibinfo {volume} {144}}
  (\bibinfo {year} {2016}),\ 10.1063/1.4953081},\ \Eprint
  {http://arxiv.org/abs/1603.07302} {arXiv:1603.07302} \BibitemShut {NoStop}%
\bibitem [{\citenamefont {Bormashenko}\ \emph {et~al.}(2018)\citenamefont
  {Bormashenko}, \citenamefont {Frenkel}, \citenamefont {Vilk}, \citenamefont
  {Legchenkova}, \citenamefont {Fedorets}, \citenamefont {Aktaev},
  \citenamefont {Dombrovsky},\ and\ \citenamefont
  {Nosonovsky}}]{voronoientBormashenko2018}%
  \BibitemOpen
  \bibfield  {author} {\bibinfo {author} {\bibfnamefont {E.}~\bibnamefont
  {Bormashenko}}, \bibinfo {author} {\bibfnamefont {M.}~\bibnamefont
  {Frenkel}}, \bibinfo {author} {\bibfnamefont {A.}~\bibnamefont {Vilk}},
  \bibinfo {author} {\bibfnamefont {I.}~\bibnamefont {Legchenkova}}, \bibinfo
  {author} {\bibfnamefont {A.~A.}\ \bibnamefont {Fedorets}}, \bibinfo {author}
  {\bibfnamefont {N.~E.}\ \bibnamefont {Aktaev}}, \bibinfo {author}
  {\bibfnamefont {L.~A.}\ \bibnamefont {Dombrovsky}}, \ and\ \bibinfo {author}
  {\bibfnamefont {M.}~\bibnamefont {Nosonovsky}},\ }\bibfield  {title}
  {\enquote {\bibinfo {title} {{Characterization of self-assembled 2D patterns
  with voronoi entropy}},}\ }\href {\doibase 10.3390/e20120956} {\bibfield
  {journal} {\bibinfo  {journal} {Entropy}\ }\textbf {\bibinfo {volume} {20}}
  (\bibinfo {year} {2018}),\ 10.3390/e20120956}\BibitemShut {NoStop}%
\bibitem [{\citenamefont {{Senthil Kumar}}\ and\ \citenamefont
  {Kumaran}(2005)}]{voronoigasSenthilKumar2005}%
  \BibitemOpen
  \bibfield  {author} {\bibinfo {author} {\bibfnamefont {V.}~\bibnamefont
  {{Senthil Kumar}}}\ and\ \bibinfo {author} {\bibfnamefont {V.}~\bibnamefont
  {Kumaran}},\ }\bibfield  {title} {\enquote {\bibinfo {title} {{Voronoi cell
  volume distribution and configurational entropy of hard-spheres}},}\ }\href
  {\doibase 10.1063/1.2011390} {\bibfield  {journal} {\bibinfo  {journal} {The
  Journal of Chemical Physics}\ }\textbf {\bibinfo {volume} {123}},\ \bibinfo
  {pages} {114501} (\bibinfo {year} {2005})}\BibitemShut {NoStop}%
\bibitem [{\citenamefont {Soper}\ and\ \citenamefont
  {Weckstr{\"{o}}m}(2006)}]{khalideSoper2006}%
  \BibitemOpen
  \bibfield  {author} {\bibinfo {author} {\bibfnamefont {A.~K.}\ \bibnamefont
  {Soper}}\ and\ \bibinfo {author} {\bibfnamefont {K.}~\bibnamefont
  {Weckstr{\"{o}}m}},\ }\bibfield  {title} {\enquote {\bibinfo {title} {{Ion
  solvation and water structure in potassium halide aqueous solutions}},}\
  }\href {\doibase 10.1016/j.bpc.2006.04.009} {\bibfield  {journal} {\bibinfo
  {journal} {Biophysical Chemistry}\ }\textbf {\bibinfo {volume} {124}},\
  \bibinfo {pages} {180--191} (\bibinfo {year} {2006})}\BibitemShut {NoStop}%
\bibitem [{\citenamefont {Mancinelli}\ \emph
  {et~al.}(2007{\natexlab{a}})\citenamefont {Mancinelli}, \citenamefont
  {Botti}, \citenamefont {Bruni}, \citenamefont {Ricci},\ and\ \citenamefont
  {Soper}}]{perturbgood_Mancinelli2007}%
  \BibitemOpen
  \bibfield  {author} {\bibinfo {author} {\bibfnamefont {R.}~\bibnamefont
  {Mancinelli}}, \bibinfo {author} {\bibfnamefont {A.}~\bibnamefont {Botti}},
  \bibinfo {author} {\bibfnamefont {F.}~\bibnamefont {Bruni}}, \bibinfo
  {author} {\bibfnamefont {M.~A.}\ \bibnamefont {Ricci}}, \ and\ \bibinfo
  {author} {\bibfnamefont {A.~K.}\ \bibnamefont {Soper}},\ }\bibfield  {title}
  {\enquote {\bibinfo {title} {{Perturbation of water structure due to
  monovalent ions in solution}},}\ }\href {\doibase 10.1039/b701855j}
  {\bibfield  {journal} {\bibinfo  {journal} {Physical Chemistry Chemical
  Physics}\ }\textbf {\bibinfo {volume} {9}},\ \bibinfo {pages} {2959--2967}
  (\bibinfo {year} {2007}{\natexlab{a}})}\BibitemShut {NoStop}%
\bibitem [{\citenamefont {Mancinelli}\ \emph
  {et~al.}(2007{\natexlab{b}})\citenamefont {Mancinelli}, \citenamefont
  {Botti}, \citenamefont {Bruni}, \citenamefont {Ricci},\ and\ \citenamefont
  {Soper}}]{perturbbad_Mancinelli2007}%
  \BibitemOpen
  \bibfield  {author} {\bibinfo {author} {\bibfnamefont {R.}~\bibnamefont
  {Mancinelli}}, \bibinfo {author} {\bibfnamefont {A.}~\bibnamefont {Botti}},
  \bibinfo {author} {\bibfnamefont {F.}~\bibnamefont {Bruni}}, \bibinfo
  {author} {\bibfnamefont {M.~A.}\ \bibnamefont {Ricci}}, \ and\ \bibinfo
  {author} {\bibfnamefont {A.~K.}\ \bibnamefont {Soper}},\ }\bibfield  {title}
  {\enquote {\bibinfo {title} {{Hydration of sodium, potassium, and chloride
  ions in solution and the concept of structure maker/breaker}},}\ }\href
  {\doibase 10.1021/jp075913v} {\bibfield  {journal} {\bibinfo  {journal}
  {Journal of Physical Chemistry B}\ }\textbf {\bibinfo {volume} {111}},\
  \bibinfo {pages} {13570--13577} (\bibinfo {year}
  {2007}{\natexlab{b}})}\BibitemShut {NoStop}%
\bibitem [{\citenamefont {Waluyo}\ \emph {et~al.}(2011)\citenamefont {Waluyo},
  \citenamefont {Huang}, \citenamefont {Nordlund}, \citenamefont {Bergmann},
  \citenamefont {Weiss}, \citenamefont {Pettersson},\ and\ \citenamefont
  {Nilsson}}]{nashell1_Waluyo2011}%
  \BibitemOpen
  \bibfield  {author} {\bibinfo {author} {\bibfnamefont {I.}~\bibnamefont
  {Waluyo}}, \bibinfo {author} {\bibfnamefont {C.}~\bibnamefont {Huang}},
  \bibinfo {author} {\bibfnamefont {D.}~\bibnamefont {Nordlund}}, \bibinfo
  {author} {\bibfnamefont {U.}~\bibnamefont {Bergmann}}, \bibinfo {author}
  {\bibfnamefont {T.~M.}\ \bibnamefont {Weiss}}, \bibinfo {author}
  {\bibfnamefont {L.~G.~M.}\ \bibnamefont {Pettersson}}, \ and\ \bibinfo
  {author} {\bibfnamefont {A.}~\bibnamefont {Nilsson}},\ }\bibfield  {title}
  {\enquote {\bibinfo {title} {{The structure of water in the hydration shell
  of cations from x-ray Raman and small angle x-ray scattering
  measurements}},}\ }\href {\doibase 10.1063/1.3533958} {\bibfield  {journal}
  {\bibinfo  {journal} {The Journal of Chemical Physics}\ }\textbf {\bibinfo
  {volume} {134}},\ \bibinfo {pages} {064513} (\bibinfo {year}
  {2011})}\BibitemShut {NoStop}%
\bibitem [{\citenamefont {Waluyo}\ \emph {et~al.}(2014)\citenamefont {Waluyo},
  \citenamefont {Nordlund}, \citenamefont {Bergmann}, \citenamefont
  {Schlesinger}, \citenamefont {Pettersson},\ and\ \citenamefont
  {Nilsson}}]{xraychaoWaluyo2014}%
  \BibitemOpen
  \bibfield  {author} {\bibinfo {author} {\bibfnamefont {I.}~\bibnamefont
  {Waluyo}}, \bibinfo {author} {\bibfnamefont {D.}~\bibnamefont {Nordlund}},
  \bibinfo {author} {\bibfnamefont {U.}~\bibnamefont {Bergmann}}, \bibinfo
  {author} {\bibfnamefont {D.}~\bibnamefont {Schlesinger}}, \bibinfo {author}
  {\bibfnamefont {L.~G.~M.}\ \bibnamefont {Pettersson}}, \ and\ \bibinfo
  {author} {\bibfnamefont {A.}~\bibnamefont {Nilsson}},\ }\bibfield  {title}
  {\enquote {\bibinfo {title} {{A different view of structure-making and
  structure-breaking in alkali halide aqueous solutions through x-ray
  absorption spectroscopy}},}\ }\href {\doibase 10.1063/1.4881600} {\bibfield
  {journal} {\bibinfo  {journal} {The Journal of Chemical Physics}\ }\textbf
  {\bibinfo {volume} {140}},\ \bibinfo {pages} {244506} (\bibinfo {year}
  {2014})}\BibitemShut {NoStop}%
\bibitem [{\citenamefont {Gaiduk}\ and\ \citenamefont
  {Galli}(2017)}]{naclrdfs_Gaiduk2017}%
  \BibitemOpen
  \bibfield  {author} {\bibinfo {author} {\bibfnamefont {A.~P.}\ \bibnamefont
  {Gaiduk}}\ and\ \bibinfo {author} {\bibfnamefont {G.}~\bibnamefont {Galli}},\
  }\bibfield  {title} {\enquote {\bibinfo {title} {{Local and Global Effects of
  Dissolved Sodium Chloride on the Structure of Water}},}\ }\href {\doibase
  10.1021/acs.jpclett.7b00239} {\bibfield  {journal} {\bibinfo  {journal} {The
  Journal of Physical Chemistry Letters}\ }\textbf {\bibinfo {volume} {8}},\
  \bibinfo {pages} {1496--1502} (\bibinfo {year} {2017})}\BibitemShut {NoStop}%
\bibitem [{\citenamefont {Soper}(2013)}]{Soper2013}%
  \BibitemOpen
  \bibfield  {author} {\bibinfo {author} {\bibfnamefont {A.~K.}\ \bibnamefont
  {Soper}},\ }\bibfield  {title} {\enquote {\bibinfo {title} {{The Radial
  Distribution Functions of Water as Derived from Radiation Total Scattering
  Experiments: Is There Anything We Can Say for Sure?}}}\ }\href {\doibase
  10.1155/2013/279463} {\bibfield  {journal} {\bibinfo  {journal} {ISRN
  Physical Chemistry}\ }\textbf {\bibinfo {volume} {2013}},\ \bibinfo {pages}
  {1--67} (\bibinfo {year} {2013})}\BibitemShut {NoStop}%
\bibitem [{\citenamefont {Zeron}, \citenamefont {Abascal},\ and\ \citenamefont
  {Vega}(2019)}]{madrid19_Zeron2019}%
  \BibitemOpen
  \bibfield  {author} {\bibinfo {author} {\bibfnamefont {I.~M.}\ \bibnamefont
  {Zeron}}, \bibinfo {author} {\bibfnamefont {J.~L.~F.}\ \bibnamefont
  {Abascal}}, \ and\ \bibinfo {author} {\bibfnamefont {C.}~\bibnamefont
  {Vega}},\ }\bibfield  {title} {\enquote {\bibinfo {title} {{A force field of
  Li + , Na + , K + , Mg 2+ , Ca 2+ , Cl − , and SO42− in aqueous solution
  based on the TIP4P/2005 water model and scaled charges for the ions}},}\
  }\href {\doibase 10.1063/1.5121392} {\bibfield  {journal} {\bibinfo
  {journal} {The Journal of Chemical Physics}\ }\textbf {\bibinfo {volume}
  {151}},\ \bibinfo {pages} {134504} (\bibinfo {year} {2019})}\BibitemShut
  {NoStop}%
\bibitem [{\citenamefont {Abascal}\ and\ \citenamefont
  {Vega}(2005)}]{t4p05Abascal2005}%
  \BibitemOpen
  \bibfield  {author} {\bibinfo {author} {\bibfnamefont {J.~L.~F.}\
  \bibnamefont {Abascal}}\ and\ \bibinfo {author} {\bibfnamefont
  {C.}~\bibnamefont {Vega}},\ }\bibfield  {title} {\enquote {\bibinfo {title}
  {{A general purpose model for the condensed phases of water: TIP4P/2005}},}\
  }\href {\doibase 10.1063/1.2121687} {\bibfield  {journal} {\bibinfo
  {journal} {The Journal of Chemical Physics}\ }\textbf {\bibinfo {volume}
  {123}},\ \bibinfo {pages} {234505} (\bibinfo {year} {2005})}\BibitemShut
  {NoStop}%
\end{thebibliography}%

\makeatletter\@input{xsupp.tex}\makeatother 

\end{document}


\preprint{AIP/123-QED}

\title[]{Supplemental: Role of Water Model on Ion Dissociation at Ambient Conditions}

\author{Alec Wills}
\email{alec.wills@stonybrook.edu}
\author{Marivi Fern\'{a}ndez-Serra}%
 \email{maria.fernandez-serra@stonybrook.edu}
\affiliation{ 
Physics and Astronomy Department, Stony Brook University. Stony Brook, New York 11794-3800, United States
}%
\affiliation{Institute for Advanced Computational Science, Stony Brook, New York 11794-3800, United States}

\date{29 January 2021}

\maketitle

\section{Complement to the discussion section}

In addition to information contained in the PMFs, the different thermodynamic quantities tell us how the ions affect the bulk as a whole.
As the different models have been parameterized to yield different behaviors, pressure changes between systems at different densities and between pure solvent and solution can tell us about the equilibrium states.

Moreover, when examining the discrepancies between models, it is important to investigate differences on a smaller scale than bulk thermodynamic values.
Thus, micropscopic solvent characteristics can be investigated by examining the radial distribution functions between the ions and neighbor oxygen atoms.
Local and long-range effects are apparent in the solvation shells that are defined by these distributions, and can tell us how the ions affect the solvent water.

In calculating system averages for the ion solutions, we use the calculated PMF values to thermodynamically average the mean system values across the ion separation range.
That is, for a given simulation average $\langle X(r_i)\rangle$ at ion constraint distance $r_i$, we find the average across simulations as \begin{equation}
    \langle X \rangle = \frac{\sum_i^N e^{-\beta U(r_i)} \langle X(r_i)\rangle}{\sum_i^N e^{-\beta U(r_i)}}, \label{eq:tdynavg}
\end{equation}
where $U(r_i)$ is the PMF value at constraint distance $r_i$.
This way, presented averages take into account the configurational preference towards the favored state.

\subsection{Pressure}

The mean pressure of the simulated systems gives key information as to the level of thermodynamic equilibrium we establish with our chosen parameters.
The thermodynamically averaged pressures for the ion solution and pure water systems, $P_\mathrm{ions}$ and $P_\mathrm{pure}$, respectively, are given in Table \ref{supp-tbl:tdynavgs}.
Additionally, the differences in pressure between the same systems with and without the ion pair is included.

We see that there is minimal, though nonzero, effect in the change of TIP4P system pressure with the inclusion of the ions, whereby their presence would induce an expansion of the box size through the small positive pressure.
However, the change in simulation density affects the pressure as we would expect.
With an equilibrium mass density of $\SI{0.994}{\gram\ \cm^{-3}}$,\cite{t4prhoeq_Vega2005} decreasing the volume (and thus density) bring the system further from equilibrium, explaining the change in magnitude of pressure.

Analagous to the classical force field simulations, we see slight differences in adding the ion pair to \textit{ab initio} water systems.
However, the equilibrium densities of PBE and the dispersion-inclusive liquid water systems are very different.
The mass density of PBE water at mechanical equilibrium (zero pressure) is around $\SI{0.87}{\gram\ \cm^{-3}}$, whereas functionals that include dispersion corrections typically equilibrate to higher densities, on the order of $\SI{1.1}{\gram\ \cm^{-3}}$.\cite{pberhoeq_Gillan2016, marivi_vdwrhoeq_Corsetti2013,marivi_dpps_vdwrhoeq_Fritz2016}
In this manner, inclusion of the ions again would increase seek the system volume for the PBE solutions to varying degrees, which our results confirm.
Addition of the ions to both PBE systems increases the density of the box, bringing it further away from equilibrium, resulting in higher pressures for the ionic solutions.
The 0.52 M system, already close to equilibrium, does not see a large pressure difference when the ions are included.
This directly contrasts with the 0.56 M solution, which is further away from equilibrium and results in a pressure change an order of magnitude greater than the 0.52 M difference.
This behavior is mirrored in the vdW-DF-cx liquid, where the 0.56 M solution is the closer to equilibrium and so reacts less to the ion addition/removal.

\subsection{Voronoi Volume}

\begin{figure*}
    \centering
    \begin{subfigure}[b]{0.32\textwidth}
    \includegraphics[scale=0.32]{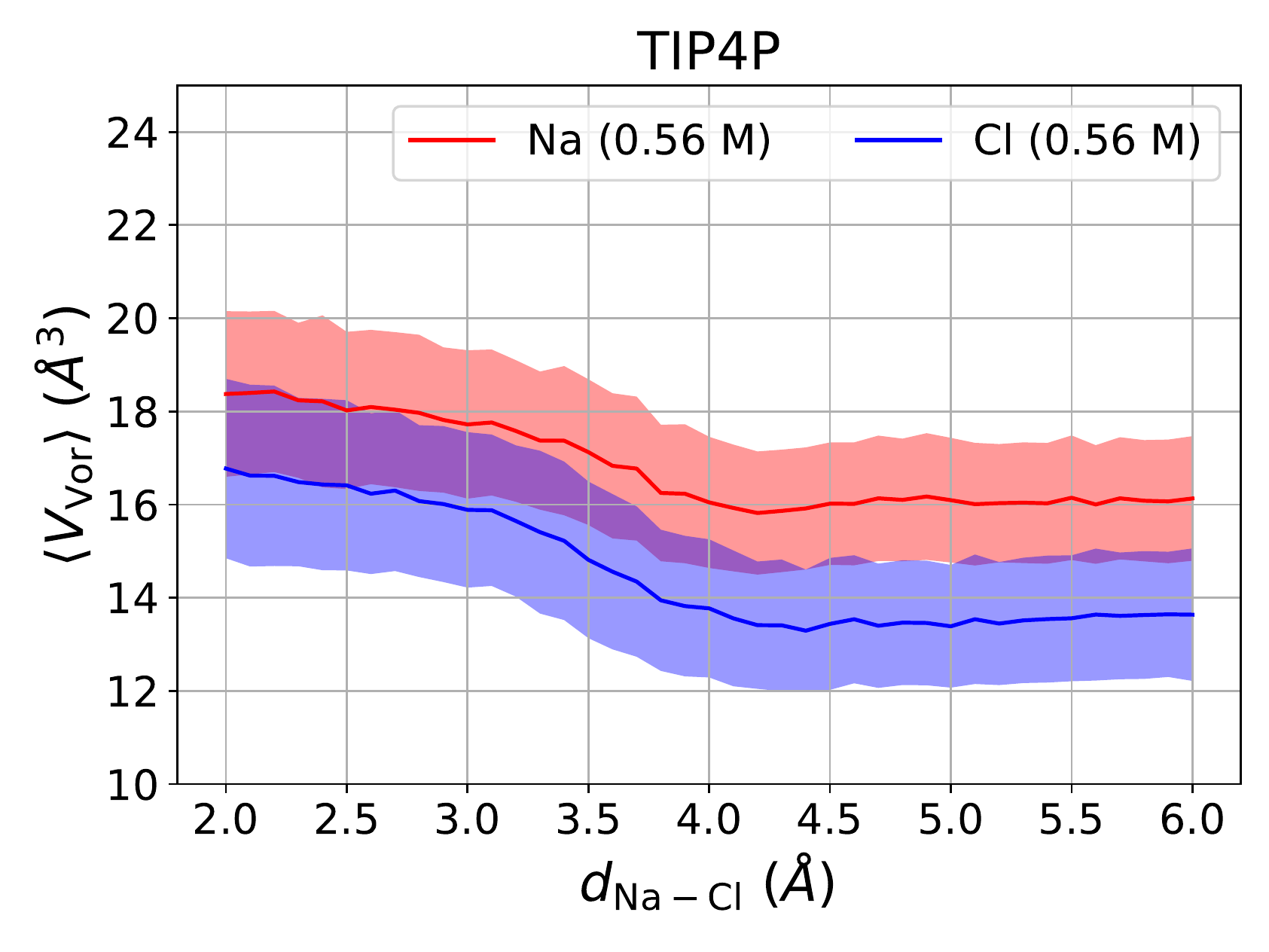}
        \caption{}
        \label{supp-fig:t4pvol}
    \end{subfigure}
    \begin{subfigure}[b]{0.32\textwidth}
    \includegraphics[scale=0.32]{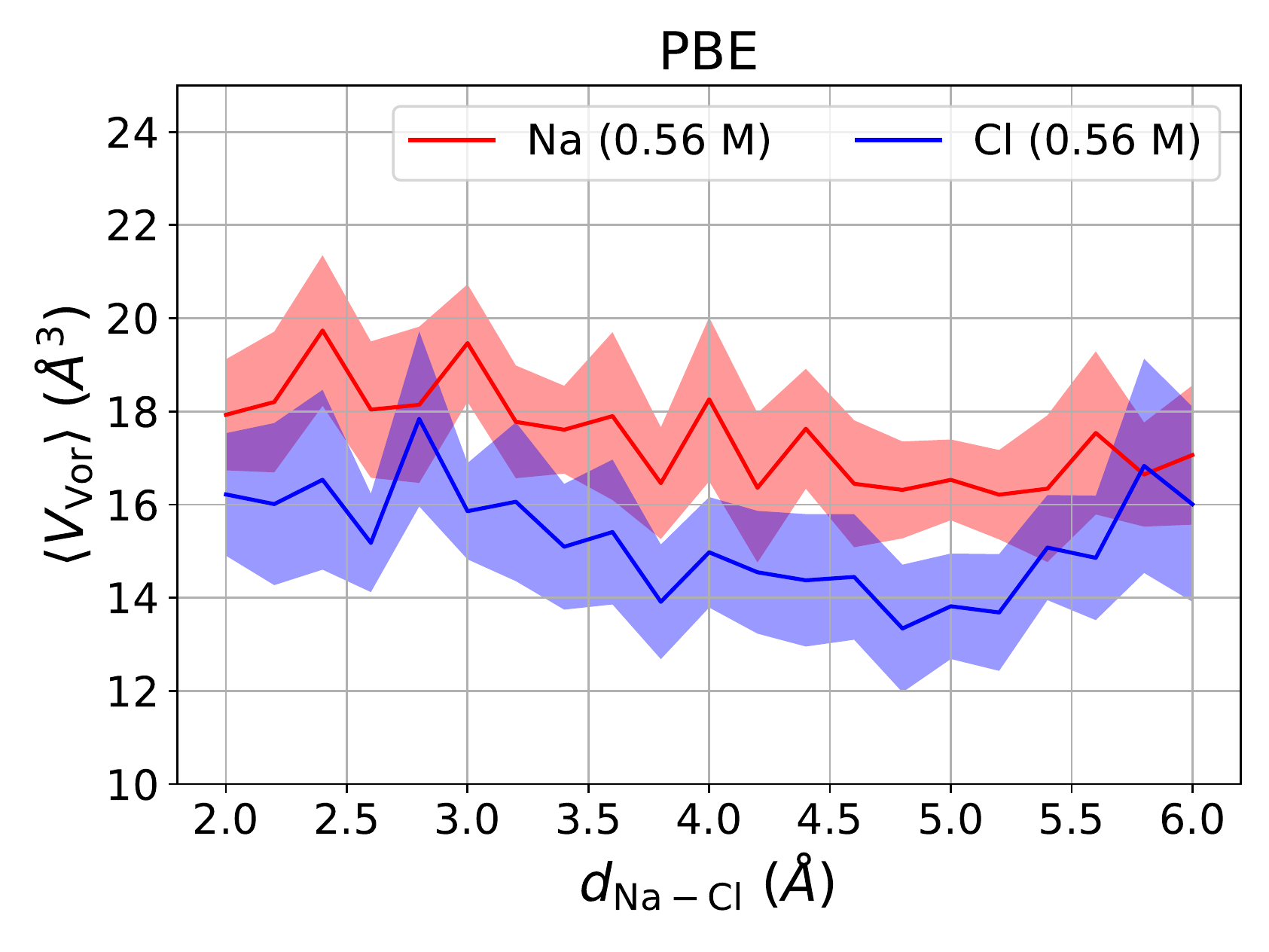}
        \caption{}
        \label{supp-fig:pbevol}
    \end{subfigure}
    ~ 
    \begin{subfigure}[b]{0.32\textwidth}
    \includegraphics[scale=0.32]{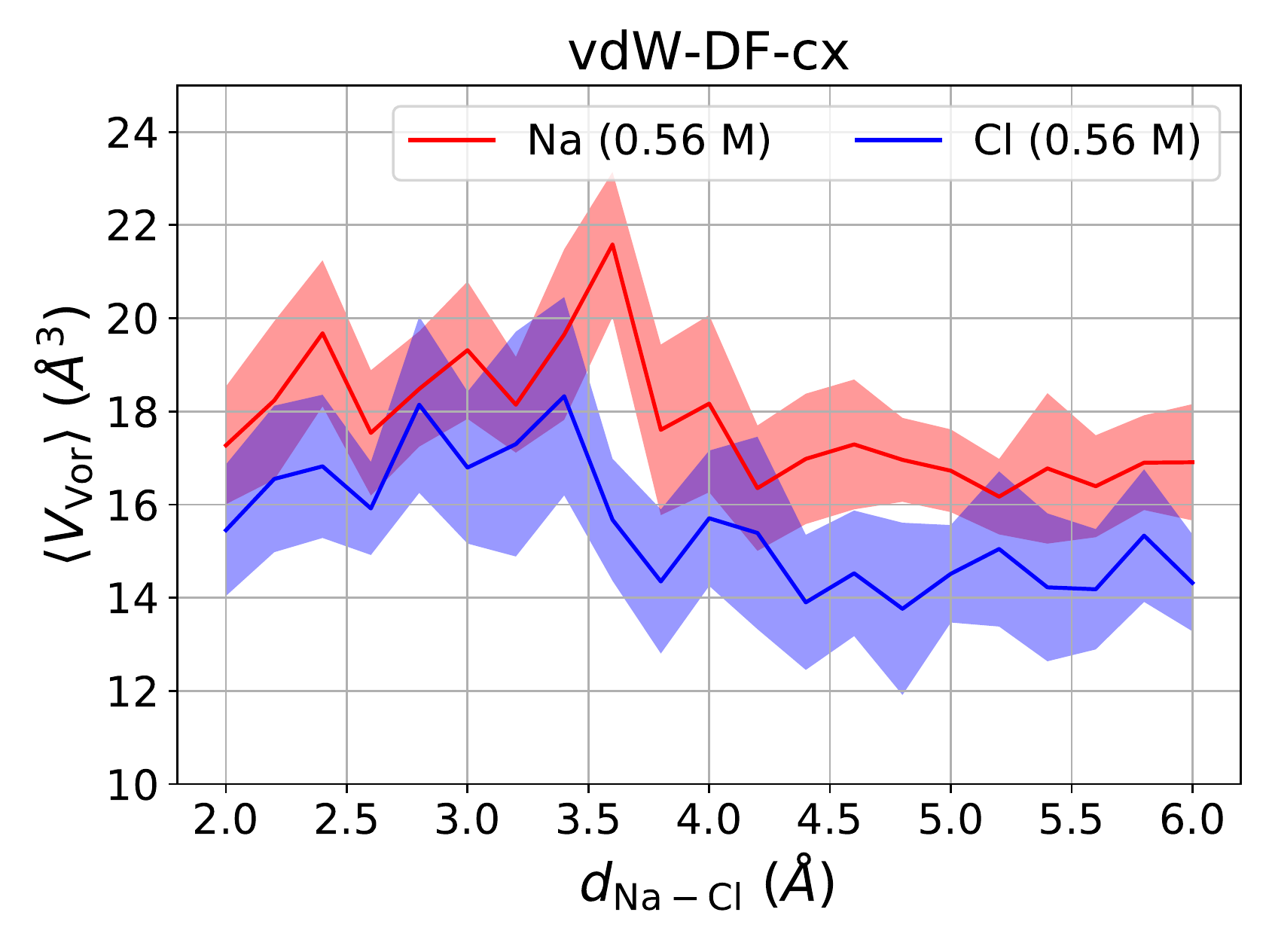}
        \caption{}
        \label{supp-fig:bhvol}
    \end{subfigure}
    \caption{\textbf{(a-c)} Voronoi volumes as a function of ionic distance for the 0.56 M TIP4P, PBE, vdW-DF-cx simulations. The red and blue lines correspond to Voronoi volume averages for Na and Cl, respectively. Average \ce{H2O} volumes are constant across the simulations and are not plotted, but are presented in Table~\ref{supp-tbl:vorvol}.}\label{supp-fig:vorvols}
\end{figure*}

Table~\ref{supp-tbl:vorvol} shows the thermodynamically averaged volumes for the Voronoi cells of the ions and water molecules.
The water averages confirm what is shown in Fig.~\ref{supp-fig:vorvols} -- namely, that the models all predict the same occupied per \ce{H2O} molecule and its value is only dependent on the density of the system.
On the other hand, the volumes of the Na and Cl ions are, within error, largely independent of density effects in both classical and \textit{ab initio} simulations, with the \textit{ab initio} volumes having slightly larger averages that agree between the different XC parameterization and molarity combinations.
There is a small increase in the \textit{ab initio} ion Voronoi volume with decreasing solution volume, but between XC functionals the averages agree and are larger than the corresponding classical density.

\subsection{Voronoi Entropy}

Tessellations like those generated in the Voronoi diagrams and their connectivity have a measure of information content. 
In particular, the partitions have an associated information entropy that characterize their disorder. 

We may use this fact to construct the information entropy of the Voronoi diagram at a given simulation step: \begin{equation}
    S_{\mathrm{Vor}} = -\sum_{i}P_i \ln P_i, \label{eq:vorent}
\end{equation} where $P_i$ is the fraction of polyhedra with $i$ faces. 
In two dimensions, the probability is chosen with respect to $i$ edges of the Voronoi polygon, \cite{voronoientBormashenko2018} the only unambiguous connection between any two polygons in the tessellation.
Analogously, in three dimensions, the unambiguous connectors between the polyhedra are their faces. 
It is for this reason we construct the probabilities in Eq.~\ref{eq:vorent} with the face number of a polyhedron, as it is the direct measure of connectivity in the tessellation.
The sum over $i$ is taken from $i=4$ faces to the maximal value of $i$ faces for a given tessellation. 
In a completely ordered system (all polyhedra the same), $S_{\mathrm{Vor}}=0,$ thus we expect increasing disorder with increasing value of the information entropy. 
It is important to note that this entropy is not the thermodynamic entropy, however they have been related in some models. \cite{voronoigasSenthilKumar2005}

\begin{figure*}
    \centering
    \begin{subfigure}[b]{0.32\textwidth}
    \includegraphics[scale=0.32]{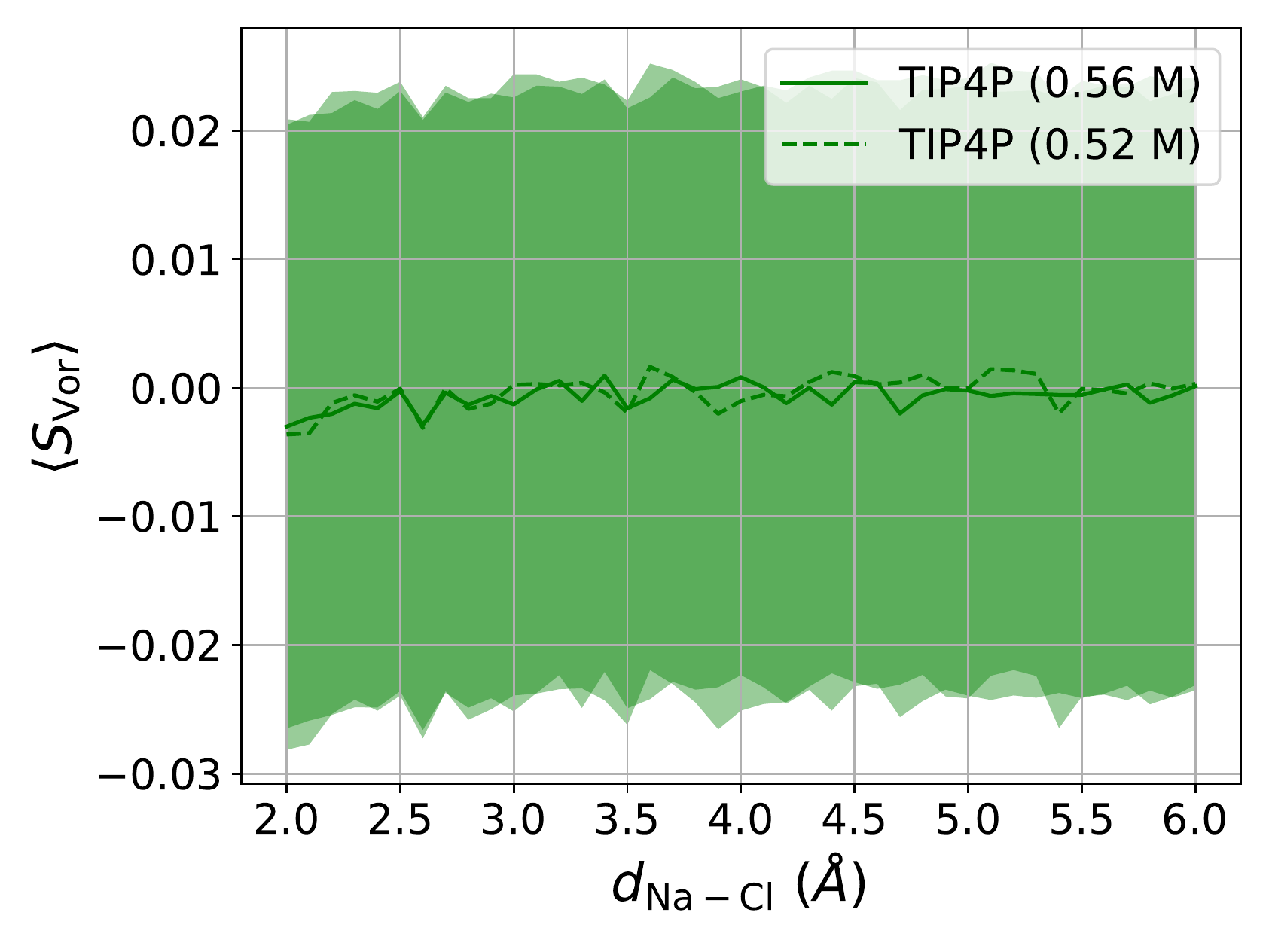}
        \caption{}
        \label{supp-fig:t4pvorent}
    \end{subfigure}
    \begin{subfigure}[b]{0.32\textwidth}
    \includegraphics[scale=0.32]{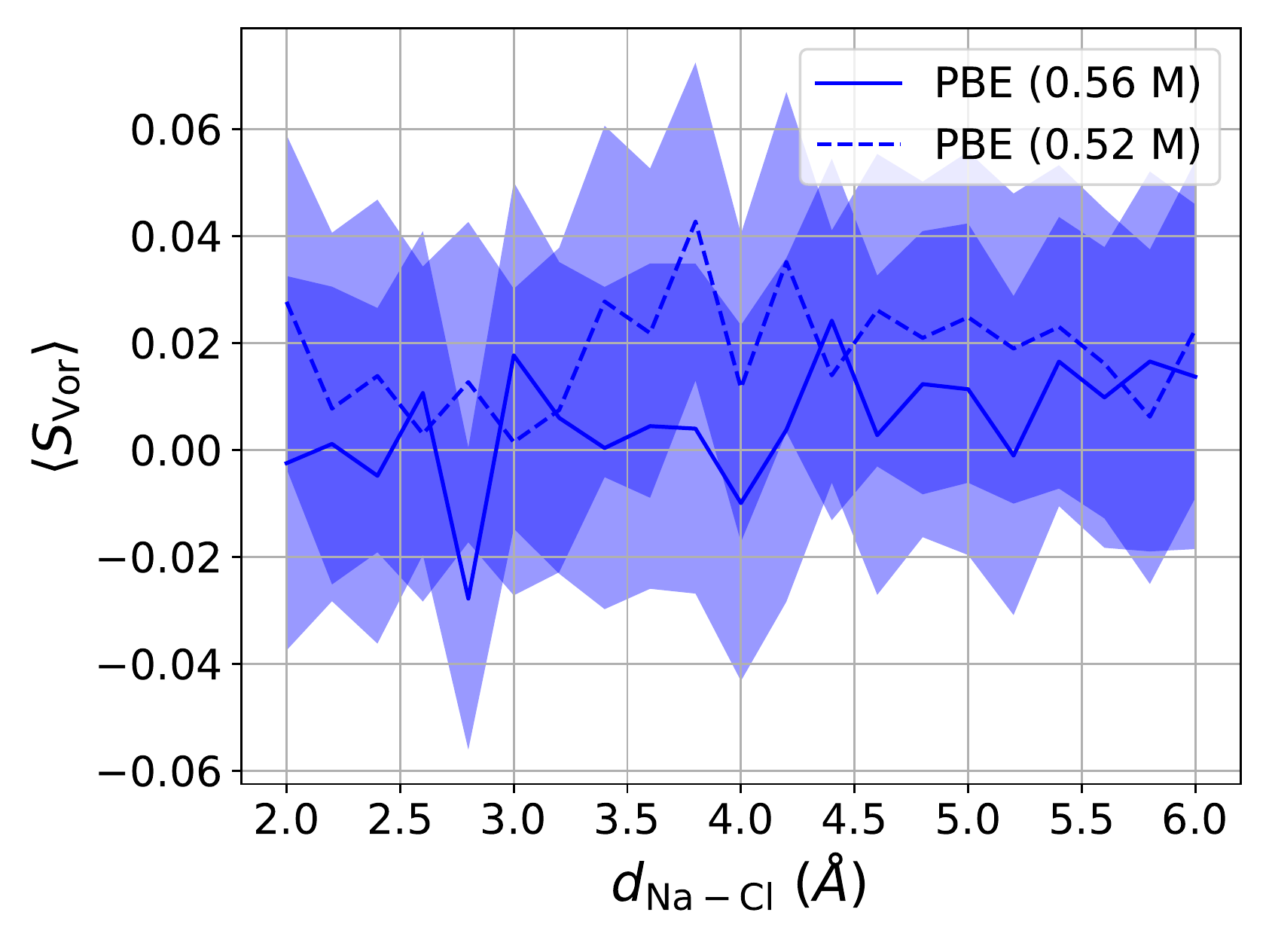}
        \caption{}
        \label{supp-fig:pbevorent}
    \end{subfigure}
    ~ 
    \begin{subfigure}[b]{0.32\textwidth}
    \includegraphics[scale=0.32]{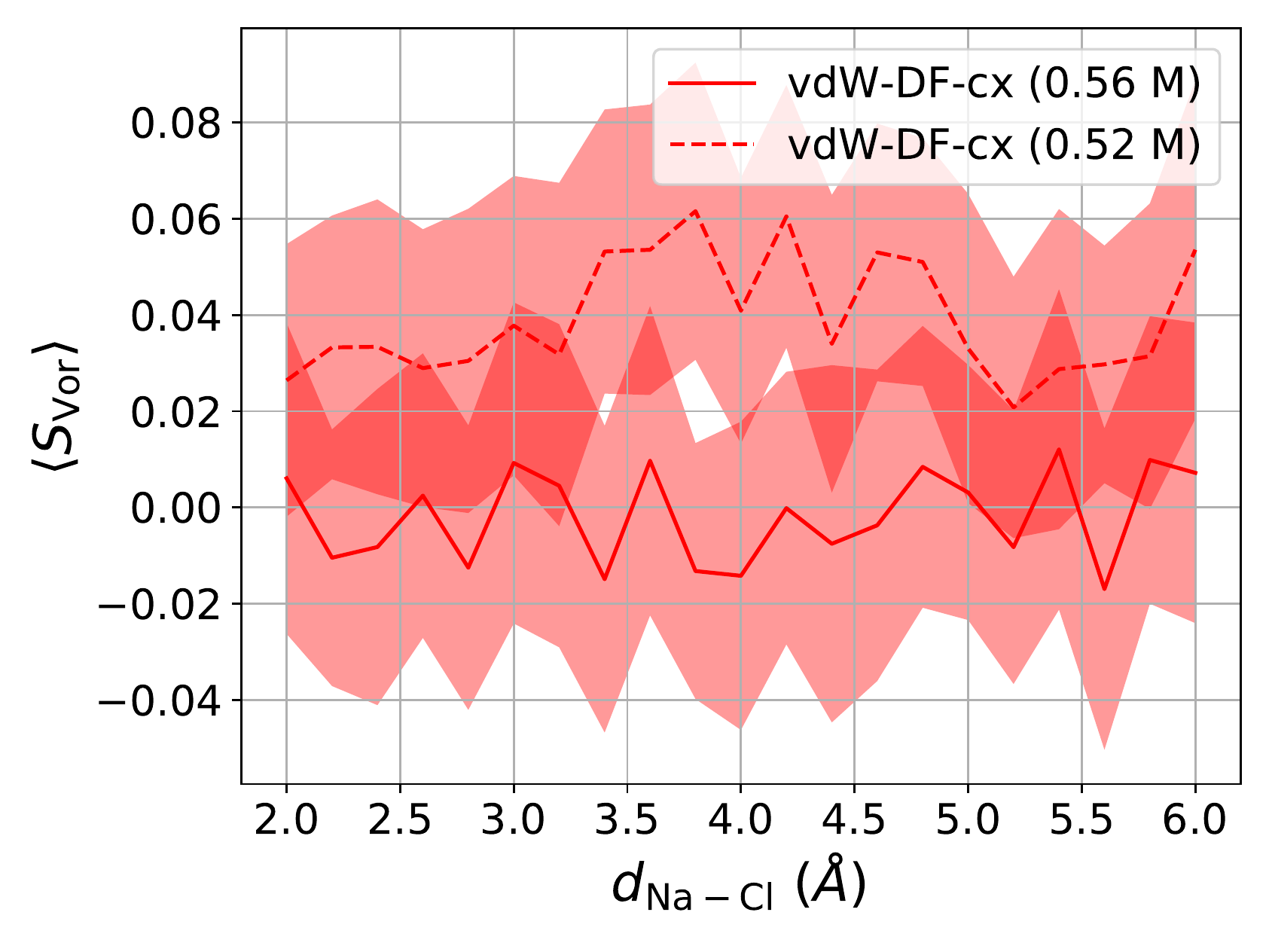}
        \caption{}
        \label{supp-fig:bhvorent}
    \end{subfigure}
    \caption{\textbf{(a-c)} Voronoi entropies as a function of ionic distance for the 0.56 M TIP4P, PBE, vdW-DF-cx simulations. The solid and dashed lines correspond to Voronoi entropy averages for 0.56 M and 0.52 M solutions, respectively.}\label{supp-fig:vorents}
\end{figure*}

Fig.~\ref{supp-fig:vorents} shows the calculated average Voronoi entropies from the sampled 0.56 M configurations from each \textit{ab initio} simulation, shifted by the average entropy from a trajectory of the corresponding system without the ions.
A similar plot including the 0.52 M trajectories is shown in Fig.~\ref{supp-fig:vorents}.
We see that the 0.52 M simulations tend to have, on average, more disorder in the respective Voronoi diagrams than the corresponding high density solutions for the \textit{ab initio} simulations.
Moreover, for each density there tends to be  marginally more disorder in the PBE simulations than in the vdW-DF-cx.
Inclusion of the ions in the classical trajectories does not induce much variability in the average Voronoi entropies.
Furthermore, we find the effect of increasing the box size has a noticeably greater effect on the average Voronoi entropies of the PBE system, varying by up to $\sim0.07$, while density changes have a less pronounced effect on the vdW-DF-cx system.
This corroborates the behavior of the drastic behavior change in the PBE PMF.

\begin{figure}
    \centering
    \begin{subfigure}[b]{0.49\textwidth}
    \includegraphics[scale=0.45]{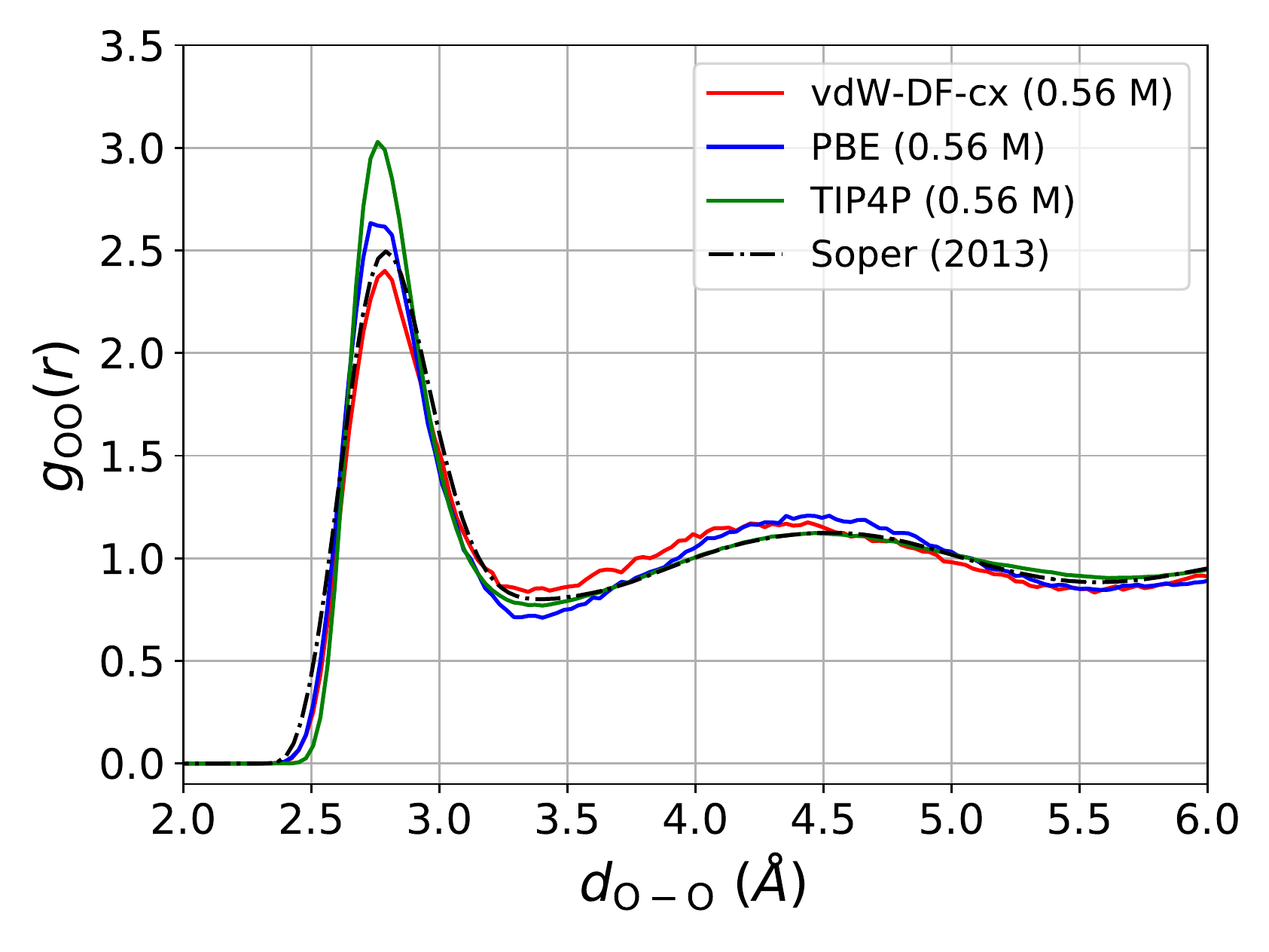}
    \caption{Pure Water Solutions}
    \label{supp-fig:hdld_goo}
    \end{subfigure}
    \begin{subfigure}[b]{0.49\textwidth}
    \includegraphics[scale=0.45]{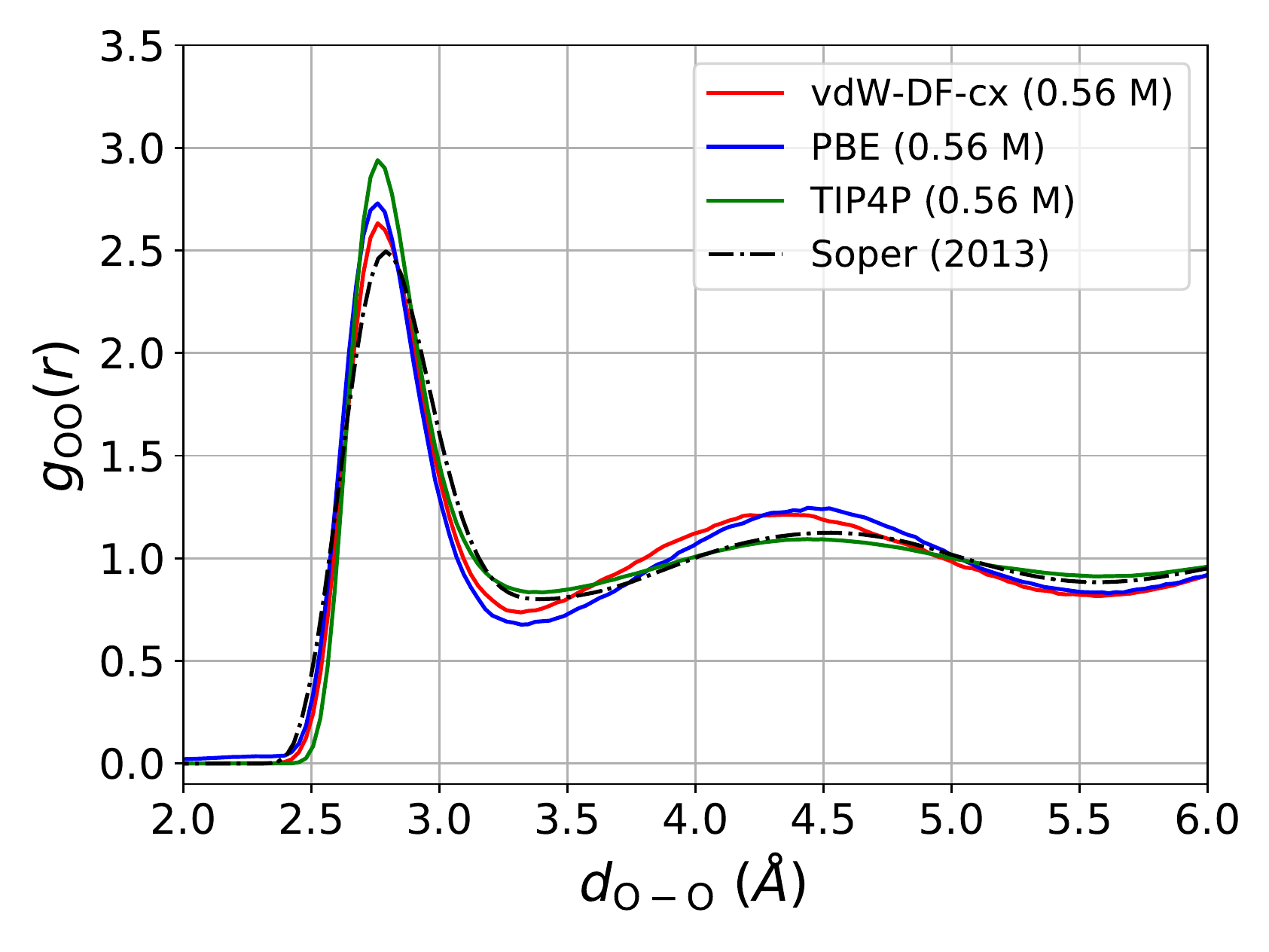}
    \caption{Ionic Solutions}
    \label{supp-fig:hdld_goo_ion_tdyn}
    \end{subfigure}
    \caption{\textbf{(a)} The \ce{O-O} radial distributions functions of pure water using TIP4P, vdW-DF-cx, and PBE (green, red, and blue, respectively) with the box size that corresponds to the 0.56 M solutions with the ions removed. \textbf{(b)} The therymodynamically averaged \ce{O-O} radial distributions functions of the ion solutions for the TIP4P, vdW-DF-cx, and PBE simulations with concentrations of 0.56 M. Both figures show experimental results at ambient conditions for comparison.\cite{Soper2013} The inclusion of the ions serves to increase the structure in the \textit{ab initio} cases.}
\end{figure}

Table \ref{supp-tbl:tdynavgs} contains the mean Voronoi entropies for the different pure water systems, $S_{\mathrm{Vor,pure}}$.
The results indicate minimal difference within the same XC for different densities.
However, we see that, despite the slightly more structured \ce{O-O} radial distribution functions for PBE shown in Fig. \ref{supp-fig:hdld_goo} when compared to the vdW-DF-cx systems, the PBE Voronoi entropy is higher.
This is surprising, given the ice-like nature of PBE water.
This indicates more variability in the polyhedra generated by the Voronoi tessellation in PBE systems.

\subsection{Radial Distribution Functions and Coordination Numbers}
Insight into solute ion effects on the solvent can be gained from examining the ion-oxygen and oxygen-oxygen radial distribution functions (RDFs).
Initial neutron diffraction analysis led to results indicating minimal affects beyond the first ion hydration shell,\cite{khalideSoper2006} while later results showed that ion effects extend beyond the first solvation shell.\cite{perturbgood_Mancinelli2007}
Even further results suggested that cosmo/chaotropic effects have minimal bearing on molecular interactions between the solute and solvent, despite the changes in solvent structure.\cite{perturbbad_Mancinelli2007} 
X-ray absorption studies have indicated cationic destabilization of the hydrogen bond network indicative of extended cationic effects on the solvent, while anionic effects from Cl$^-$ indicated insignificant structural effects.\cite{nashell1_Waluyo2011, xraychaoWaluyo2014}

However, GGA-level \textit{ab initio} simulations have suggested that Na$^+$ affects the solvent only locally (not beyond the first solvation shell), while Cl$^-$ has longer-range effects,\cite{naclrdfs_Gaiduk2017} directly contrasting with interpretations of experimental results.
Specifically, it was shown that the Cl$^-$ ion weakens hydrogen bonding up to the third solvation shell.
We summarize our simulation results in Table \ref{supp-tbl:rdfs}, where we list the coordination numbers and distances of the first two radial minima in the ion-oxygen and oxygen-oxygen RDFs at different ion separation distances and the thermodynamic averages.

\begin{figure}
    \centering
    \begin{subfigure}[b]{0.49\textwidth}
    \includegraphics[scale=0.45]{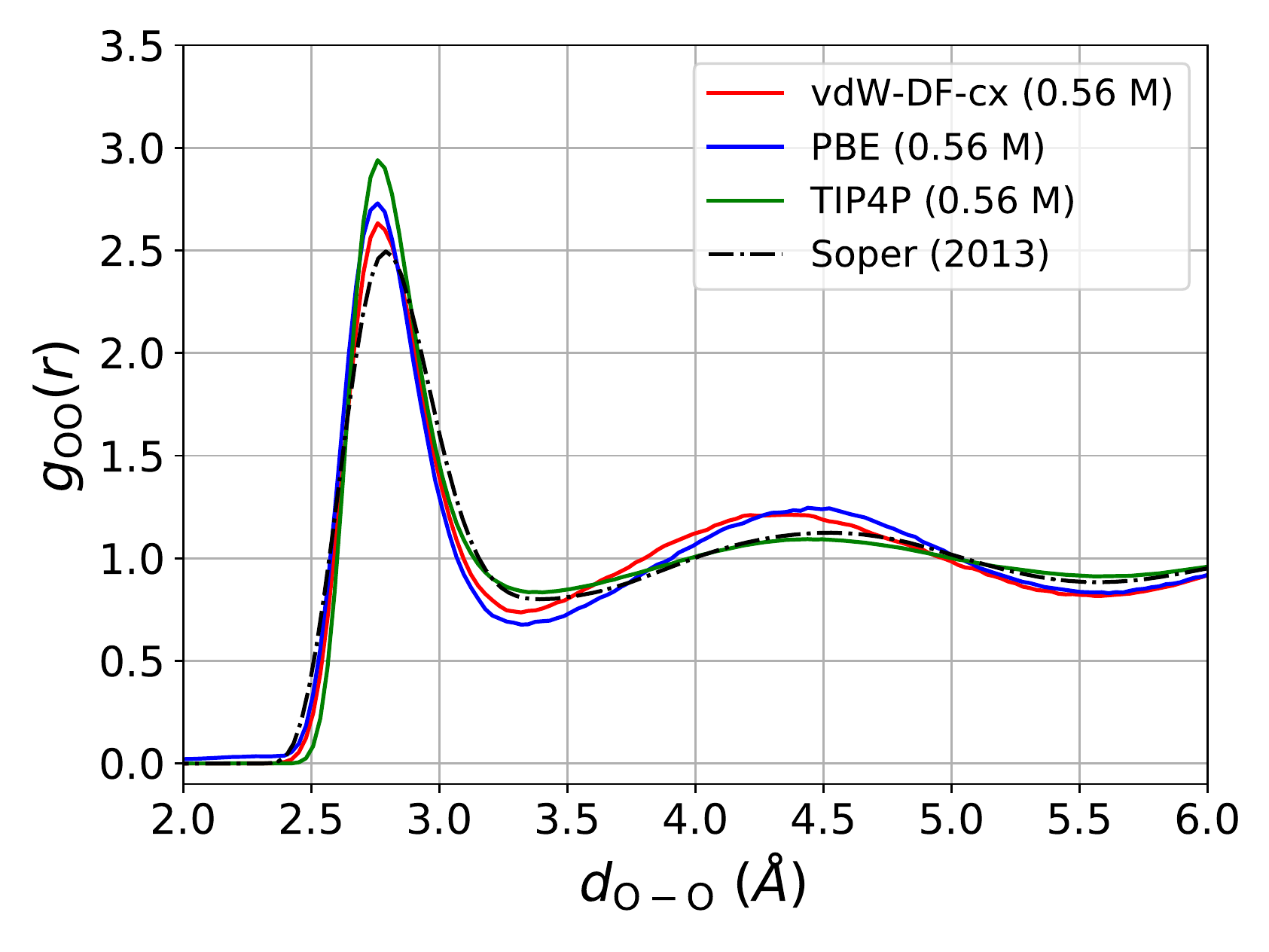}
    \caption{\ce{O-O} radial distribution functions.}
    \label{supp-fig:gmx_aimd_goo_ions_tdyn}
    \end{subfigure}
    \begin{subfigure}[b]{0.49\textwidth}
    \includegraphics[scale=0.45]{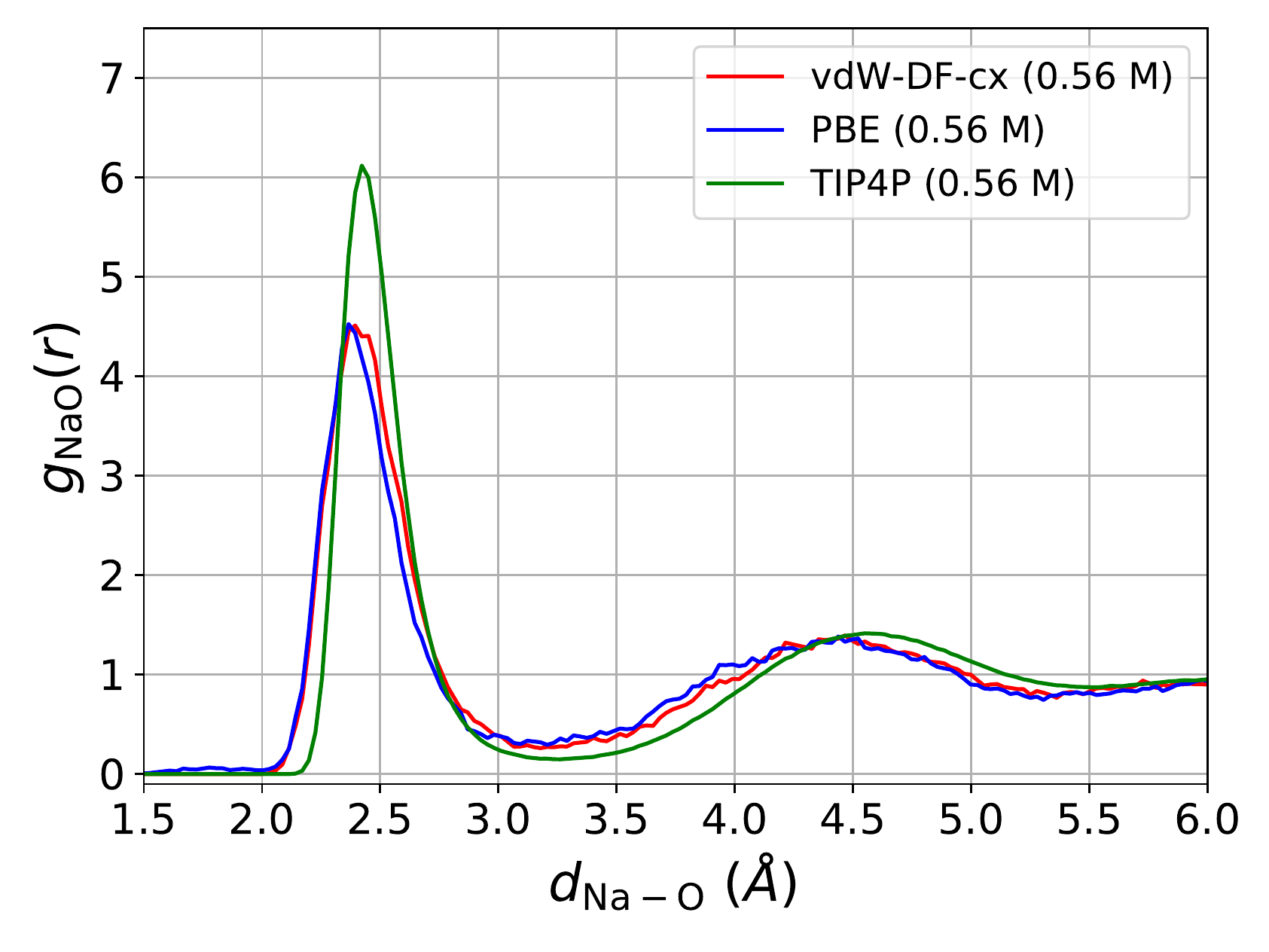}
    \caption{\ce{Na-O} radial distribution functions.}
    \label{supp-fig:gmx_aimd_gnao_ion_tdyn}
    \end{subfigure}
    \begin{subfigure}[b]{0.49\textwidth}
    \includegraphics[scale=0.45]{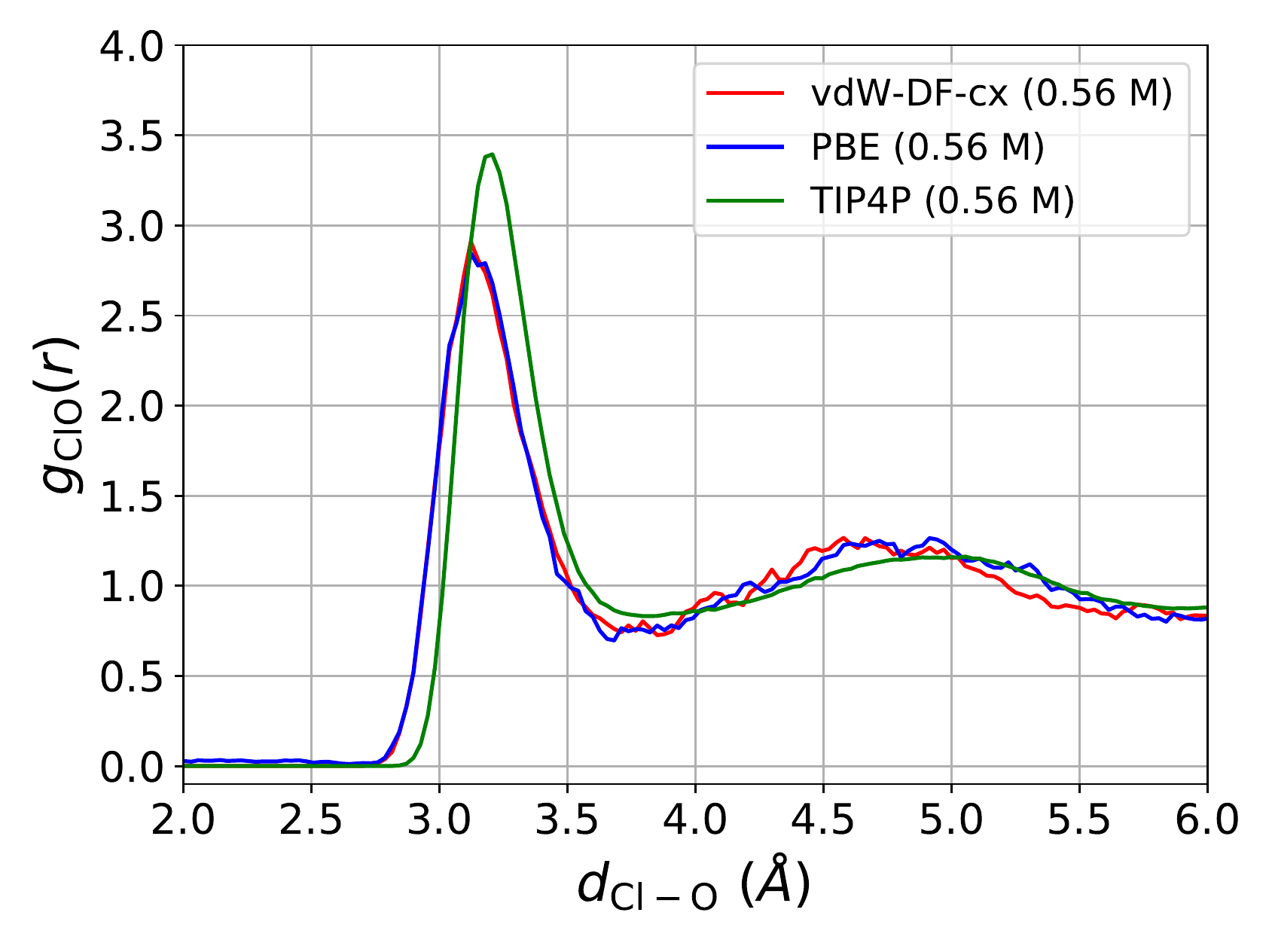}
    \caption{\ce{Cl-O} radial distribution functions.}
    \label{supp-fig:gmx_aimd_gclo_ion_tdyn}
    \end{subfigure}
    \caption{Thermodynamically averaged \ce{X-O} radial distribution functions for the 0.56 M solutions. In each, the vdW-DF-cx, PBE, and TIP4P solutions are red, blue, and green, respectively. \textbf{(a)} Plotted with the simulated \ce{O-O} radial distribution functions is experimental data.\cite{Soper2013} \textbf{(b)} The \ce{Na-O} radial distribution functions. Notable is the clear presence of a second shell, indicative of cosmotropic effects beyond the first solvation shell. \textbf{(c)} The \ce{Cl-O} radial distribution functions, indicating the similar structural effects as seen in the \ce{Na-O} case, although not as significant.} \label{supp-fig:tdynrdfs}
\end{figure}

Notably, we see in Fig.~\ref{supp-fig:tdynrdfs} the effects the ions have on the neighboring solution through the thermodynamically averaged radial distribution functions of the ion-oxygen pairs.
The presence of a clear second shell in the \ce{Na-O} distributions (Fig.~\ref{supp-fig:gmx_aimd_gnao_ion_tdyn}) is indicative of slightly extended perturbations on the surrounding solvent molecules, what one might see referred to as a ``cosmotropic'' effect in that it introduces structure into the surrounding solvent.
An analogous effect is present in the \ce{Cl-O} distributions (Fig.~\ref{supp-fig:gmx_aimd_gclo_ion_tdyn}), although not as strongly as is the case with the \ce{Na-O} pair.

Additionally, indications of local cosmotropic effects are present in Table~\ref{supp-tbl:rdfs}, where the first hydration shell radius ($r_1$) for the \ce{Na-O} RDF is notably pulled in to lower distances.
Moreover, the PBE \ce{Na-O} second coordination shell is markedly more contracted inwards than the vdW-DF-cx simulations, for both molarities.
This indicates that the PBE solution is more prone to nonlocal cosmotropic perturbations induced by the \ce{Na+} cation.
Table~\ref{supp-tbl:rdfs} further shows that the \ce{Cl-O} RDFs show extended perturbation in solvation structure, in agreement with previous results.\cite{naclrdfs_Gaiduk2017}
The solvation radii for both the first and second solvation shells ($r_1$ and $r_2$, respectively) are notably shifted to higher distances, indicative of a weakening of the hydrogen bond network in the vicinity of the \ce{Cl-} ion -- a ``chaotropic'' response.

Fig.~\ref{supp-fig:gmx_aimd_goo_ions_tdyn} also shows the \ce{O-O} pair distributions of the ionic solutions compared to experimental data.
In all cases besides the TIP4P water model, the second shell is overstructured compared to experiment, further evidence for induced structural effects in \textit{ab initio} simulations that might bring about conflicting results in the PMF energetic ordering.
Bulk structural effects will influence the strength with which the solvent interacts with the solute ions, affecting the stability of either given solvation state.

\bibliography{__refs}

\begin{table*}
	\caption{\label{supp-tbl:tdynavgs} Thermodynamic averages of the Voronoi entropies, pressures, and temperatures of different systems.}
	\begin{ruledtabular}
		\begin{tabular}{cc|ccccccc}
			Model  & Molarity (M) &  $S_{\mathrm{Vor, ions}}$ & $S_{\mathrm{Vor, pure}}$ & $\Delta S_{\mathrm{Vor}}$ & $P_{\mathrm{ions}}$ (kBar) & $P_{\mathrm{pure}}$ (kBar) & $\Delta P$ (kBar) & T (K) \\ \hline
			\multirow{2}{*}{TIP4P} & 0.56& $3.06\pm0.02$ & $3.06\pm0.02$ & $0\pm0.03$ & $-.33\pm1.2$ & $-0.50 \pm 1.15$ & $0.17\pm 1.6$ & $298.3 \pm 17.0$ \\
			& 0.52 & $3.07\pm0.02$ & $3.07\pm0.02$ & $0\pm 0.03$ & $-1.44\pm1.1$ & $-1.42 \pm 1.06$ & $-0.2\pm 1.5$ & $298.0\pm17.3$ \\ \hline
			\multirow{2}{*}{PBE} & 0.56& $2.86\pm0.03$ & $2.86\pm0.03$ & $0.00\pm0.04$ & $2.72\pm3.1$ & $2.32 \pm 3.16$ & $0.4\pm4.4$ & $362.6\pm14.5$\\
			& 0.52 & $2.90\pm0.03$ & $2.88\pm0.03$ & $0.02\pm0.03$ & $0.25\pm2.9$ & $0.20 \pm 3.14$ & $0.05\pm4.2$ & $367.9\pm15.1$ \\ \hline
			\multirow{2}{*}{vdW-DF-cx} & 0.56& $2.84\pm0.03$ & $2.84\pm0.03$ & $0\pm0.04$ & $-3.75\pm3.0$ & $-3.79 \pm 2.99$ & $0.04\pm4.2$ & $350.2\pm14.8$ \\
			& 0.52 & $2.88\pm0.03$ & $2.84\pm0.03$ & $0.04\pm 0.04$ & $-5.71\pm2.8$ & $-5.23 \pm 3.05$ & $-0.48\pm4.1$ & $336.9\pm 15.8$ \\
			
		\end{tabular}
		
		{\footnotesize Ion solution weighted averages are computed using Boltzmann factors weighted by the PMF values at a given distance. Individual trajectory averages for pressure and temperature values are calculated with time averaging 100 equally spaced subtrajectories to estimate error. The parameters for 0.56 M and 0.52 M simulations when used in the pure water simulations (no ions) correspond to 0.97 \si[inter-unit-product=\ensuremath{\cdot}]{\gram\per\cubic\centi\meter} and 0.90 \si[inter-unit-product=\ensuremath{\cdot}]{\gram\per\cubic\centi\meter}, respectively.}
		
	\end{ruledtabular}
\end{table*}

\begin{table*}
  \caption{Thermodynamically averaged Voronoi volumes for the atomic ions and water molecules.}
  \label{supp-tbl:vorvol}
  \begin{ruledtabular}
  \begin{tabular}{cc|ccc}
  Model  & Molarity (M) &  $\langle V_{\mathrm{Vor, Na}} \rangle$ (\AA$^{3}$) & $\langle V_{\mathrm{Vor, Cl}} \rangle$ (\AA$^{3}$) & $\langle V_{\mathrm{Vor, water}} \rangle$ (\AA$^{3}$) \\ \hline
    \multirow{2}{*}{TIP4P} & 0.56& $17.2 \pm 1.5$ & $15.1 \pm 1.5$ & $30.592 \pm 0.024$ \\
    & 0.52 & $17.9\pm 1.7$ & $15.8\pm1.7$ & $32.906\pm 0.027$ \\ \hline
    \multirow{2}{*}{PBE} & 0.56& $17.1\pm1.2$ & $14.9\pm 1.4$ & $30.596\pm 0.019$ \\
    & 0.52 & $18.0\pm 1.6$ & $15.3\pm 1.4$ & $32.911\pm0.024$ \\ \hline
    \multirow{2}{*}{vdW-DF-cx} & 0.56& $16.9\pm 1.1$ & $14.8\pm 1.5$ & $30.598\pm0.021$ \\
    & 0.52 & $18.4\pm1.6$ & $16.5\pm1.7$ & $32.894 \pm 0.026$\\ 
  \end{tabular}
  \end{ruledtabular}
\end{table*}

\begin{table*}
\caption{\label{supp-tbl:pureh2oavgs} Average temperatures for the pure water systems.}
\begin{tabular}{ccc|c}
\hline\hline
    Model  & Concentration$^*$ (M) &  $\rho$ (\si[inter-unit-product=\ensuremath{\cdot}]{\gram\per\cubic\centi\meter}) & T (K) \\ \hline
    \multirow{2}{*}{TIP4P} & 0.56& $0.97$ & $297.70 \pm 19.69$ \\
    & 0.52 & $0.90$ & $300.50 \pm 18.61$ \\ \hline
    \multirow{2}{*}{PBE} & 0.56& $0.97$ & $399.50 \pm 15.78$  \\
    & 0.52 & $0.90$ & $423.25 \pm 16.96$  \\ \hline
    \multirow{2}{*}{vdW-DF-cx} & 0.56& $0.97$ & $379.02 \pm 14.73$ \\
    & 0.52 & $0.90$ & $382.24 \pm 14.21$  \\ \hline\hline

\end{tabular}

$^*${\footnotesize Concentrations are given to clarify that the box parameters used for the pure water simulations are the same as those of the given molarity when the ions are included.}
\end{table*}

\begin{table*}
\caption{Characteristics of the \ce{Cl-O}, \ce{Na-O}, \ce{O-O} radial distribution functions at ion separations of 2.8, 4.8, and 6.0 \AA\ and the thermodynamic averages of the values across all separations. $N$ is the coordination number in the first solvation shell, while $r_1$ and $r_2$ are the minima (in \AA) defining the first and second shell, respectively.}
\label{supp-tbl:rdfs}
\begin{ruledtabular}
\begin{tabular}{cc|ccc|ccc|ccc|ccc}
    \multicolumn{2}{c|}{\textbf{\ce{Cl-O}}} & \multicolumn{3}{c|}{2.8 \AA} & \multicolumn{3}{c|}{4.8 \AA} & \multicolumn{3}{c
    |}{6.0 \AA} & \multicolumn{3}{c}{Avg.}\\ \hline
    Model & Molarity (M) & $r_1$ & $r_2$ & $N$ & $r_1$ & $r_2$ & $N$ & $r_1$ & $r_2$ & $N$ & $r_1$ & $r_2$ & $N$ \\ \hline
    \multirow{2}{*}{TIP4P} & 0.56 & 3.63 & 5.87 & 5.36 & 3.85 & 5.98 & 7.30 & 3.88 & 6.03 & 7.30 & 3.79 & 5.95 & 6.62  \\
    & 0.52 & 3.68 & 5.92 & 5.46 & 3.88 & 6.01 & 7.21 & 3.88 & 6.12 & 7.08 & 3.85 & 5.98 & 6.58 \\ \hline
    \multirow{2}{*}{PBE} & 0.56 & 3.21 & 5.50 & 2.21 & 4.07 & 6.15 & 7.74 & 3.99 & 6.03 & 7.04 & 3.15 & 5.92 & 2.02 \\
    & 0.52 & 3.21 & 5.36 & 1.89 & 3.77 & 5.89 & 5.14 & 3.57 & 5.87 & 4.43 & 3.79 & 5.89 & 5.62 \\ \hline
    \multirow{2}{*}{vdW-DF-cx} & 0.56 & 3.18 & 5.53 & 1.85 & 3.77 & 6.03 & 6.18 & 3.99 & 6.15 & 6.93 & 3.71 & 5.73 & 5.81 \\
    & 0.52 & 3.77 & 5.84 & 4.82 & 3.71 & 6.06 & 5.07 & 3.77 & 5.95 & 5.17 & 3.77 & 5.84 & 5.51 \\ \hline \hline 
    \multicolumn{2}{c|}{\textbf{\ce{Na-O}}} & $r_1$ & $r_2$ & $N$ & $r_1$ & $r_2$ & $N$ & $r_1$ & $r_2$ & $N$ & $r_1$ & $r_2$ & $N$ \\ \hline
    \multirow{2}{*}{TIP4P} & 0.56 & 3.32 & 5.53 & 4.67 & 3.23 & 5.53 & 5.94 & 3.29 & 5.53 & 5.97 & 3.26 & 5.53 & 5.13 \\
    & 0.52 & 3.23 & 5.53 & 4.46 & 3.35 & 5.56 & 5.89 & 3.23 & 5.59 & 5.84 & 3.23 & 5.53 & 4.86  \\ \hline
    \multirow{2}{*}{PBE} & 0.56 & 3.01 & 5.19 & 3.59 & 3.23 & 5.25 & 4.94 & 3.18 & 5.22 & 5.04 & 3.21 & 5.31 & 4.64 \\
    & 0.52 & 3.40 & 5.64 & 3.75 & 3.15 & 5.56 & 4.12 & 3.26 & 5.31 & 4.80 & 3.21 & 5.56 & 4.44 \\ \hline
    \multirow{2}{*}{vdW-DF-cx} & 0.56 & 3.15 & 5.28 & 3.80 & 3.07 & 5.31 & 5.25 & 3.09 & 5.39 & 4.57 & 3.18 & 5.39 & 4.93  \\
    & 0.52 & 3.09 & 5.28 & 3.55 & 3.40 & 5.42 & 4.85 & 3.15 & 5.42 & 4.43 & 3.29 & 5.39 & 4.63 \\ \hline \hline
    \multicolumn{2}{c|}{\textbf{\ce{O-O}}} & $r_1$ & $r_2$ & $N$ & $r_1$ & $r_2$ & $N$ & $r_1$ & $r_2$ & $N$ & $r_1$ & $r_2$ & $N$ \\ \hline
    \multirow{2}{*}{TIP4P} & 0.56 & 3.40 & 5.61 & 4.91 & 3.35 & 5.59 & 4.68 & 3.35 & 5.61 & 4.68 & 3.40 & 5.61 & 4.91 \\
    & 0.52 & 3.40 & 5.70 & 4.60 & 3.40 & 5.67 & 4.60 & 3.40 & 5.67 & 4.59 & 3.40 & 5.70 & 4.60 \\ \hline
    \multirow{2}{*}{PBE} & 0.56 & 3.43 & 5.64 & 4.87 & 3.32 & 5.59 & 4.33 & 3.26 & 5.61 & 4.18 & 3.32 & 5.59 & 4.33 \\
    & 0.52 & 3.29 & 5.53 & 4.03 & 3.35 & 5.59 & 4.15 & 3.26 & 5.50 & 3.90 & 3.35 & 5.59 & 4.15 \\ \hline
    \multirow{2}{*}{vdW-DF-cx} & 0.56 & 3.29 & 5.47 & 4.27 & 3.35 & 5.59 & 4.45 & 3.32 & 5.56 & 4.27 & 3.32 & 5.56 & 4.29 \\
    & 0.52 & 3.26 & 5.53 & 3.87 & 3.32 & 5.47 & 3.95 & 3.29 & 5.56 & 3.91 & 3.29 & 5.53 & 3.96 \\ \hline
    Experiment\cite{Soper2013} & -- & -- & -- & -- & -- & -- & -- & -- & -- & -- & 3.39 & 5.58 & 4.724 \\

\end{tabular}
\end{ruledtabular}
\end{table*}

\begin{figure}
    Timestep redshifting.
    \centering
    \begin{subfigure}[b]{\textwidth}  \includegraphics[scale=0.5]{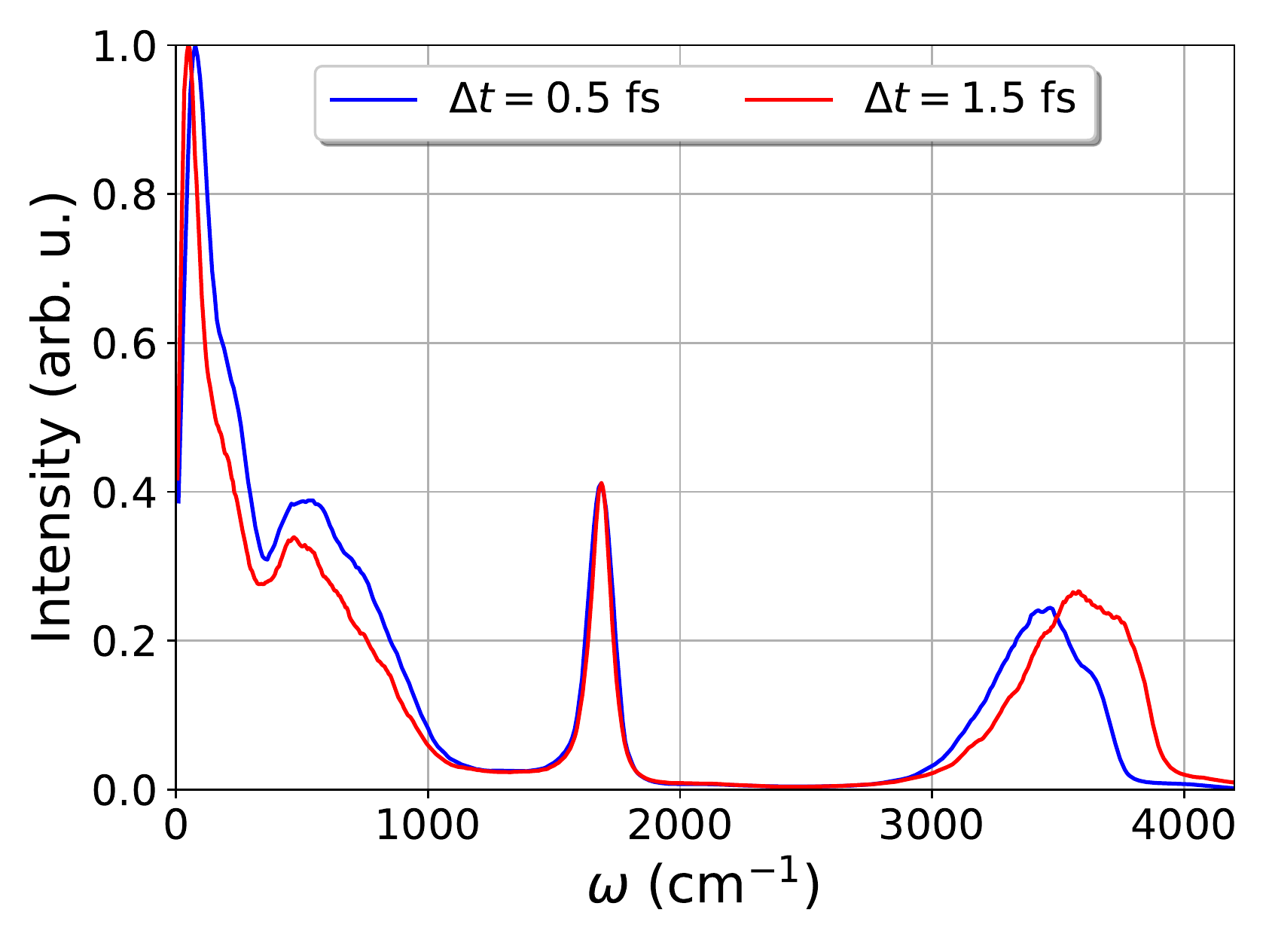}
    \caption{}\label{vibspec}
    \end{subfigure}
    \begin{subfigure}[b]{\textwidth}
    \includegraphics[scale=0.5]{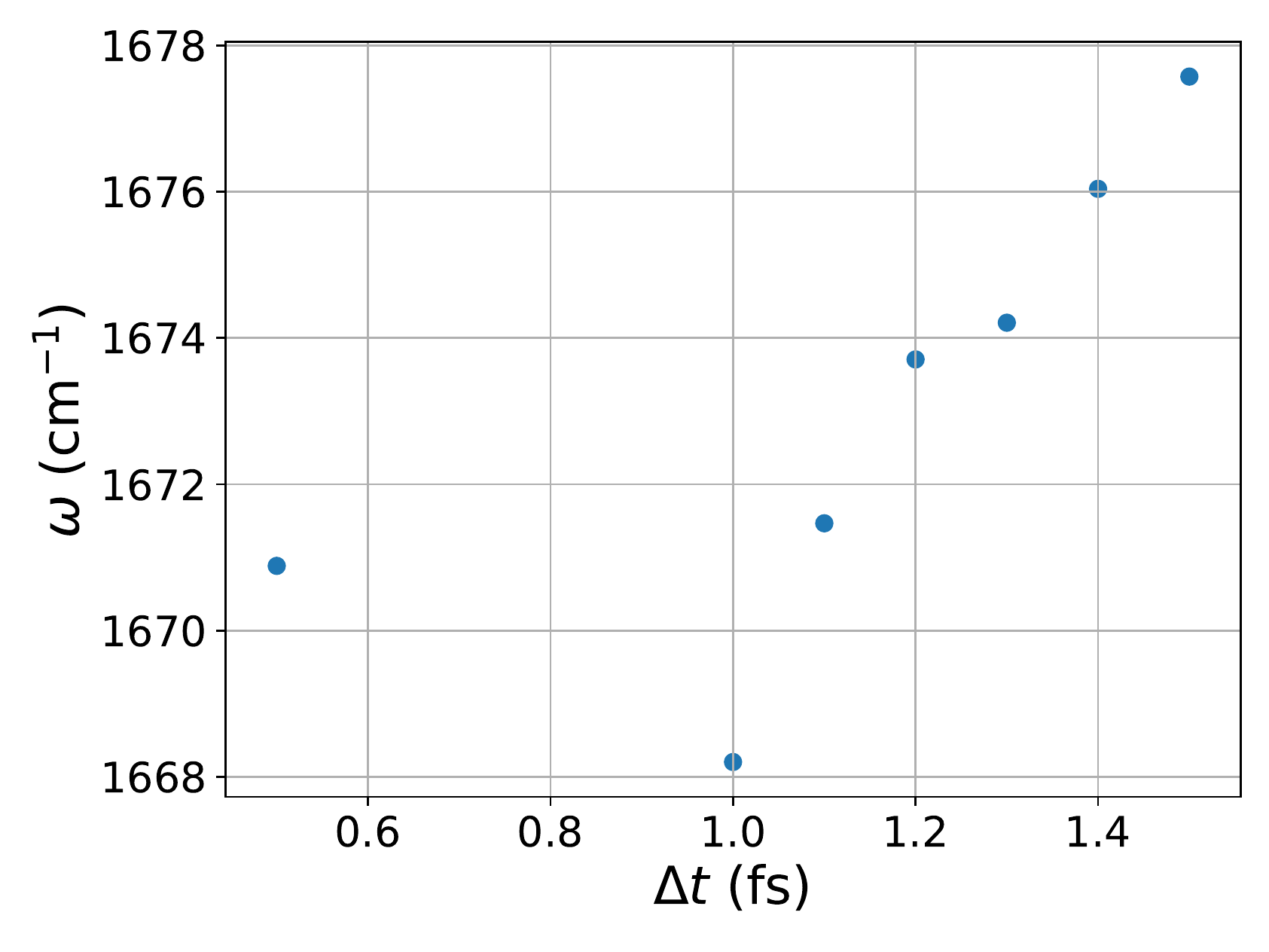} 
    \caption{}\label{nu2}
    \end{subfigure}
    \begin{subfigure}[b]{\textwidth}
    \includegraphics[scale=0.5]{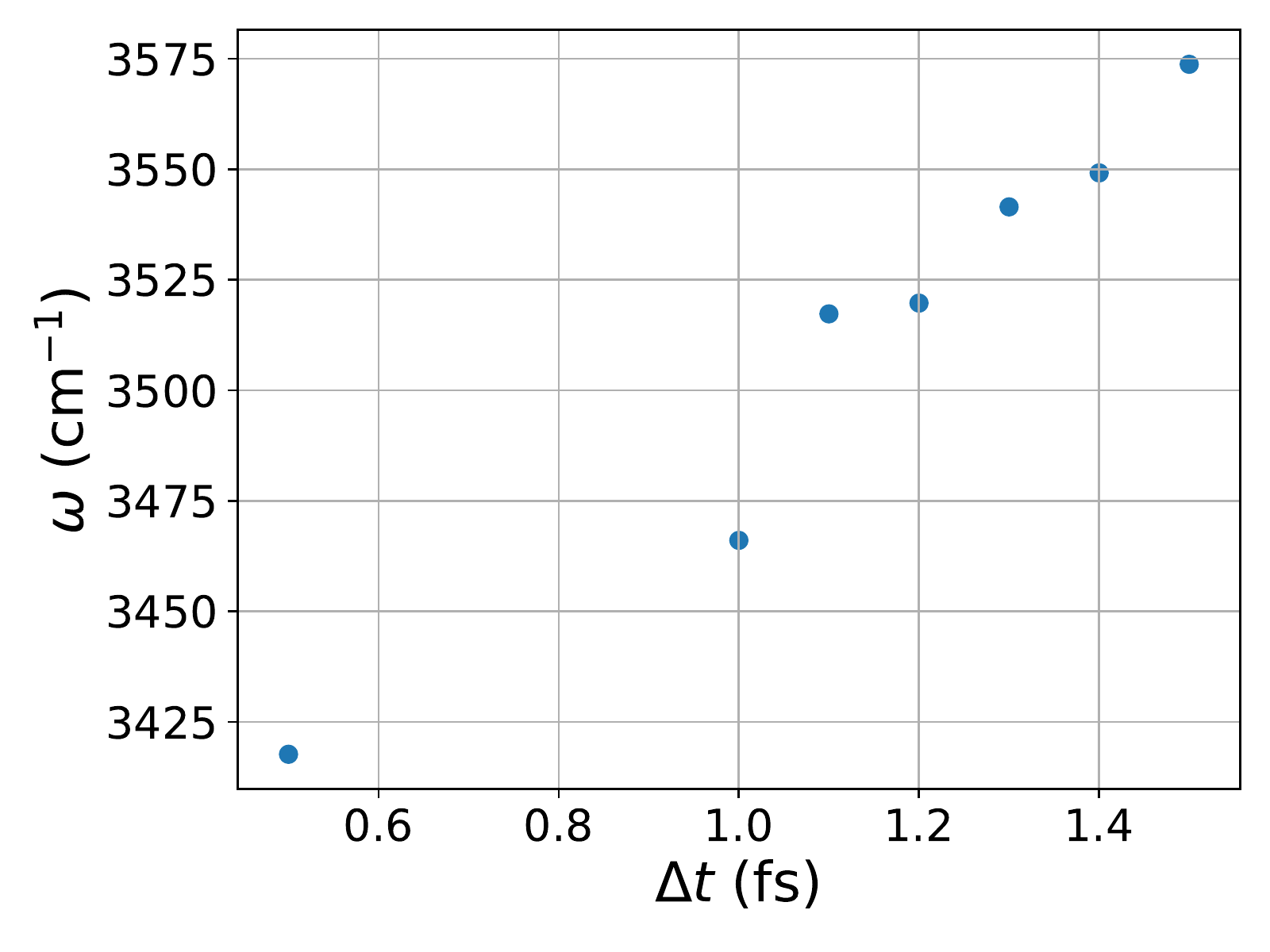}
    \caption{}\label{nu3}
    \end{subfigure}
        \caption{\textbf{(a)}: The vibrational spectra of water, simulated with a 0.5 fs timestep (blue) and a 1.5 fs timestep (red). \textbf{(b)}: The resonance peak of the $\nu_2$ (HOH angle bending) mode as time a function of simulation timestep. \textbf{(c)}: The resonance peak of the $\nu_{1+3}$ (stretching) modes as a function of the timestep. A large redshift is evident as the timestep increases.}
        \label{supp-fig:convtimestep}
\end{figure}

\begin{figure}
    \centering
    \includegraphics[scale=0.5]{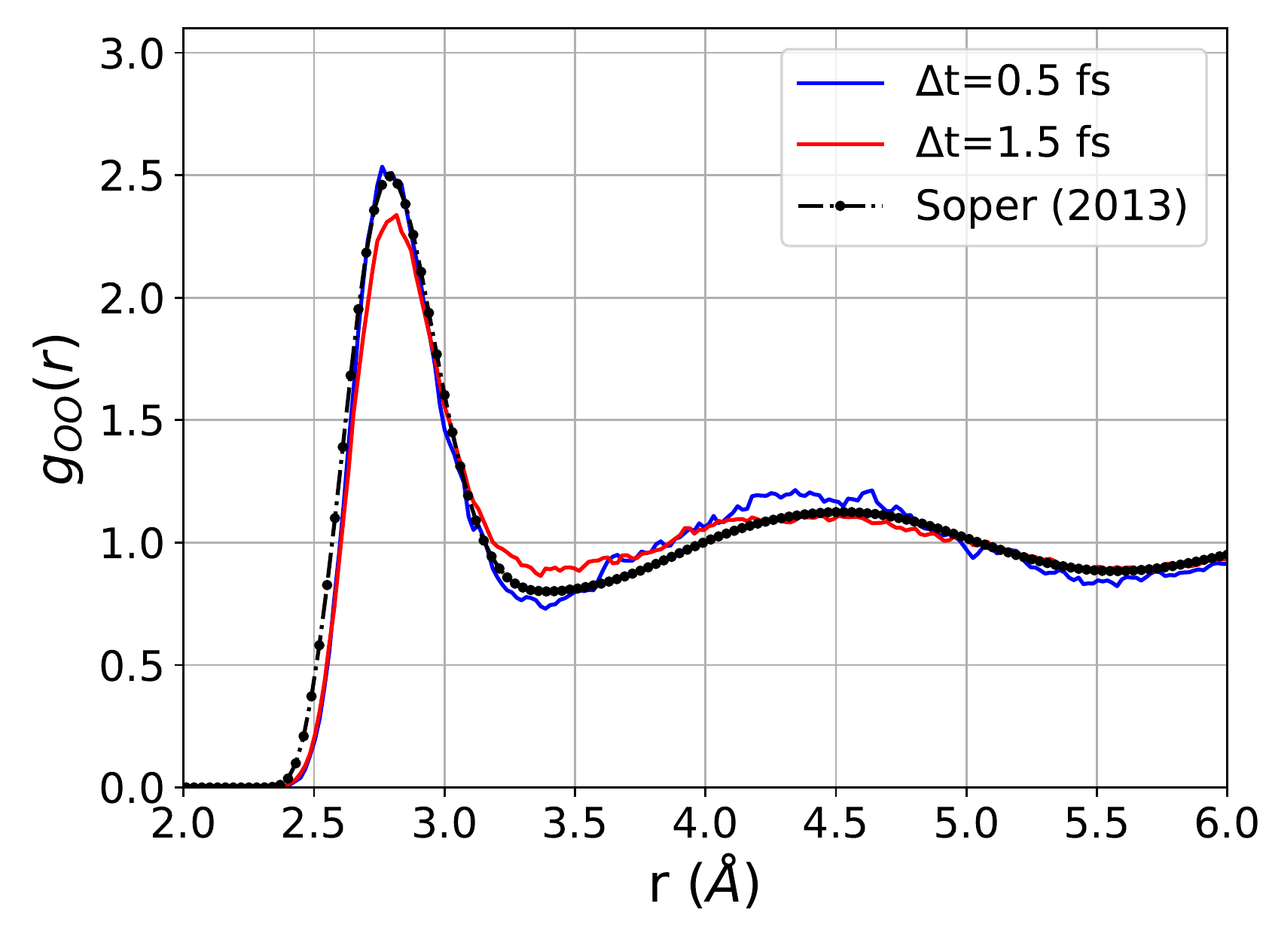}
    \caption{The \ce{O-O} radial distribution functions for pure water simulated with different timesteps at the same mass density, plotted alongside experimental results. The first shell of the 1.5 fs simulation is clearly destructured compared to experiment, while the 0.5 fs shell accurately captures the first shell peak.}
    \label{supp-fig:dt_goo}
\end{figure}

\begin{figure}
    \centering
    \includegraphics[scale=0.5]{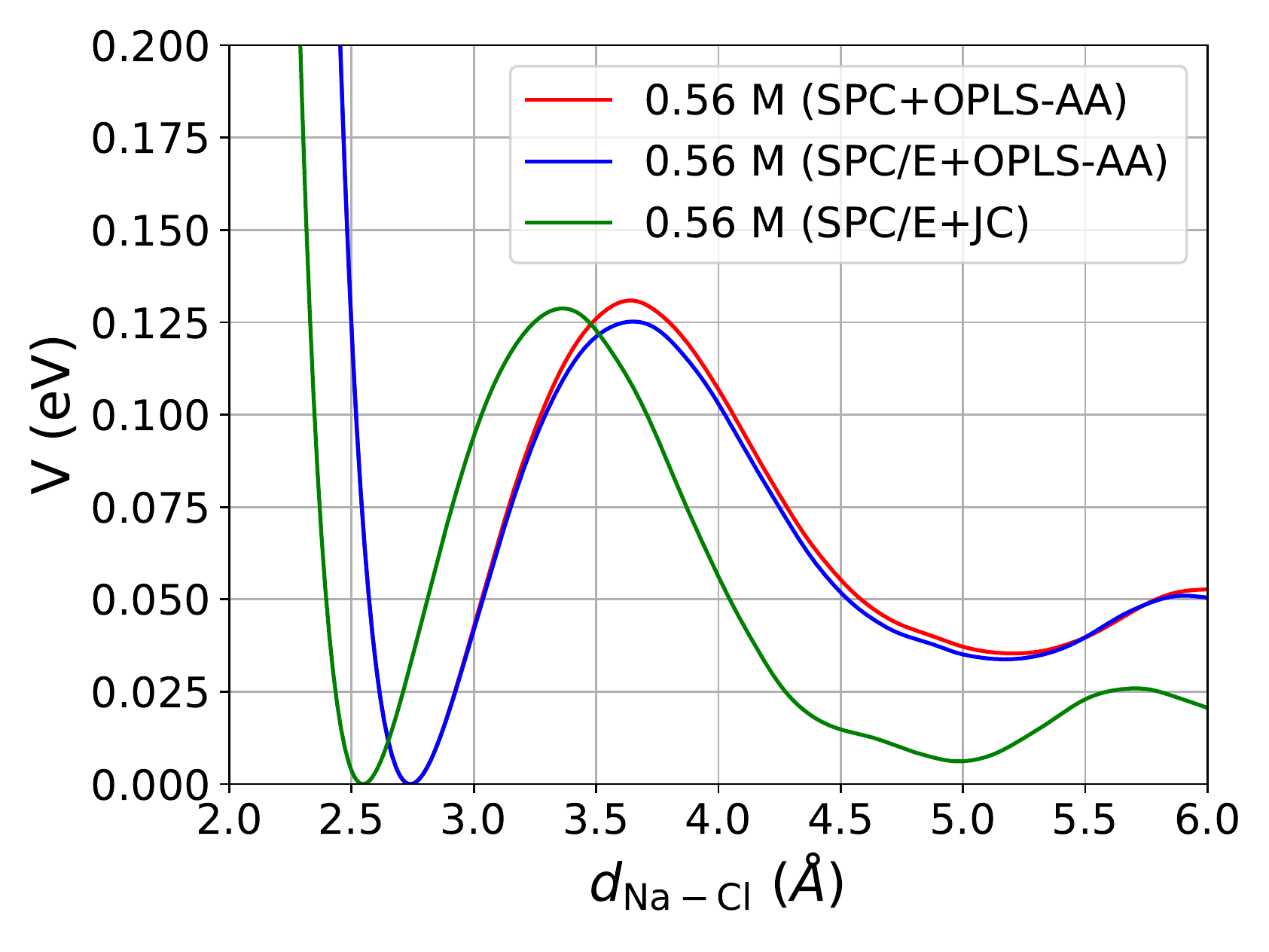}
    \caption{PMFs generated with different parameter choices in 0.56 M solution. There is no siginificant difference in the predicted behavior between SPC/E and SPC solvents with the same ion parameters. Note the enhanced stability of the SPC/E (blue) and SPC (red) SSIP states when compared to those in main body TIP4P simulations. Moreover, changing from the OPLS-AA parameters in the SPC/E solvent to the JC parameters (green) further enhances the stability of the SSIP, and affects the CIP location as well.}
    \label{supp-fig:gmxdcspc}
\end{figure}

\begin{figure}
	\centering
	\includegraphics[scale=0.5]{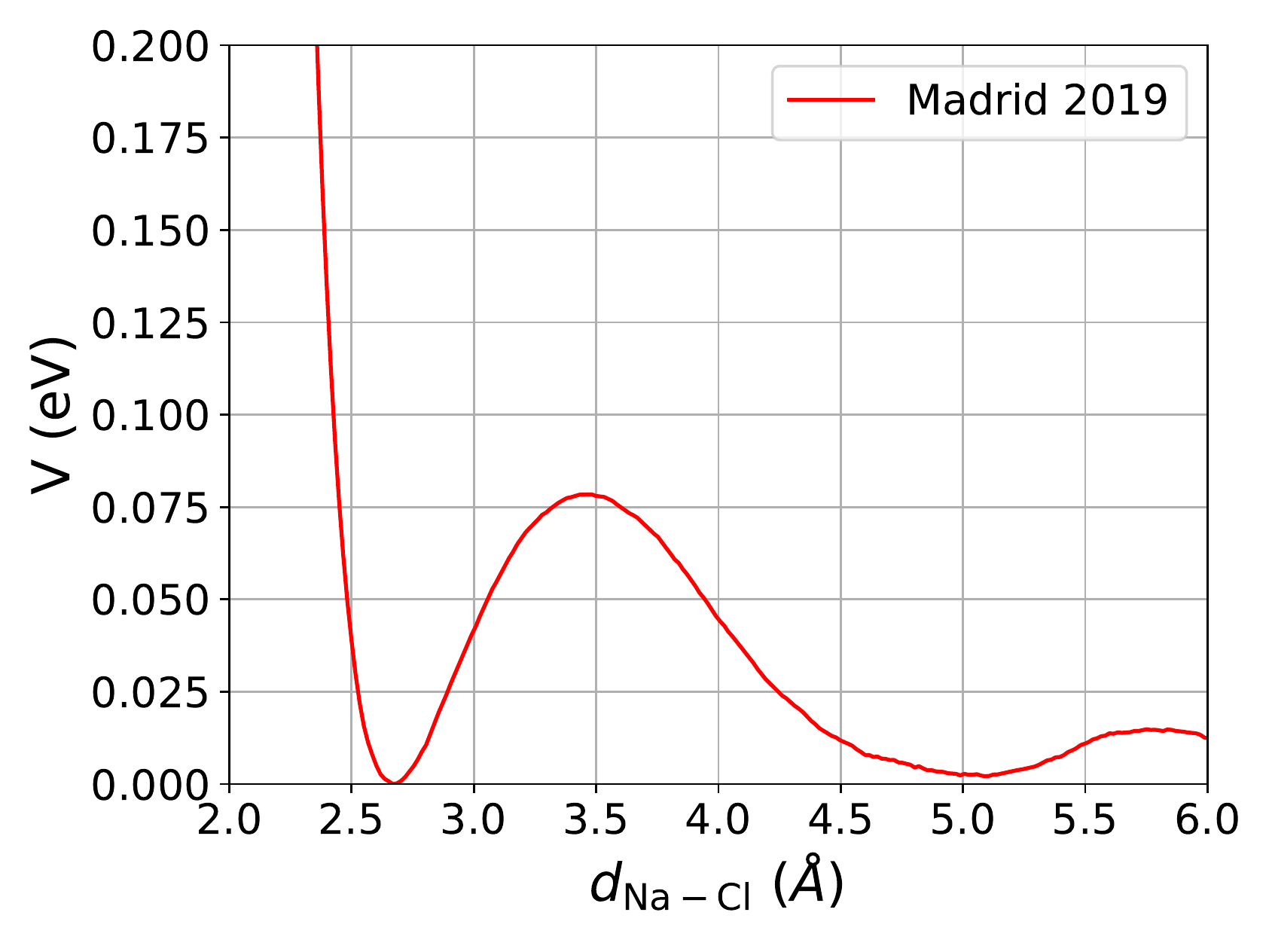}
	\caption{The PMF generated using ion parameters from a partial-charge classical force field\cite{madrid19_Zeron2019} using TIP4P/2005\cite{t4p05Abascal2005} solvent molecules. Of note is the almost equal stability of the CIP and SSIP states.}
	\label{supp-fig:naclmadrid}
\end{figure}

\begin{figure}
	\centering
	\includegraphics[scale=0.5]{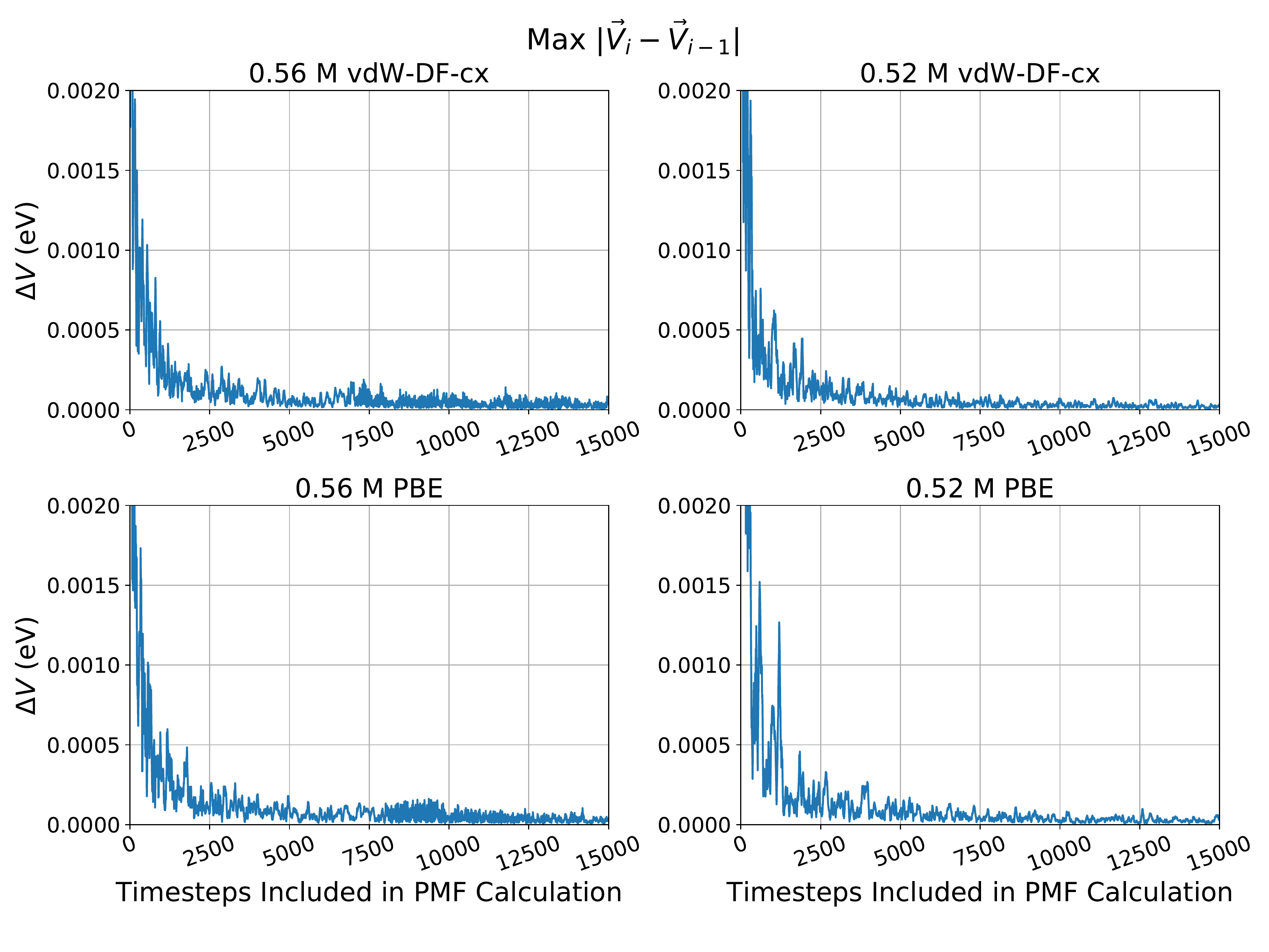}
	\caption{Convergence figures for the PMFs of the \textit{ab initio} systems. The maximum of the absolute error between successive PMF estimates is taken as a convergence parameter.}
	\label{supp-fig:aimdconv}
\end{figure}

\begin{figure*}
    \centering
    \begin{subfigure}[b]{0.32\textwidth}
    \includegraphics[scale=0.32]{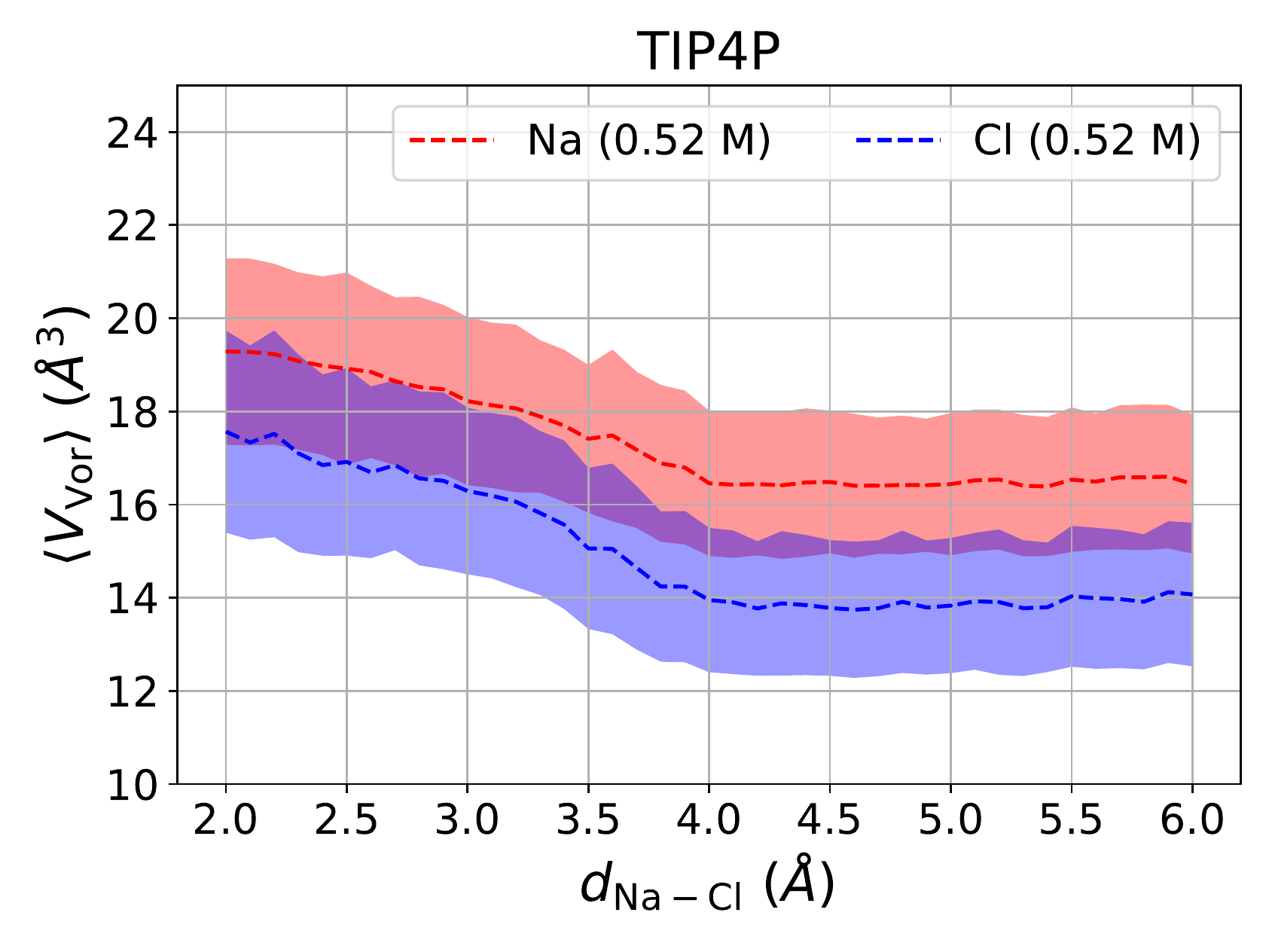}
        \caption{}
        \label{supp-fig:t4pvol52}
    \end{subfigure}
    \begin{subfigure}[b]{0.32\textwidth}
    \includegraphics[scale=0.32]{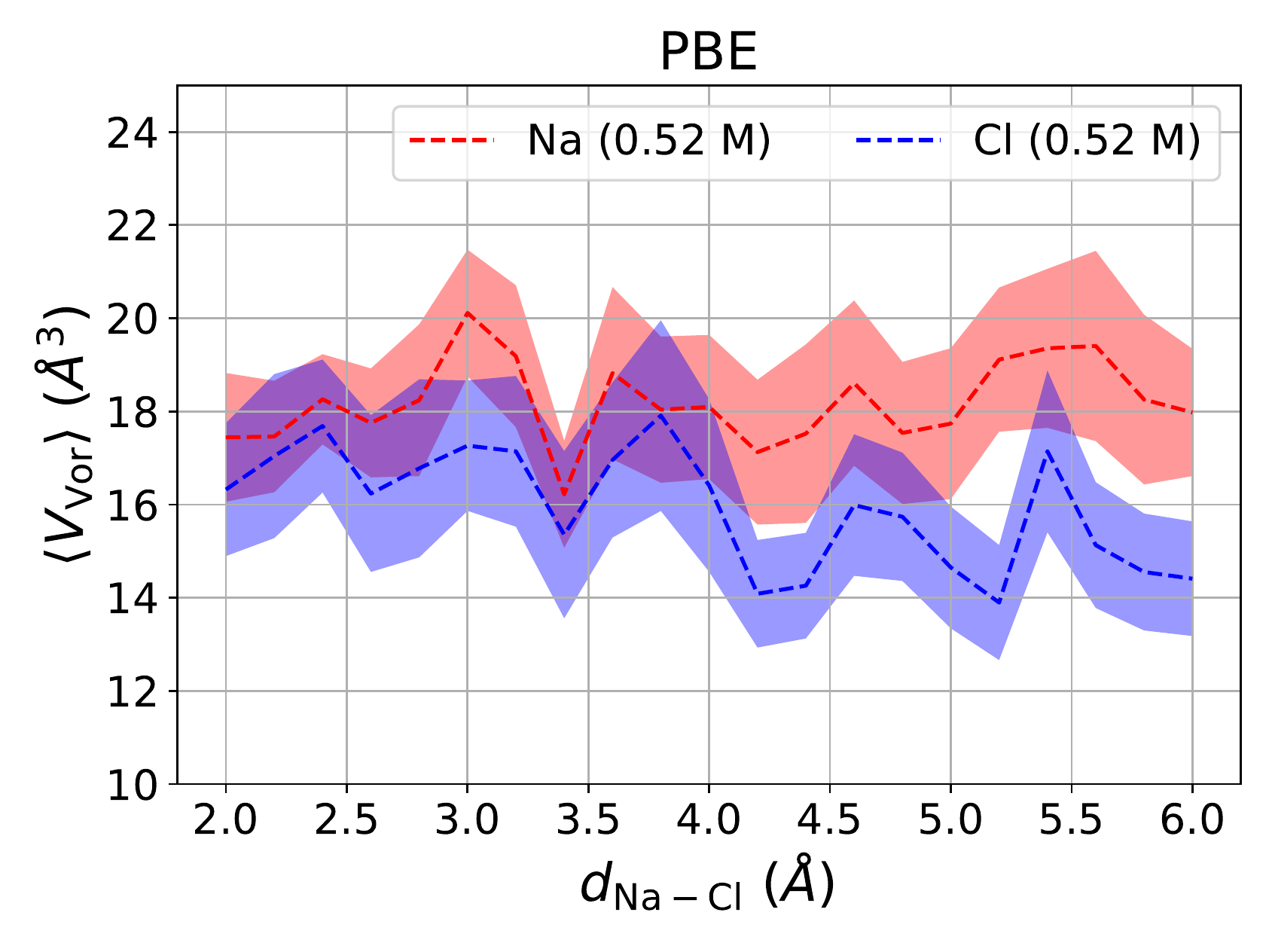}
        \caption{}
        \label{supp-fig:pbevol52}
    \end{subfigure}
    ~ 
    \begin{subfigure}[b]{0.32\textwidth}
    \includegraphics[scale=0.32]{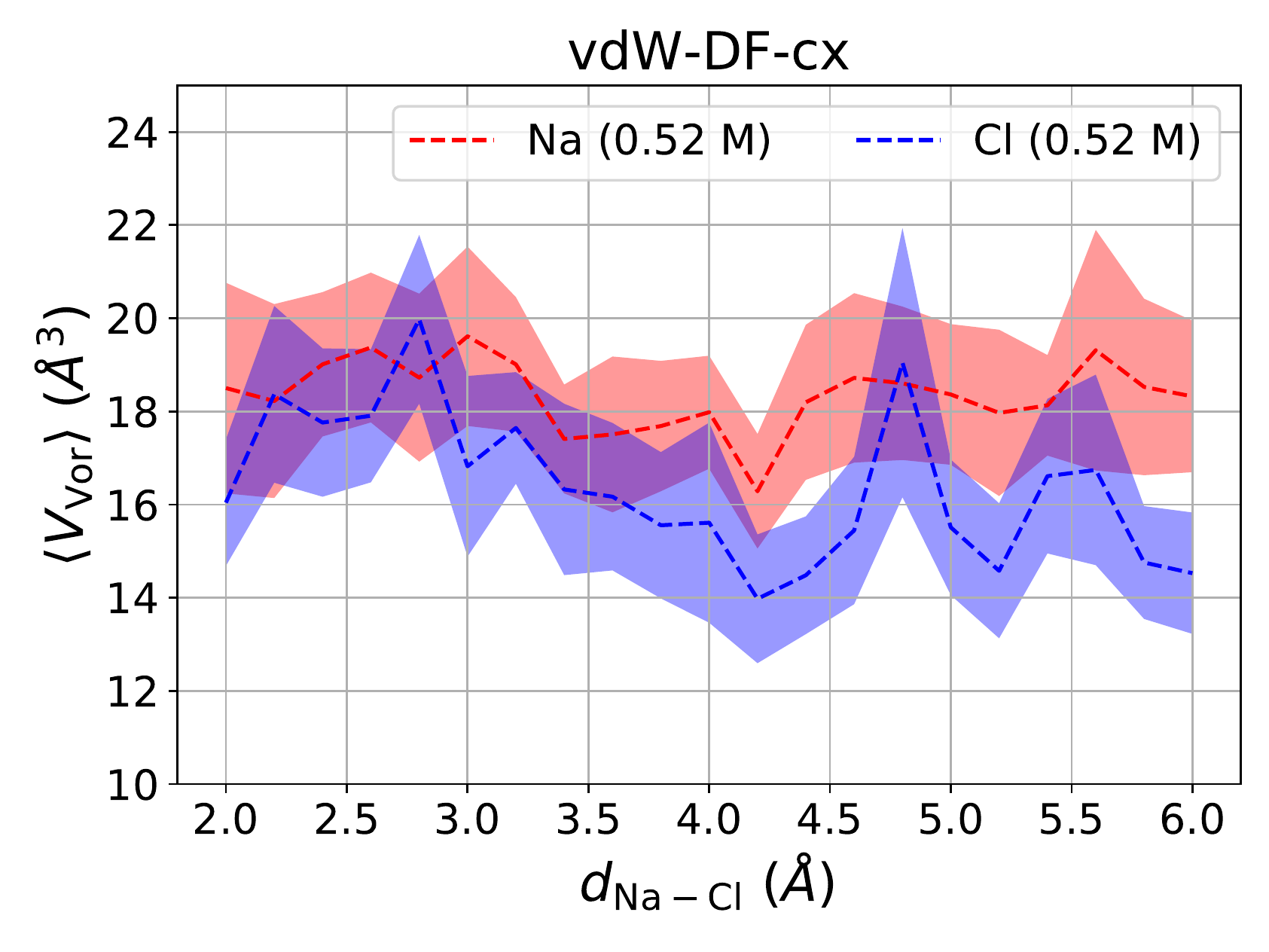}
        \caption{}
        \label{supp-fig:bhvol52}
    \end{subfigure}
    \caption{\textbf{(a-c)} Voronoi volumes as a function of ionic distance for the 0.52 M TIP4P, PBE, vdW-DF-cx simulations. The red and blue lines correspond to Voronoi volume averages for Na and Cl, respectively. Average \ce{H2O} volumes are constant across the simulations and are not plotted, but are presented in Table~\ref{supp-tbl:vorvol}.}\label{supp-fig:vorvols52}
\end{figure*}

